# Dating the fall of Babylon and Ur

*Abstract. The traditional date of 1595 BCE for the destruction of Babylon by the Hittite king Mursili I is accepted by most historians for many years despite notable controversies. This pivotal date is considered crucial to the various calculations of the early chronology of the ancient Near East. According to the Venus Tablet (Enuma Anu Enlil 63) which describes the rising and setting of Venus during the reign of Ammisaduqa, there are only four possibilities implying four dates for the destruction of Ur: 1912, 1944, 2004, 2064 BCE. However, a tablet of astronomical omens (Enuma Anu Enlil 20) mentions a lunar eclipse, dated 14/III/48, at the end of the reign of Šulgi and another (Enuma Anu Enlil 21) mentions a lunar eclipse, dated 14/XII/24, at the end of the reign of Ibbi-Sin. Over the period 2200-1850 BCE there are only three pairs of eclipses, spaced by 42 years, matching the description of astronomical omens but only one agreeing with the previous four dates. Despite the excellent agreement the date of 1499 BCE is considered too low compared to Kassite and Hittite chronologies.*

*The second way to check the date for the fall of Babylon is to rebuild the chronology of this period thanks to synchronisms dated by astronomy from Assyrian, Babylonian, Egyptian and Israelite chronologies which provide synchronisms that can be dated independently. The Mesopotamian chronology of this period is reconstructed using the number of Assyrian eponyms (one a year) and the length of Babylonian reigns combined with the set of synchronisms among Assyrian and Babylonian kings in Annals. Consequently, the reign of Kassite King Gandaš (1661-1635), obtained from average durations, coincides with the reign of the Assyrian king Samsu-iluna (1654-1616) and Sealand king Ilum-maz-ilî (1664-1594). In addition, the reign of Kassite King Agum II (1503-1487) and Sealand King Ayadaragalama (1498-1482) are consistent with a fall of Babylon in 1499 BCE. During the reign of Aššur-dân I (1179-1133) eponyms began on 1st Nisan instead of 1 Ṣippu, but Assyrian lunar years without intercalation remained the norm until Tiglath-pileser I. As the Babylonian year began on 1st Nisan (shortly after the spring equinox), Assyrian years coincided with Babylonian lunar years with intercalation, thus the period between Aššur-dân I and Tiglath-pileser I was therefore transitional.*

*Owing to the Babylonian chronology and synchronisms it is possible to date other chronologies (Egyptian, Elamite, Hittite and Mitannian). As lunar day 1, called psḏntyw "shining ones", has played a major role in Egyptian religious celebrations, it is regularly quoted in ancient documents, which sometimes also date it in the civil calendar. This double-dating then allows an absolute dating, on condition that provided proper identification of the moon phase for that particular day. Present specialists rely on the work of Parker (in 1950) who defined this day as a first invisibility, that is to say the day (invisible!) just before the first lunar crescent. However in the papyrus Louvre 7848 containing a double date, lunar and civil, in the year 44 of Amasis, the first date (II Shemu 13) is lunar and the second (I Shemu 15) is civil and as the civil date fell on 21 September 558 BCE the lunar date fell on 9 (= 21 – 12) September 558 BCE which was a full moon day according to astronomy, not first invisibility! The lunar calendar at Elephantine with its system of double dates (Egyptian and Babylonian) used by Persians officials and Jewish scribes from 500 to 400 BCE confirms that the Egyptian lunar day 1 was a full moon (see Dating the Reign of Xerxes and Artaxerxes).*

The date of the sack of Babylon by the Hittite king Mursili I is considered crucial to the various calculations of the early chronology of the ancient Near East[1]. According to the Venus Tablet, there are only four possible dates for the sack of Babylon. This astronomical tablet (*Enuma Anu Enlil 63*)[2], copied in 7th century BCE, describes the rising and setting of Venus during the reign of Ammisaduqa (a descendant of Hammurabi):

| |
|---|
| Year 1 inferior Venus sets on Shabatu 15 and after 3 days rises on Shabatu 18 |
| Year 2 superior Venus vanishes E. on Arahsamnu 21 and after 1 month 25 days appears W. on Tebetu 16 |
| Year 3 inferior Venus sets on Ululu 29 and after 16 days rises on Tashritu 15 |
| Year 4 superior Venus vanishes E. on Dumuzi 3 and after 2 months 6 days appears W. on Ululu 9 |
| Year 5 inferior Venus sets on Nisan 29 and after 12 days rises on Ayar 11 |
| Year 5 superior Venus vanishes E. on Kislimu 27 and after 2 months 3 days appears W. on Shabatu 30 |
| Year 6 inferior Venus sets on Arahsamnu 28 and after 3 days rises on Kislimu 1 |
| Year 7 superior Venus vanishes E. on Abu 30 and after 2 months appears W. on Tashritu 30 |
| Year 8 inferior Venus sets on Dumuzi 9 and after 17 days rises on Dumuzi 26 |
| Year 8 superior Venus vanishes E. on Adar 27 and after 2 months 16 days appears W. on Simanu 13 |
| Year 9 inferior Venus sets on Adar 12 and after 2 days rises on Adar 14 |
| Year 10 superior Venus vanishes E. on Arahsamnu 17 and after 1 month 25 days appears W. on Tebetu 12 |
| Year 11 inferior Venus sets on Ululu 25 and after 16 days rises on II Ululu 11 |
| Year 12 superior Venus vanishes E. on Ayar 29 and after 2 months 6 days appears W. on Abu 5 |
| Year 13 inferior Venus sets on Nisan 25 and after 12 days rises on Ayar 7 |
| Year 13 superior Venus vanishes E. on Tebetu 23 and after 2 months 3 days appears W. on Adar 26 |
| Year 14 inferior Venus sets on Arahsamnu 24 and after 3 days rises on Arahsamnu 27 |
| Year 15 superior Venus vanishes E. on Abu 26 and after 2 months appears W. on Tashritu 26 |
| Year 16 inferior Venus sets on Dumuzi 5 and after 16 days rises on Dumuzi 21 |
| Year 16 superior Venus vanishes E. on Adar 24 and after 2 months 15 days appears W. on Simanu 9 |
| Year 17 inferior Venus sets on Adar 8 and after 3 days rises on Adar 11 |
| Year 18 superior Venus vanishes E. on Arahsamnu 13 and after 1 month 25 days appears W. on Tebetu 8 |
| Year 19 inferior Venus sets on II Ululu 20 and after 17 days rises on Tashritu 8 |
| Year 20 superior Venus vanishes E. on Simanu 25 and after 2 months 6 days appears W. on Ululu 1 |
| Year 21 inferior Venus sets on Nisan 22 and after 11 days rises on Ayar 3 |
| Year 21 superior Venus vanishes E. on Tebetu 19 and after 2 months 3 days appears W. on Adar 22 |

Although the interpretation of this astronomical tablet is difficult[3], because many data appear to have been poorly copied, the fall of Babylon can be dated to the period 1500-1700 only according to four possibilities[4]:

| Chronology (BCE): | Ultra-Low | Low | Middle | High |
|---|---|---|---|---|
| Fall of Ur | 1912 | 1944 | 2004 | 2064 |
| Reign of Hammurabi | 1697-1654 | 1729-1686 | 1793-1750 | 1849-1806 |
| Reign of Ammisaduqa | 1551-1530 | 1583-1562 | 1647-1626 | 1703-1682 |
| Fall of Babylon | 1499 | 1531 | 1595 | 1651 |

The date 1595 is chosen mainly as it is consistent with the chronology accepted by most historians to the late 20th century, hence the name of "Middle chronology". However, other lunar eclipses are used for dating the fall of Babylon[5]. A tablet of astronomical omens (*Enuma Anu Enlil 20*) mentions a lunar eclipse, dated 14 Siwanu, at the end of the reign of

[1] R. PRUZSINSKY – Mesopotamian Chronology of the 2nd Millennium B.C.
Wien 2009 Ed. Österreichischen Akademie der Wissenschaften pp. 17-104.
[2] E. REINER, D. PINGREE – Babylonian Planetary Omens. Part 1. The Venus Tablet of Ammisaduqa
Malibu 1975 Ed. Undena Publications pp. 17-62.
[3] http://arxiv.org/pdf/physics/0311036
V.G. GURZADYAN – The Venus Tablet and Refraction
in: *Akkadica 124* (2003) pp. 13-17.
[4] V.G. GURZADYAN – On the Astronomical Records and Babylonian Chronology
in: *Akkadica* 119-120 (2000) pp. 175-184.
[5] B. BANJEVIC – Ancient Eclipses and Dating the Fall of Babylon
in: *Publ. Astron. Obs. Belgrade* N° 80 (2006) pp. 251-257.

Šulgi (14/III/**48**) and another (*Enuma Anu Enlil 21*) mentions a lunar eclipse, dated 14 Addaru, at the end of the Ur III dynasty ending with the reign of Ibbi-Sin (14/XII/**24**). These two lunar eclipses are separated by 42 years of reign (= 9 years of Amar-Sin + 9 years of Šu-Sîn + 24 years of Ibbi-Sin). Moreover, in a tablet of Mari, a scribe mentions a [total] lunar eclipse during the eponymy of Asqudum[6] (= year 12/13 of Hammurabi).

Over the period 2200-1850 there are only three pairs of eclipses, spaced by 42 years, matching the description of astronomical omens[7]:

| 1st eclipse (14/III/**48**) | Magnitude | 2nd eclipse (14/XII/**24**) | Magnitude | gap (1st - 2nd) | |
|---|---|---|---|---|---|
| 13/08/2189 | 1.21 | 12/03/2107 | 1.00 | 82 years | |
| 12/05/2175 | 1.80 | " | " | 68 years | |
| 04/07/2150 | 1.32 | " | " | 43 years | ≈ |
| 25/07/2095 | 1.32 | 04/05/2063 | 1.78 | 32 years | |
| " | " | 13/04/2053 | 0.63 | **42** years | **OK** |
| " | " | 11/02/2031 | 1.14 | 64 years | |
| 26/06/2019 | 1.07 | 24/04/2016 | 1.84 | 3 years | |
| " | " | 15/03/1977 | 0.82 | **42** years | **OK** |
| " | " | 04/03/1976 | 1.47 | 43 years | ≈ |
| 25/05/2008 | 0.96 | 15/04/1969 | 1.84 | 39 years | |
| 18/07/2002 | 1.08 | 23/02/1929 | 1.63 | 73 years | |
| 27/06/1954 | 1.39 | 16/03/1912 | 0.58 | **42** years | **OK** |
| 18/07/1937 | 0.75 | 14/02/1901 | 0.94 | 36, years | |
| 18/05/1915 | 1.47 | 14/02/1882 | 1.58 | 33 years | |
| " | " | 27/03/1875 | 1.82 | 40 years | |
| 28/06/1908 | 1.04 | " | " | 33 years | |
| 19/06/1861 | 1.04 | | | | |
| 31/07/1854 | 0.73 | | | | |

| Lunar eclipse matching the: | | | Fall of Babylon | Chronology |
|---|---|---|---|---|
| Last year of Šulgi (14/III/**48**) | Fall of Ur III (14/XII/**24**) | Year 12/13 of Hammurabi | According to Venus Tablet | |
| [2106]# | [2064]# | [1836]# | 1651 | High |
| 25/07/2095 | 13/04/2053 | | # | |
| [2046]# | [2004]# | [1780]# | 1595 | Middle |
| 26/06/2019 | 15/03/1977 | | # | |
| [1986]# | [1944]# | 03/09/1716 | 1531 | Low |
| 27/06/1954 | 16/03/1912 | 07/12/1684 | **1499** | Ultra-Low |

Despite the excellent agreement among all these astronomical data[8], the date of 1499 is considered too low compared to Kassite and Hittite chronologies. This criticism is unfounded, because these two chronologies are very approximate: most durations of reigns are unknown and they have no anchor in astronomy. In addition, the dendrochronological dating of the Acemhöyüke palace requires to locate the death of Šamši-Adad I after -1752[9] eliminating the Middle Chronology dating this reign 1807-1775.

The second way to set a date for the fall of Babylon is to rebuild the chronology of this period thanks to synchronisms dated by astronomy. Assyrian, Babylonian, Egyptian and Israelite chronologies provide synchronisms that can be dated independently. For example, Assyrian chronology may rebuilt for the period 911-609 only thanks to eponyms. The list of Assyrian eponyms is anchored on the solar eclipse occurred on Simanu (month III, day 30) in the eponymy of Bur-Sagale (dated June 15, 763 BCE). The Assyrian period 911-648 is dated owing to its canonical eponyms[10] and the period 648-609 by a prosopography of its eponyms[11]. A few eponyms are non canonical because they died during the year of their eponymy and there are also some gaps of 1 year between eponym dates and regnal years in tablet with double dates because the first Assyrian regnal year (accession) was reckoned in either system: year 0 (Babylonian) or year 1 (Assyrian). Thus, as there are exactly 154 canonical eponyms between Gargamisaiu and Bur-Sagale, which is dated 763 BCE, that involves to date the one of Gargamisaiu into 609 (= 763 – 154).

[6] W. HEIMPEL – Letters to the King of Mari: A New Translation, With Historical Introduction, Notes, and Commentary Leiden 2003 Ed. Eisenbrauns pp. 209-210.
[7] P.J. HUBER – Astronomy and Ancient Chronology in: *Akkadica* 119-120 (2000) pp. 159-176.
[8] H. GASCHE – La fin de la première dynastie de Babylone : une chute difficile in: *Akkadica* 124 (2003) pp. 205-221.
[9] C. MICHEL, P. ROCHER – La chronologie du IIe millénaire revue à l'ombre d'une éclipse de soleil in: *Jaarbericht (...) Ex Oriente Lux* N° 35/36 (1997-2000) Chicago pp. 111-126.

The only solar eclipse over Assyria during the period 800-750 is the total eclipse dated June 15, 763 BCE. The partial solar eclipses dated June 4, 800 BCE and June 24, 791 BCE were not able to be viewed over Assyria.

- The fall of the Assyrian empire, which took place in October 609 BCE after the battle of Harran, is characterized by a quadruple synchronisms, since the year of Assur-uballit II corresponds to year 17 of Nabopolassar to Josiah's year 31 and year 1 of Necho II.
- According to the biography of Adad-Guppi[12], mother of Nabonidus, Nabopolassar reigned 21 years, then Nebuchadnezzar 43 years, Amel-Marduk 2 years, Neriglissar 4 years just before Nabonidus. According to the Hillah's stele[13] there were 54 years between the destruction of the temple of Sin, in Harran, and the beginning of the reign of Nabonidus. According to a Babylonian chronicle (BM 21901)[14] and Adad-Guppi's stele, the temple of Harran was destroyed in the year 16 of Nabopolassar.
- Dated lunar eclipses[15] are: year 1 and 2 of Merodachbaladan (March 19/20 721 BCE, March 8/9 and September 1/2 720 BCE); year 5 of Nabopolassar (April 21/22 621 BCE); year 2 of Šamaš-šuma-ukîn (April 10/11 666 BCE); year 42 of Nebuchadnezzar (March 2/3 562 BCE). A diary (VAT 4956)[16] contains numerous astronomical conjunctions in years 37 and 38 of Nebuchadnezzar dated from astronomy in 568 and 567 BCE. An astronomical journal (BM 38462)[17] list some lunar eclipses in the years 1 to 27 of Nebuchadnezzar which are dated from 604 to 578 BCE.

The chronology of the Saite period (663-525) may be reckoned only thanks to "biographies[18] of prominent men or Apis bulls":

1. *Grave stele of Psammetichus son of Genefbahorek*. Date of birth: Year 3 of Necho II, month 10, day 1. Date of death: Year 35 of Amasis, month 2, day 6. Length of life: 71 years, 4 months, 6 days (see column A. 1st Thot matches the beginning of Egyptian year).

[10] S. PARPOLA – Assyrian Chronology 681-648 BC.
in: Letters from Assyrian Scholars to the Kings Esarhaddon and Assurbanipal Part II Winona Lake 2007 Ed. Eisenbrauns pp. 381-430.
[11] S. PARPOLA – The Prosopography of the Neo-Assyrian Empire
Helsinki 1998 University of Helsinki pp. XVIII-XX.
[12] J.B. PRITCHARD - Ancient Near Eastern Texts
Princeton 1969 Ed. Princeton University Press p. 560-561.
[13] P.A. BEAULIEU – The Reign of Nabonidus, King of Babylon 556-539 B.C.
in: *Yale Near Eastern Research* 10 (1989) n°2.
[14] J.J. GLASSNER – Chroniques mésopotamiennes n°22
Paris 1993 Éd. Belles Lettres pp. 193-197.
[15] F.R. STEPHENSON - Historical Eclipses and Earth's Rotation
Cambridge 1997 Ed. Cambridge University Press pp. 99-100, 151-152, 206.
[16] A.J. SACHS, H. HUNGER - Astronomical Diaries and Related Texts from Babylonia vol. I
Wien 1988 Ed. Akademie der Wissenschaften (n° -567).
[17] H. HUNGER - Astronomical Diaries and Related Texts from Babylonia vol. V n° 6
Wien 2001 Ed. Akademie der Wissenschaften pp. 27-30,396.
[18] H. GAUTHIER – Le livre des rois d'Égypte
Le Caire 1915 Éd. Institut Français d'Archéologie Orientale pp. 74, 87-88, 92-93, 106, 115, 119.
F.K. KIENITZ – Die politische Geschichte Ägyptens vom 7. bis zum 4. Jahrhundert vor der Zeitwende
Berlin 1953 Ed. Akademie-Verlag pp. 154-156.
J.H. BREASTED – Ancient records of Egypt: Historical documents from the earliest times to the Persian conquest. Vol. IV
Chicago 1906 (1962) Ed. The University of Chicago Press pp. 497-498, 501-505, 518-520.

2. *Grave stele of the priest Psammetichus son of Iahuben*. Date of birth: Year 1 of Necho II, month 11, day 1. Date of death: Year 27 [of Amasis], month 8, day 28. Length of life: 65 years, 10 months, 2 days (see column B).
3. *Grave stele of the 4th Apis of the 26th Dynasty*. Date of birth: Year 16 of Necho II, month 2, day 7. Installation: Year 1 of Psammetichus II, month 11, day 9. Date of death: Year 12 of Apries, month 8, day 12. Date of burial: Year 12 of Apries, month 10, day 21. Length of life: 17 years, 6 months, 5 days (see column C).
4. *Grave stele of the 3rd Apis of the 26th Dynasty*. Date of birth: Year 53 of Psammetichus I, month 6, day 19. Installation: Year 54 of Psammetichus I, month 3, day 12. Date of death: Year 16 of Necho II, month 2, day 6. Date of burial: Year 16 of Necho II, month 4, day 16. Length of life: 16 years, 7 months, 17 days (see column D).
5. *Epitaph of Apis bull from Cambyses*[19]. Date of birth: Year 27 [of Amasis]. Date of death: Year 6 of Cambyses II. Length of life unknown, but the average life-span of Apis bulls is from 16 to 19 years during the 26th Dynasty[20] (see column E).
6. Pharaoh Apries was still living according to a stele[21] dated year 3 of Amasis (which was beginning on January 12, 567 BCE).

| Egyptian King | | | Year | 1st Thot | A | B | C | D | E |
|---|---|---|---|---|---|---|---|---|---|
| **Psammetichus I** | 1 | | 663 | 5-Feb | | | | | |
| | 2 | | **662** | 5-Feb | | | | | |
| | 3 | | **661** | 4-Feb | | | | | |
| | 4 | | **660** | 4-Feb | | | | | |
| | 5 | | **659** | 4-Feb | | | | | |
| | 6 | | **658** | 4-Feb | | | | | |
| | 7 | | **657** | 3-Feb | | | | | |
| | 8 | | **656** | 3-Feb | | | | | |
| | 9 | | **655** | 3-Feb | | | | | |
| | 10 | | **654** | 3-Feb | | | | | |
| | 11 | | **653** | 2-Feb | | | | | |
| | 12 | | **652** | 2-Feb | | | | | |
| | 13 | | **651** | 2-Feb | | | | | |
| | 14 | | **650** | 2-Feb | | | | | |
| | 15 | | **649** | 1-Feb | | | | | |
| | 16 | | **648** | 1-Feb | | | | | |
| | 17 | | **647** | 1-Feb | | | | | |
| | 18 | | **646** | 1-Feb | | | | | |
| | 19 | | **645** | 30-jan | | | | | |
| | 20 | | **644** | 31-jan | | | | | |
| | 21 | | **643** | 31-jan | | | | | |
| | 22 | | **642** | 31-jan | | | | | |
| | 23 | | **641** | 30-jan | | | | | |
| | 24 | | **640** | 30-jan | | | | | |
| | 25 | | **639** | 30-jan | | | | | |
| | 26 | | **638** | 30-jan | | | | | |
| | 27 | | **637** | 29-jan | | | | | |
| | 28 | | **636** | 29-jan | | | | | |
| | 29 | | **635** | 29-jan | | | | | |
| | 30 | | **634** | 29-jan | | | | | |
| | 31 | | **633** | 28-jan | | | | | |
| | 32 | | **632** | 28-jan | | | | | |

[19] A. KUHRT – The Persian Empire
London 2010 Ed. Routledge pp. 122-124.
[20] M. MALININE, G. POSENER, J. VERCOUTER – Catalogue des stèles du Sérapéum de Memphis I
Paris 1968 Éd. Imprimerie Nationale p. XIII.
[21] A. SPALINGER - Egypt and Babylonia: A Survey
Hamburg 1977, in: *Studien Zur Altägyptischen Kultur* Band 5 pp. 241-242.

| | | | | | | | | | |
|---|---|---|---|---|---|---|---|---|---|
| | 33 | | **631** | 28-jan | | | | | |
| | 34 | | **630** | 28-jan | | | | | |
| | 35 | | **629** | 27-jan | | | | | |
| | 36 | | **628** | 27-jan | | | | | |
| | 37 | | **627** | 27-jan | | | | | |
| | 38 | | **626** | 27-jan | | | | | |
| | 39 | | **625** | 26-jan | | | | | |
| | 40 | | **624** | 26-jan | | | | | |
| | 41 | | **623** | 26-jan | | | | | |
| | 42 | | **622** | 26-jan | | | | | |
| | 43 | | **621** | 25-jan | | | | | |
| | 44 | | **620** | 25-jan | | | | | |
| | 45 | | **619** | 25-jan | | | | | |
| | 46 | | **618** | 25-jan | | | | | |
| | 47 | | **617** | 24-jan | | | | | |
| | 48 | | **616** | 24-jan | | | | | |
| | 49 | | **615** | 24-jan | | | | | |
| | 50 | | **614** | 24-jan | | | | | |
| | 51 | | **613** | 23-jan | | | | | |
| | 52 | | **612** | 23-jan | | | | | |
| | **53** | | **611** | 23-jan | | | | 0 | |
| | 54 | | **610** | 23-jan | | | | 1 | |
| **Necho II** | **1** | | **609** | 22-jan | | 0 | | 2 | |
| | 2 | | **608** | 22-jan | | 1 | | 3 | |
| | **3** | | **607** | 22-jan | 0 | 2 | | 4 | |
| | 4 | | **606** | 22-jan | 1 | 3 | | 5 | |
| | 5 | | **605** | 21-jan | 2 | 4 | | 6 | |
| | 6 | | **604** | 21-jan | 3 | 5 | | 7 | |
| | 7 | | **603** | 21-jan | 4 | 6 | | 8 | |
| | 8 | | **602** | 21-jan | 5 | 7 | | 9 | |
| | 9 | | **601** | 20-jan | 6 | 8 | | 10 | |
| | 10 | | **600** | 20-jan | 7 | 9 | | 11 | |
| | 11 | | **599** | 20-jan | 8 | 10 | | 12 | |
| | 12 | | **598** | 20-jan | 9 | 11 | | 13 | |
| | 13 | | **597** | 19-jan | 10 | 12 | | 14 | |
| | 14 | | **596** | 19-jan | 11 | 13 | | 15 | |
| | 15 | | **595** | 19-jan | 12 | 14 | | 16 | |
| **Psammetichus II** | **16** | **1** | **594** | 19-jan | 13 | 15 | **0** **0** | **16y7m** | |
| | | 2 | **593** | 18-jan | 14 | 16 | 1 | | |
| | | 3 | **592** | 18-jan | 15 | 17 | 2 | | |
| | | 4 | **591** | 18-jan | 16 | 18 | 3 | | |
| | | 5 | **590** | 18-jan | 17 | 19 | 4 | | |
| | | 6 | **589** | 17-jan | 18 | 20 | 5 | | |
| **Apries** | 1 | 7 | **588** | 17-jan | 19 | 21 | 6 | | |
| | 2 | | **587** | 17-jan | 20 | 22 | 7 | | |
| | 3 | | **586** | 17-jan | 21 | 23 | 8 | | |
| | 4 | | **585** | 16-jan | 22 | 24 | 9 | | |
| | 5 | | **584** | 16-jan | 23 | 25 | 10 | | |
| | 6 | | **583** | 16-jan | 24 | 26 | 11 | | |
| | 7 | | **582** | 16-jan | 25 | 27 | 12 | | |
| | 8 | | **581** | 15-jan | 26 | 28 | 13 | | |
| | 9 | | **580** | 15-jan | 27 | 29 | 14 | | |
| | 10 | | **579** | 15-jan | 28 | 30 | 15 | | |
| | 11 | | **578** | 15-jan | 29 | 31 | 16 | | |
| | **12** | | **577** | 14-jan | 30 | 32 | **17y6m** | | |
| | 13 | | **576** | 14-jan | 31 | 33 | | | |
| | 14 | | **575** | 14-jan | 32 | 34 | | | |
| | 15 | | **574** | 14-jan | 33 | 35 | | | |
| | 16 | | **573** | 13-jan | 34 | 36 | | | |
| | 17 | | 572 | 13-jan | 35 | 37 | | | |
| | 18 | | 571 | 13-jan | 36 | 38 | | | |
| | 19 | | 570 | 13-jan | 37 | 39 | | | |
| **Amasis** | 20 | 1 | 569 | 12-jan | 38 | 40 | | | |
| | 21 | 2 | 568 | 12-jan | 39 | 41 | | | |
| | 22 | 3 | 567 | 12-jan | 40 | 42 | | | |
| | | 4 | 566 | 12-jan | 41 | 43 | | | |
| | | 5 | 565 | 11-jan | 42 | 44 | | | |
| | | 6 | 564 | 11-jan | 43 | 45 | | | |
| | | 7 | 563 | 11-jan | 44 | 46 | | | |
| | | 8 | 562 | 11-jan | 45 | 47 | | | |
| | | 9 | 561 | 10-jan | 46 | 48 | | | |
| | | 10 | 560 | 10-jan | 47 | 49 | | | |
| | | 11 | 559 | 10-jan | 48 | 50 | | | |
| | | 12 | 558 | 10-jan | 49 | 51 | | | |
| | | 13 | 557 | 9-jan | 50 | 52 | | | |
| | | 14 | 556 | 9-jan | 51 | 53 | | | |
| | | 15 | 555 | 9-jan | 52 | 54 | | | |
| | | 16 | 554 | 9-jan | 53 | 55 | | | |
| | | 17 | 553 | 8-jan | 54 | 56 | | | |
| | | 18 | 552 | 8-jan | 55 | 57 | | | |
| | | 19 | 551 | 8-jan | 56 | 58 | | | |
| | | 20 | 550 | 8-jan | 57 | 59 | | | |
| | | 21 | 549 | 7-jan | 58 | 60 | | | |
| | | 22 | 548 | 7-jan | 59 | 61 | | | |
| | | 23 | 547 | 7-jan | 60 | 62 | | | |
| | | 24 | 546 | 7-jan | 61 | 63 | | | |
| | | 25 | 545 | 6-jan | 62 | 64 | | | |
| | | 26 | 544 | 6-jan | 63 | 65 | | | |
| | | 27 | 543 | 6-jan | 64 | 65y10m | | | 0 |
| | | 28 | 542 | 6-jan | 65 | | | | 1 |
| | | 29 | 541 | 5-jan | 66 | | | | 2 |
| | | 30 | 540 | 5-jan | 67 | | | | 3 |
| | | 31 | 539 | 5-jan | 68 | | | | 4 |
| | | 32 | 538 | 5-jan | 69 | | | | 5 |
| | | 33 | 537 | 4-jan | 70 | | | | 6 |
| | | 34 | 536 | 4-jan | 71 | | | | 7 |
| | | 35 | 535 | 4-jan | 71y4m | | | | 8 |
| | | 36 | 534 | 4-jan | | | | | 9 |
| | | 37 | 533 | 3-jan | | | | | 10 |
| | | 38 | 532 | 3-jan | | | | | 11 |
| | | 39 | 531 | 3-jan | | | | | 12 |
| | | 40 | 530 | 3-jan | | | | | 13 |
| | | 41 | 529 | 2-jan | | | | | 14 |
| | | 42 | 528 | 2-jan | | | | | 15 |
| | | 43 | 527 | 2-jan | | | | | 16 |
| **Psammetichus III** | 1 | 44 | 526 | 2-jan | | | | | 17 |
| **Cambyses II** | 2 | 5 | 525 | 1-jan | | | | | 18 |
| | 3 | 6 | 524 | 1-jan | | | | | (19y) |
| | 4 | 7 | 523 | 1-jan | | | | | |

Several historical synchronisms with the Egyptian chronology are anchored on astronomical data as lunar eclipses:

- The partial eclipse in year 7 of Cambyses II (tablet BM 33066) may be dated 523 BCE July 16/17 [magnitude = 0.54] and the total eclipse 522 BCE January 9/10. Claudius Ptolemy had to know the original tablet because he gave the right magnitude of 0.50 for the partial eclipse (Almagest V:14). Another astronomical tablet (BM 36879) describes eclipses in years 1-4 of Cambyses II, dated by astronomy 529-526 BCE[22]. A diary (VAT 4956)[23] contains numerous astronomical conjunctions in years 37 and 38 of Nebuchadnezzar dated from astronomy in 568 and 567 BCE. An astronomical journal (BM 38462)[24] list some lunar eclipses in the years 1 to 27 of Nebuchadnezzar which are dated from 604 to 578 BCE. Other dated lunar eclipses[25] are these of: year 1 and 2 of Merodachbaladan (March 19/20 721 BCE, March 8/9 and September 1/2 720 BCE); year 5 of Nabopolassar (April 21/22 621 BCE); year 2 of Šamaš-šuma-ukîn (April 10/11 666 BCE); year 42 of Nebuchadnezzar (March 2/3 562 BCE).

- Cambyses II defeated Egypt in his 5th year, month 2 (May -525), which is also dated year 2, month 5, of Psammetichus III (May -525).
- According to the biography of Adad-Guppi[26], mother of Nabonidus, Nabopolassar reigned 21 years, then Nebuchadnezzar 43 years, Amel-Marduk 2 years, Neriglissar 4 years just before Nabonidus. According to the Hillah's stele[27] there were 54 years between the destruction of the temple of Sin, in Harran, and the beginning of the reign of Nabonidus. According to a Babylonian chronicle (BM 21901)[28] and Adad-Guppi's stele, the temple of Harran was destroyed in the year 16 of Nabopolassar.
- After the fall of the Assyrian empire in October 609 BCE, Babylonian domination lasted exactly 70 years until its fall in October 539 BCE, according to Jeremiah 25:11,12.
- The Assyrian period 911-648 is dated owing to its eponyms[29] and the period 648-609 by a prosopography of its eponyms[30].
- Year 1 of Amel Marduk (in 561 BCE) corresponds to year 37 of Jehoiachin's exile (2 Kings 25:27). This exile began just after the attack on Jerusalem by Nebuchadnezzar II in the year 7 of his reign (in 598 BCE).
- The fall of the Assyrian empire, which took place in October 609 BCE after the battle of Harran, is characterized by four synchronisms, since the year 3 of Assur-uballit II corresponds to year 17 of Nabopolassar to Josiah's year 31 and year 1 of Necho II.
- Year 6 of Assurbanipal corresponds to year 1 of Psammetichus I[31].

Chronology of the Saite period:

| Pharaoh | Reign (from Apis) | Length of reign | Highest year | Synchronism with: |
|---|---|---|---|---|
| Psammetichus I | 02/**663**-01/609 | 54 years | 54 | year **6** of Assurbanipal |
| Necho II | 02/**609**-10/594 | 15 years 10 months | 16 | year **17** of Nabopolassar |
| Psammetichus I | 11/594-01/588 | 6 years 1 month | 7 | |
| Apries | 02/588-12/570 | 19 years | 17 | |
| Apries/ Amasis | 01/569-12/567 | [3 years co-regency] | [3] | |
| Amasis | 01/569-10/526 | 43 years 10 months | 44 | |
| Psammetichus III | 11/526-04/**525** | 6 months | 2 | year **5** of Cambyses II |

---

22 P.J. HUBER, S. DE MEIS – Babylonian Eclipse Observations from 750 BC to 1 BC
Milano 2004 Ed. Mimesis pp. 94-96.
23 A.J. SACHS, H. HUNGER - Astronomical Diaries and Related Texts from Babylonia vol. I
Wien 1988 Ed. Akademie der Wissenschaften (n° -567).
24 H. HUNGER - Astronomical Diaries and Related Texts from Babylonia vol. V n° 6
Wien 2001 Ed. Akademie der Wissenschaften pp. 27-30,396.
25 F.R. STEPHENSON - Historical Eclipses and Earth's Rotation
Cambridge 1997 Ed. Cambridge University Press pp. 99-100, 151-152, 166-167, 206.
26 J.B. PRITCHARD - Ancient Near Eastern Texts
Princeton 1969 Ed. Princeton University Press p. 560,561.
27 P.A. BEAULIEU – The Reign of Nabonidus, King of Babylon 556-539 B.C.
in: *Yale Near Eastern Research* 10 (1989) n°2.
28 J.J. GLASSNER – Chroniques mésopotamiennes n°22
Paris 1993 Éd. Belles Lettres pp. 193-197.
29 S. PARPOLA – Assyrian Chronology 681-648 BC.
in: Letters from Assyrian Scholars to the Kings Esarhaddon and Assurbanipal Part II Winona Lake 2007 Ed. Eisenbrauns pp. 381-430.
30 S. PARPOLA – The Prosopography of the Neo-Assyrian Empire
Helsinki 1998 University of Helsinki pp. XVIII-XX.
31 A.K. GRAYSON – The Chronology of the Reign of Ashurbanipal
in: *Zeitschrift für Assyriologie und Vorderasiatische Archäologie* 0 (1980) pp. 227-245.

| BCE | | | Assyrian king / Egyptian king | *Assyrian eponym* | | | | Babylonian king | | |
|---|---|---|---|---|---|---|---|---|---|---|
| 708 | 5 | | Sargon II | *Šamaš-upaḫḫir* | 14 | [7] | 2 | Sargon II | | |
| 707 | 6 | | Chabataka/ | *Ša-Aššur-dubbu* | 15 | [8] | 3 | | | |
| 706 | 7 | | Taharqa | *Mutakkil-Aššur* | 16 | [9] | 4 | | | |
| 705 | 8 | | | *Nashru-Bêl* | 17 | [10] | 5 | | | |
| 704 | 9 | | Sennacherib | *Nabû-deni-epuš* | 1 | | 18 | (Sargon II) | | |
| 703 | 10 | | | *Nuḫšaya* | 2 | | 19 | Marduk-zakir-šumi II | | |
| 702 | 11 | | | *Nabû-lē'i* | 3 | | 1 | Bêl-ibni | | |
| 701 | 12 | | | *Hananu* | 4 | | 2 | | | |
| 700 | 13 | | | *Metunu* | 5 | | 3 | | | |
| 699 | 14 | | | *Bêl-šarrani* | 6 | | 1 | Aššur-nâdin-šumi II | | |
| 698 | 15 | | Arda-Mulissu | *Sulmu-šarri* | 7 | [1] | 2 | | | |
| 697 | 16 | | | *Nabû-dûru-usur* | 8 | [2] | 3 | | | |
| 696 | 17 | | | *Šulmu-bêli* | 9 | [3] | 4 | | | |
| 695 | 18 | | | *Aššur-bêlu-usur* | 10 | [4] | 5 | | | |
| 694 | 19 | | | *Ilu-issîya* | 11 | [5] | 6 | | | |
| 693 | 20 | | | *Iddin-aḫḫê* | 12 | [6] | 1 | Nergal-ušêzib | | |
| 692 | 21 | | | *Zazâya* | 13 | [7] | 1 | Mušêzib-Marduk | | |
| 691 | 22 | | | *Bêl-êmuranni* | 14 | [8] | 2 | | | |
| 690 | 23 | | | *Nabû-kênu-usur* | 15 | [9] | 3 | | | |
| 689 | 1 | | Taharqa | *Gihilu* | 16 | [10] | 4 | | | |
| 688 | 2 | | | *Iddin-aḫḫê* | 17 | [11] | 1 | Sennacherib | | |
| 687 | 3 | | | *Sin-aḫḫê-erîba* | 18 | [12] | 2 | | | |
| 686 | 4 | | | *Bêl-êmuranni* | 19 | [13] | 3 | | | |
| 685 | 5 | | | *Aššur-da''inanni* | 20 | [14] | 4 | | | |
| 684 | 6 | | | *Manzernê* | 21 | [15] | 5 | | | |
| 683 | 7 | | | *Mannu-kî-Adad* | 22 | [1] | 6 | | | |
| 682 | 8 | | | *Nabû-sharru-usur* | 23 | [2] | 7 | | | |
| 681 | 9 | | | *Nabû-aḫḫê-êreš* | 24 | [3] | 8 | | | |
| 680 | 10 | | Esarhaddon | *Danânu* | 1 | | 1 | Esarhaddon | | |
| 679 | 11 | | | *Issi-Adad-anênu* | 2 | | 2 | | | |
| 678 | 12 | | | *Nergal-šarru-uṣur* | 3 | | 3 | | | |
| 677 | 13 | | | *Abî-râmu* | 4 | | 4 | | | |
| 676 | 14 | | | *Banbâ* | 5 | | 5 | | | |
| 675 | 15 | | | *Nabû-aḫḫê-iddin* | 6 | | 6 | | | |
| 674 | 16 | | | *Šarru-nûrî* | 7 | | 7 | | | |
| 673 | 17 | | | *Atar-ilu* | 8 | | 8 | | | |
| 672 | 18 | | | *Nabû-bêlu-uṣur* | 9 | [1] | 9 | | | |
| 671 | 19 | | | *Kanûnâyu* | 10 | [2] | 10 | | | |
| 670 | 20 | | | *Šulmu-bêli-lašme* | 11 | [3] | 11 | | | |
| 669 | 21 | | | *Šamash-kâšid-ayâbi* | 12 | [4] | 12 | | | |
| 668 | 22 | | Assurbanipal | *Marlarim* | 1 | | 1 | Aššurbanipal | | |
| 667 | 23 | | | *Gabbaru* | 2 | | 1 | Šamaš-šuma-ukîn | | |
| 666 | 24 | | | *Kanûnâyu* | 3 | | 2 | *Tablet BM 45640* | | |
| 665 | 25 | | | *Mannu-kî-šarri* | 4 | | 3 | | | |
| 664 | 26 | | *Thebes devastated* | *Šarru-lû-dâri* | 5 | | 4 | | | |
| 663 | 1 | | Psammetichus I | *Bêl-na'id* | 6 | | 5 | | | *1* |
| 662 | 2 | | | *Tab-šar-Sîn* | 7 | | 6 | | | *2* |
| 661 | 3 | | | *Arba'ilâyu* | 8 | | 7 | | | *3* |
| 660 | 4 | | | *Girsapûnu* | 9 | | 8 | | | *4* |
| 659 | 5 | | | *Silim-Aššur* | 10 | | 9 | | | *5* |
| 658 | 6 | | | *Ša-Nabû-šû* | 11 | | 10 | | | *6* |
| 657 | 7 | | | *Lâ-bâši* | 12 | | 11 | | | *7* |
| 656 | 8 | | | *Milkî-râmu* | 13 | | 12 | | | *8* |
| 655 | 9 | | | *Amyânu* | 14 | | 13 | | | *9* |
| 654 | 10 | | | *Assur-nâsir* | 15 | | 14 | | | *10* |
| 653 | 11 | | | *Assur-ilâya* | 16 | | 15 | | | *11* |
| 652 | 12 | | | *Assur-dûru-uṣur* | 17 | | 16 | | | *12* |
| 651 | 13 | | | *Sagabbu* | 18 | | 17 | | | *13* |
| 650 | 14 | | | *Bêl-Ḫarrân-šadûa* | 19 | | 18 | | | *14* |
| 649 | 15 | | | *Aḫu-ilâya* | 20 | | 19 | | | *15* |
| 648 | 16 | | | *Belshunu* | 21 | | 20 | | | *16* |
| 647 | 17 | | | *Nabû-nadin-ahi* | 22 | | 1 | Kandalanu | | *17* |
| **646** | 18 | | | *Nabû-shar-ahhešu* | 23 | | 2 | | | *18* |
| **645** | 19 | | | *Šamaš-da''inanni of Babylon* | 24 | | 3 | | | *19* |
| **644** | 20 | | | *Nabû-sharru-uṣur* | 25 | | 4 | | | *20* |
| **643** | 21 | | | *Nabû-sharru-uṣur of Marash* | 26 | | 5 | | | *21* |
| **642** | 22 | | | *Šamaš-da''inanni of Que* | 27 | | 6 | | | *22* |
| **641** | 23 | | | *Aššur-garu'a-nere* | 28 | | 7 | | | *23* |
| **640** | 24 | | | *Šarru-metu-uballit* | 29 | | 8 | | | *24* |
| **639** | 25 | | | *Mušallim-Aššur* | 30 | | 9 | | | *25* |
| **638** | 26 | | | *Aššur-gimilli-tere* | 31 | | 10 | | | *26* |
| **637** | 27 | | | *Zababa-eriba* | 32 | | 11 | | | *27* |
| **636** | 28 | | | *Sin-šarru-uṣur* | 33 | | 12 | | | *28* |
| **635** | 29 | | | *Bel-lu-dari* | 34 | | 13 | | | *29* |
| **634** | 30 | | | *Bullutu* | 35 | | 14 | | | *30* |
| **633** | 31 | | | *Upaqa-ana-Arbail* | 36 | | 15 | | | *31* |
| **632** | 32 | | | *Tab-sil-Sin* | 37 | | 16 | | | *32* |
| **631** | 33 | | | *Adad-remanni* | 38 | | 17 | | | *33* |
| **630** | 34 | | | *Salmu-šarri-iqbi* | 39 | | 18 | | | *34* |
| **629** | 35 | | **Aššur-etel-ilâni** | *Nabû-šarru-uṣur* | [40] | 1 | 19 | | | *35* |
| **628** | 36 | | | *?Nur-salam-sarpi?* | [41] | 2 | 20 | | | *36* |
| **627** | 37 | | | *Marduk-šarru-uṣur* | [42] | 3 | 21 | **Sin-šum-lišir** | | *37* |
| **626** | 38 | | **Sin-šar-iškun** | *Iqbi-ilani* / *Marduk-remanni* | **0** | 4 | 22) | **Sin-šar-iškun** | | *38* |
| **625** | 39 | | | *Sin-šarru-uṣur* | 1 | | 1 | **Nabopolassar** | | *39* |
| **624** | 40 | | | *Kanunaiu* | 2 | | 2 | | | *40* |
| **623** | 41 | | | *Aššur-matu-taqqin* | 3 | | 3 | | | *41* |
| **622** | 42 | | | *Daddî* | 4 | | 4 | | | *42* |
| **621** | 43 | | | *Bel-iqbi* | 5 | | 5 | | | *43* |
| **620** | 44 | | | *Sa'ilu* | 6 | | 6 | | | *44* |
| **619** | 45 | | | *Mannu-ki-ahhe* | 7 | | 7 | | | *45* |
| **618** | 46 | | | *Nabû-sakip* | 8 | | 8 | | | *46* |
| **617** | 47 | | | *Assur-remanni* | 9 | | 9 | | | *47* |
| **616** | 48 | | | *Bel-ahu-uṣur* | 10 | | 10 | | | *48* |
| **615** | 49 | | | *Sin-alik-pani* | 11 | | 11 | | | *49* |
| **614** | 50 | | | *Paši* | 12 | | 12 | | | *50* |
| **613** | 51 | | | *Nabû-tapputi-alik* | 13 | | 13 | | | *51* |
| **612** | 52 | | | *Shamash-šarru-ibni* | 14 | | 14 | | | *52* |
| **611** | 53 | | **Aššur-uballit II** | *Nabû-mar-šarri-uṣur* | 1 | | 15 | | | *53* |
| **610** | 54 | | | *Nabû-šarru-uṣur* | 2 | | 16 | *Temple of Harran wrecked* | | *54* |
| **609** | 1 | | **Necho II** | *Gargamisaiu* | 3 | [0] | 17 | *Stele of Adad-Guppi* | *1* | *55* |
| **608** | 2 | | | | | [1] | 18 | | *2* | *56* |
| **607** | 3 | | | | | [2] | 19 | | *3* | *57* |
| **606** | 4 | | | | | [3] | 20 | | *4* | *58* |
| **605** | 5 | | | | | [4] | 21 | | *5* | *59* |
| **604** | 6 | | | | | | 1 | **Nebuchadnezzar II** | *6* | *60* |
| **603** | 7 | | | | | | 2 | | *7* | *61* |
| **602** | 8 | | | | | | 3 | | *8* | *62* |
| **601** | 9 | | | | | | 4 | | *9* | *63* |
| **600** | 10 | | | | | | 5 | | *10* | *64* |
| **599** | 11 | | | | | | 6 | | *11* | *65* |
| **598** | 12 | | | | | | 7 | | *12* | *66* |
| **597** | 13 | | | | | | 8 | | *13* | *67* |
| **596** | 14 | | | | | | 9 | | *14* | *68* |
| **595** | 15 | | | | | | 10 | | *15* | *69* |
| **594** | 16 | 1 | **Psammetichus II** | | | | 11 | | *16* | *70* |
| **593** | | 2 | | | | | 12 | | *17* | *71* |
| **592** | | 3 | | | | | 13 | | *18* | *72* |
| **591** | | 4 | | | | | 14 | | *19* | *73* |
| **590** | | 5 | | | | | 15 | | *20* | *74* |
| **589** | | 6 | | | | | 16 | | *21* | *75* |
| **588** | 1 | 7 | **Apries** | | | | 17 | | *22* | *76* |
| **587** | 2 | | | | | | 18 | | *23* | *77* |
| **586** | 3 | | | | | | 19 | | *24* | *78* |
| **585** | 4 | | | | | | 20 | | *25* | *79* |
| **584** | 5 | | | | | | 21 | | *26* | *80* |
| **583** | 6 | | | | | | 22 | | *27* | *81* |
| **582** | 7 | | | | | | 23 | | *28* | *82* |
| **581** | 8 | | | | | | 24 | | *29* | *83* |
| **580** | 9 | | | | | | 25 | | *30* | *84* |
| **579** | 10 | | | | | | 26 | | *31* | *85* |
| **578** | 11 | | | | | | 27 | | *32* | *86* |
| **577** | 12 | | | | | | 28 | | *33* | *87* |
| **576** | 13 | | | | | | 29 | | *34* | *88* |
| **575** | 14 | | | | | | 30 | | *35* | *89* |
| **574** | 15 | | | | | | 31 | | *36* | *90* |
| **573** | 16 | | | | | | 32 | | *37* | *91* |
| **572** | 17 | | | | | | 33 | | *38* | *92* |
| **571** | 18 | | | | | | 34 | | *39* | *93* |
| **570** | 19 | | | | | | 35 | | *40* | *94* |
| **569** | [20] | 1 | **Amasis** | | | | 36 | | *41* | *95* |
| **568** | [21] | 2 | | | | | 37 | *Tablet VAT 4956* | *42* | *96* |
| **567** | [22] | 3 | | | | | 38 | | *43* | *97* |
| **566** | | 4 | | | | | 39 | | *44* | *98* |
| **565** | | 5 | | | | | 40 | | *45* | *99* |
| **564** | | 6 | | | | | 41 | | *46* | *100* |
| **563** | | 7 | | | | | 42 | | *47* | *101* |
| **562** | | 8 | | | | 0 | 43 | | *48* | *102* |
| **561** | | 9 | | | | | 1 | **Amel-Marduk** | *49* | *103* |
| **560** | | 10 | | | | 0 | 2 | | *50* | *104* |
| **559** | | 11 | | | | | 1 | **Neriglissar** | *51* | *105* |
| **558** | | 12 | *pap. Louvre 7848* | **Cyrus II** | [1] | | 2 | | *52* | *106* |
| **557** | | 13 | | | [2] | | 3 | | *53* | *107* |
| **556** | | 14 | | | [3] | | 4 | | *54* | *108* |
| | | | | | | 0 | 0 | **Lâbâši-Marduk** | | |
| **555** | | 15 | | *stele of Hillah* | [4] | | 1 | **Nabonidus** | | *109* |
| **554** | | 16 | | | [5] | | 2 | | | *110* |
| **553** | | 17 | | | [6] | [0] | 3 | **Belshazzar** | | *111* |
| **552** | | 18 | | | [7] | [1] | 4 | | | *112* |
| **551** | | 19 | | | [8] | [2] | 5 | | | *113* |
| **550** | | 20 | | | [9] | [3] | 6 | | | *114* |
| **549** | | 21 | | | [10] | [4] | 7 | | | *115* |
| **548** | | 22 | | | [11] | [5] | 8 | | | *116* |
| **547** | | 23 | | | [12] | [6] | 9 | | | *117* |
| **546** | | 24 | | | [13] | [7] | 10 | | | *118* |
| **545** | | 25 | | | [14] | [8] | 11 | | | *119* |
| **544** | | 26 | | | [15] | [9] | 12 | | | *120* |
| **543** | | 27 | | | [16] | [10] | 13 | | | *121* |
| **542** | | 28 | | | [17] | [11] | 14 | | | *122* |
| **541** | | 29 | | | [18] | [12] | 15 | | | *123* |
| **540** | | 30 | | | [19] | [13] | 16 | | | *124* |
| **539** | | 31 | | | [20] | [14] | 17 | *Fall of Babylon* | | *125* |
| **538** | | 32 | | **Cyrus II** | 1 | [1] | | **Ugbaru** | | *126* |
| **537** | | 33 | | | 2 | 1 | | **Cambyses II** | | *127* |
| **536** | | 34 | | | 3 | [2] | | | | *128* |
| **535** | | 35 | | | 4 | [3] | | | | *129* |
| **534** | | 36 | | | 5 | [4] | | | | *130* |
| **533** | | 37 | | | 6 | [5] | | | | *131* |
| **532** | | 38 | | | 7 | [6] | | | | *132* |
| **531** | | 39 | | | 8 | [7] | | | | *133* |
| **530** | | 40 | | | 9 | [8] | | | | *134* |
| **529** | | 41 | | **Cambyses II** | 1 | | | | | *135* |
| **528** | | 42 | | | 2 | | | | | *136* |
| **527** | | 43 | | | 3 | | | | | *137* |
| **526** | *1* | 44 | | | 4 | | | | | *138* |
| | | 1 | **Psammetichus III** | | | | | | | |
| **525** | *2* | 2 | *Stele IM.4187* | | 5 | | | | | |
| | | 5 | **Cambyses II** | | | | | | | |
| **524** | *3* | 6 | | | 6 | | | | | |
| **523** | *4* | 7 | | *Tablet BM 33066* | 7 | | | | | |
| **522** | *5* | 8 | | | 8 | | | | | |

Year 44 of Amasis, the last of his reign, should be dated 526 BCE. The solution proposed by Parker of a year 45 of Amasis dated 526 BCE is not possible, as recognized by Depuydt[32] who prefers to date the death of Amasis in 527 BCE in his 44th year, assuming that the 4th year of Cambyses (at 526 BCE) was a period of disorder without pharaoh! But this choice leads to an implausible result, contrary to the accounts of all the ancient historians (Herodotus was close to events, and Manetho, an Egyptian priest, was to know the history of his country): the throne of Egypt would have been vacuum for one year after the disappearance of Psammetichus III, from May 526 to May 525 BCE, when Cambyses was recognized Pharaoh. In fact, the end of the ancient Egyptian empire was an important milestone that has been recounted by the following historians:

- According to Diodorus Siculus: *After a reign of 55 years*[33] *he [Amasis] ended his days at the time when Cambyses, the king of the Persians, attacked Egypt, in the 3rd year of the 63rd Olympiad* (Historical Library I:68:6). Thus Amasis died between July -526 and July -525.
- According to the Egyptian priest Manetho[34]: *Cambyses, in the 5th year of his reign over the Persians [in -525] became king of Egypt and led it for 3 years [from spring -525 to spring -522].*
- According to Herodotus (around -450): *On the death of Cyrus, Cambyses his son by Cassandane daughter of Pharnaspes took the kingdom (...) Amasis was the Egyptian king against whom Cambyses, son of Cyrus, made his expedition; and with him went an army composed of the many nations under his rule, among them being included both Ionic and Aeolic Greeks (...) One of the mercenaries of Amasis, a Halicarnassian, Phanes by name, a man of good judgment, and a brave warrior, dissatisfied for some reason or other with his master, deserted the service, and taking ship, fled to Cambyses, wishing to get speech with him (...) Psammenitus, son of Amasis, lay encamped at the mouth of the. Nile, called the Pelusiac, awaiting Cambyses. For Cambyses, when he went up against Egypt, found Amasis no longer in life: he had died after ruling Egypt 44 years, during all which time no great misfortune had befallen him (...) The Egyptians who fought in the battle, no sooner turned their backs upon the enemy, than they fled away in complete disorder to Memphis (...) 10 days after the fort had fallen, Cambyses resolved to try the spirit of Psammenitus, the Egyptian king, whose whole reign had been but 6 months (...) Psammenitus plotted evil, and received his reward accordingly. He was discovered to be stirring up revolt in Egypt, wherefore Cambyses, when his guilt clearly appeared, compelled him to drink bull's blood, which presently caused his death. Such was the end of Psammenitus* (The Histories II:1; III:1,4,10-16).

The Egyptian priest Manetho indicates the same values as Herodotus, 44 years for Amasis and 6 months for Psammetichus III. By combining this information with data from the reign of Persian King Cambyses who became Egypt to in May 525 BCE, the death of Amasis can be fixed around October 526 BCE. Fixing the date of the conquest of Egypt in 525 BCE is also confirmed since the 5th year of Cambyses began the 1st Nisan (March 29) in the Persian system, and the 1st Thoth (January 2) in the Egyptian system. The account of these historians is confirmed by several archaeological finds:

- The narrative of Udjahorresnet[35], the Egyptian general who led the naval fleet under Amasis, then under Psammetichus III and finally under Cambyses, authenticates the version of Herodotus. This war probably lasted at least six months because, according to the historian Polyaenus: *When Cambyses attacked Pelusium, which guarded the entrance into Egypt, the Egyptians defended it with great resolution. They advanced formidable engines against the besiegers, and hurled missiles, stones, and fire at them from their catapults.* (Stratagems of war

[32] L. DEPUYDT - Egyptian Regnal Dating under Cambyses and the Date of the Persian Conquest
1996 in: Studies in Honor of William Kelly Simpson pp. 179-190.
[33] The reign of Amasis is counted from the revolt after the attack of Nebuchadnezzar II in -582.
[34] W.G. WADDELL - Manetho (Loeb Classical Library 350)
Cambridge 1956 Ed. Harvard University Press pp. 169-177.
[35] P. BRIANT - Histoire de l'empire perse. De Cyrus à Alexandre
Paris 1996 Éd. Fayard pp. 63-65.

VII:9). These narrative overlap exactly and give the following chronological scheme: war of Cambyses against Egypt beginning in the year 44, the last year of Amasis, which ends after the brief reign of 6 months of Psammetichus III, his successor or in the 5th year of Cambyses.

- According to the stele IM.4187 in the Louvre, an Apis bull was born at month 5, day 29, year 5 of Cambyses, died on month 9, day 4, year 4 of Darius I and was buried month 11, day 13, of the same year, covering a total period of 7 years 3 months and 5 days (reading 8 years less likely). This computation is consistent (between the month9, day 4, and the month 11, day 13, there are exactly 70 days for the period of embalming bull) gives the following dates in the Julian calendar: May 29, 525, August 31, 518 and November 8, 518 BCE. This stele proves that Cambyses reigned in Egypt from May 525 BCE because at the end of this month, an Apis bull is dedicated to him. Thus the conquest of Egypt had to be completed in early May 525 as the last text referring to Psammetichus III (below) is dated I Peret year 2 (May 525). That Psammetichus III was the son of Amasis is confirmed by the stele No. 309 of the Serapeum (Louvre). It is indeed Psammetichus III because one of the contracting parties cited in the text is still alive in the year 35 of Darius I[36]. 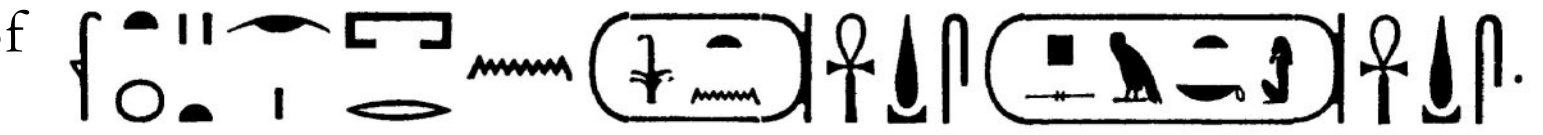

Before his conquest Cambyses was a Persian leader but thereafter he also became an Egyptian pharaoh. This new situation has created a dual system of counting the reign.

- Egyptian documents of the time of Darius I mention the events of years 3 and 4 of Cambyses, apparently before the conquest of Egypt. A papyrus dated 9th year of Darius says: *In his 2nd year, therefore, Cambyses conquered Egypt really, and in 5th year he died.* This demotic text (Papyrus Rylands IX 21), entitled *Peteisis petition* spoke of a conflict in a family of priests of the temple of Amon at Teuzoi (El-Hibeh) between the 4th year of Psammetichus I and the 4th year of Cambyses[37]. It ends with the following dates: *Until the Year 44 of Amasis. In Year 3 of Cambyses, Hor son of Psammet-kmenempe, the prophet of Amon (...) in Year 4 of Cambyses.* A second Egyptian papyrus known as the *Demotic Chronicle*, confirmed the year 44 of Amasis as last year[38]. The source said Darius I in the 3rd year of his reign would have given the satrap of Egypt the order that together a committee of wise men from among the Egyptian warriors, priests and scribes in order: *that they put in writing that Egyptian law was in force until the 44th year of the reign of Amasis.*
- Cambyses died in 522 BCE, it was therefore his 5th year in Egypt, the 2nd corresponded to 525 BCE and the 1st in 526 BCE. This conquest began in 526 BCE, since Herodotus (The Histories III:1,10) states that the war began with the death of Amasis. Years 2 to 5 of Cambyses refer to his years of domination in Egypt. It is not logical to assume that the Egyptians used a counting system reserved for their pharaohs rather than to foreign leaders[39], what was Cambyses before his conquest (though, after 525 BCE, Persian leaders will be considered as Pharaohs).

The year 5 of Cambyses (in 525 BCE) began on Nisan 1st, that is March 28, and Year 44 of Amasis (in 526 BCE) began on Thot 1st, that is January 2. Thus, as the reign of Psammeticus III was 6 months length, his year 1 (in 526 BCE) began near November and his year 2 began on Thot 1st, that is January 1st, 525 BCE, and ended around April.

[36] H. GAUTHIER – Le livre des rois d'Égypte
Le Caire 1915 Éd. Institut Français d'Archéologie Orientale pp. 131-132).
[37] P. BRIANT - Histoire de l'empire perse. De Cyrus à Alexandre
Paris 1996 Éd. Fayard p. 92.
[38] A. KUHRT - The Persian Empire
London 2010 Ed. Routeledge pp. 124-125.
[39] R.A. PARKER - Persian and Egyptian Chronology
in: *The American Journal of Semitic Languages and Literatures* LVIII/3 (1941) pp. 298-301.

| King | Egypt | | BCE | | | Persia | | King |
|---|---|---|---|---|---|---|---|---|
| **Amasis** | **43** | | **525** | 11 | VIII | **3** | | **Cambyses II** |
| | | | | 12 | IX | | | |
| *Pap. Rylands IX* | **44** | | **526** | 1 | X | | | |
| ---------------------------------- | | 1 | | 2 | XI | | | *Egypt conquest* |
| | | | | 3 | XII | | | |
| | | | | 4 | I | **4** | | |
| | | | | 5 | II | | | |
| | | | | 6 | III | | | |
| | | | | 7 | IV | | | |
| | | | | 8 | V | | | |
| | | | | 9 | VI | | | |
| | | | | 10 | VII | | | |
| **Psammetichus III** | **1** | | | 11 | VIII | | | |
| | | | | 12 | IX | | | |
| | **2** | 2 | **525** | 1 | X | | | |
| | | | | 2 | XI | | | |
| | | | | 3 | XII | | | |
| | | | | 4 | I | **5** | | |
| *stele IM.4187* | **5** | | | 5 | II | | | *Egypt defeated* |
| | | | | 6 | III | | | |
| | | | | 7 | IV | | | |
| | | | | 8 | V | | | |
| | | | | 9 | VI | | | |
| | | | | 10 | VII | | | |
| | | | | 11 | VIII | | | |
| | | | | 12 | IX | | | |
| | **6** | 3 | **524** | 1 | X | | | |
| | | | | 2 | XI | | | |
| | | | | 3 | XII | | | |
| | | | | 4 | I | **6** | | |
| | | | | 5 | II | | | |
| | | | | 6 | III | | | |
| | | | | 7 | IV | | | |
| | | | | 8 | V | | | |
| | | | | 9 | VI | | | |
| | | | | 10 | VII | | | |
| | | | | 11 | VIII | | | |
| | | | | 12 | IX | | | |
| | **7** | 4 | **523** | 1 | X | | | |
| | | | | 2 | XI | | | |
| | | | | 3 | XII | | | |
| | | | | 4 | I | **7** | | |
| | | | | 5 | II | | | |
| | | | | 6 | III | | | |
| | | | | 7 | IV | | | |
| | | | | 8 | V | | | |
| | | | | 9 | VI | | | |
| | | | | 10 | VII | | | |
| | | | | 11 | VIII | | | |
| | | | | 12 | IX | | | |
| | **8** | 5 | **522** | 1 | X | | | |
| | | | | 2 | XI | | | |
| | | | | 3 | XII | | **0** | **Cambyses II / Bardiya** |
| | | | | 4 | I | **8** | **1** | |
| | | | | 5 | II | **0** | | **Bardiya** |
| | | | | 6 | III | | | |
| | | | | 7 | IV | | | |
| | | | | 8 | V | | | |
| | | | | 9 | VI | | | |
| | | | | 10 | VII | **0** | | **Nebuchadnezzar III / Bardiya** |
| | | | | 11 | VIII | | | |
| | | | | 12 | IX | **0** | **0** | **Darius I / Nebuchadnezzar III** |
| | **1** | | **521** | 1 | X | | | |
| | | | | 2 | XI | | **[0]** | **Darius I / Nebuchadnezzar IV** |
| | | | | 3 | XII | | | |
| | | | | 4 | I | **1** | **1** | |
| | | | | 5 | II | | | |
| | | | | 6 | III | | | |
| | | | | 7 | IV | | | |
| | | | | 8 | V | | | |

It is interesting to notice that the Israelite chronology fits very well the previous chronologies (Egyptian, Assyrian and Babylonian). For example, the text of 2Kings 18:9 in which the fall of Samaria began in the 4th year of King Hezekiah, that is, the 7th year of Hoshea, that Shalmaneser the king of Assyria came against Samaria and began to lay siege to it, which lasted 3 years. According to a Babylonian chronicle the fall of Samaria began on the 5th and last year of Shalmaneser V and was achieved 3 years later on the 2nd year of Sargon II (Annals of Sargon). According to the Bible, there are many dated synchronisms between kings of Judah (Ahaz, Hezechiah) and kings of Israel (Pekah, Hosheah) with Assyrian kings (Tiglath-pileser III, Salmanazar V, Sargon II, Sennacherib) and one Babylonian king (Merodachbaladan II).

In addition, there were four dated lunar eclipses during this period: one on year 1 of Nabû-mukîn-zêri (April 9, 731 B.C.E.), one on year 1 of Merodachbaladan II (March 19, 721 B.C.E.) and two on his year 2 (March 8, September 1st, 720 B.C.E.).

| | ASSYRIA | BABYLONIA | JUDEA | ISRAEL | EGYPT | *reference* |
|---|---|---|---|---|---|---|
| **796** | 15 **Adad-nêrari III** | | 14 **Azariah** | **Jeroboam** | | *2Ki 14:23* |
| **795** | 16 | | **15** / [Uziah] | 28 | | *2Ki 15:1,2* |
| **794** | 17 | | 16 | 29 | | *2Chr 26:3* |
| **793** | 18 | | 17 | 30 | | |
| **792** | 19 | | 18 | 31 | | |
| **791** | 20 | | 19 | 32 | | |
| **790** | 21 | | 20 | 33 | | |
| **789** | 22 | | 21 | 34 | | |
| **788** | 23 | | 22 | 35 | | |
| **787** | 24 | | 23 | 36 | | |
| **786** | 25 | | 24 | 37 | | |
| **785** | 26 | | 25 | 38 | | |
| **784** | 27 | | 26 | 39 | | |
| **783** | 28 **Shalmaneser** IV | | 27 | 40 | | |
| **782** | 1 [0] **Bar Ga'ah** | | 28 | 41 | Pi(ank)e | |
| **781** | 2 [1] (**Pulu**) | | 29 | 1 **Zekariah** | 1 | *2Ki 14:29* |
| **780** | 3 [2] | | 30 | [2] | 2 | |
| **779** | 4 [3] | | 31 | [3] | 3 | |
| **778** | 5 [4] | | 32 | [4] | 4 | |
| **777** | 6 [5] | | 33 | [5] | 5 | |
| **776** | 7 [6] | | 34 | [6] | 6 | |
| **775** | 8 [7] | | 35 | [7] | 7 | |
| **774** | 9 [8] | | 36 | [8] | 8 | |
| **773** | 10 **Aššur-dân III** | | 37 | [9] | 9 | |
| **772** | 1 [10] | | 38 | [10] | 10 | *2Ki 15:8* |
| **771** | 2 [11] | | 39 | [11]**Shallum** | 11 | *2Ki 15:13* |
| **770** | 3 [12] | Eriba-Marduk | 40 | 1**Menahem** | 12 | *2Ki 15:17* |
| **769** | 4 [13] | 1 | 41 | 1 | 13 | |
| **768** | 5 [14] | 2 | 42 | 2 | 14 | |
| **767** | 6 [15] | 3 | 43 | 3 | 15 | |
| **766** | 7 [16] | 4 | 44 | 4 | 16 | (*Isa 10:5-8*) |
| **765** | 8 [17] | 5 | 45 | 5 (**Pulu**) | 17 | *2Ki 15:19-20* |
| **764** | 9 [18] | 6 | 46 | 6 | 18 | |
| **763** | 10 (*total solar eclipse*) | 7 | 47 | 7 | 19 | *Bur-Sagale* |
| **762** | 11 [20] | 8 | 48 | 8 | 20 | |
| **761** | 12 [21] | 9 Nabû-šum-iškun | 49 | 9 | 21 | |
| **760** | 13 [22] | 1 | 50 | 10 **Pekayah** | 22 | *2Ki 15:22-23* |
| **759** | 14 [23] | 2 | 51 | 1 | 23 | |
| **758** | 15 [24] | 3 | 52 **Jotham** | 2 **Pekah** | 24 | *2Ki 15:27-33* |
| **757** | 16 [25] | 4 | 1 | 1 | 25 | |
| **756** | 17 [26] | 5 | 2 | 2 | 26 | |
| **755** | 18 **Aššur-nêrari V** | 6 | 3 | 3 | 27 | |
| **754** | 1 [28] | 7 | 4 | 4 | 28 | |
| **753** | 2 [29] | 8 | 5 | 5 | 29 | |
| **752** | 3 [30] | 9 | 6 | 6 | 30 | |
| **751** | 4 [31] | 10 | 7 | 7 | 31 | |
| **750** | 5 [32] | 11 | 8 | 8 | 32 | |
| **749** | 6 [33] | 12 | 9 | 9 | 33 | |
| **748** | 7 [34] | 13 Nabû-nâsir | 10 | 10 | 34 | |
| **747** | 8 [35] | 1 | 11 | 11 | 35 | |
| **746** | 9 [36] | 2 | 12 | 12 | 36 | |
| **745** | 10 [0] | 3 | 13 | 13 | 37 | |
| **744** | 1**Tiglath-pileser** III | 4 | 14 | 14 | 1 Osorkon IV | |
| **743** | 2 | 5 | 15 | 15 | 2 (= **So**) | |
| **742** | 3 | 6 | 16 | 16 | 3 | |
| **741** | 4 | 7 | **1 Ahaz** *17* | **17** | 4 | *2Ki 16:1,7-10* |
| **740** | 5 [1] Shalmaneser V | 8 | 2 *18* | 18 | 5 | *2Chr 28:16* |
| **739** | 6 [2] | 9 | 3 *19* | 19 | 6 | *2Ki 16:5,6* |
| **738** | **7** [3] | 10 | [4] *20* | **20 Hoshea** | 7 | *2Ki 15:27-30* |
| **737** | 8 [4] | 11 | [5] | [1] | 8 | |
| **736** | 9 [5] | 12 | 6 | [2] | 9 | |
| **735** | 10 [6] | 13 | 7 | [3] | 10 | |
| **734** | 11 [7] | 14 Nabû-nâdîn-zêri | 8 | [4] | 11 | *2Ki 16:7-9* |
| **733** | 12 [8] | 1 | 9 | [5] | 12 | |
| **732** | 13 [9] | 2 Nabû-mukîn-zêri | 10 | [6] | 13 | |
| **731** | 14 [10] | **1** *(lunar eclipse April 9)* | 11 | [7] | 14 | |
| **730** | 15 [11] | 2 | 12 | [8] | 15 | |
| **729** | 16 [12] | 3 Pulu | 13 | [9] | 16 | *2Ki 17:1* |
| **728** | 17 [14] | **1** | 14 | **1** [10] | 17 | |
| **727** | 18 [15] | 2 Ulûlaiu | 15 | 2 [11] | 18 | |
| **726** | 1 **Shalmaneser** V | 1 (Shalmaneser V) | 16 **Ezechias** | 3 [12] | 19 | *2Ki 18:1* |
| **725** | 2 | 2 | 1 | 4 [13] | 20 | |
| **724** | 3 | 3 | 2 | 5 [14] | 21 | |
| **723** | 4 | 4 | 3# | 6 [15] | 22 #*(alliance)* | *2Ki 17:2-5* |
| **722** | **5 Sargon** II | 5 **Merodachbaladan II** | **4** | **7** [16] | 23 | *2Ki 18:9* |
| **721** | **1** | **1** *(lunar eclipse March 19)* | 5 | 8 [17] | 24 | |
| **720** | **2** *Fall of Samaria* | **2** *(March 8; September 1st)* | **6** | **9** [18] | 25 | *2Ki 18:10* |
| **719** | 3 | 3 | 7 | [19] | 26 | |
| **718** | 4 | 4 | 8 | [20] | 27 | |
| **717** | 5 | 5 | 9 | [21] | 28 | |
| **716** | 6 | 6 | 10 | [22] | 29 | |
| **715** | 7 #*(alliance)* | 7 | 11 | [23] | 30 #*(alliance)* | |
| **714** | 8 -[1] Sennacherib | 8 | 12 | [24] | 31 | |
| **713** | 9 -[2] | 9 | 13 | [25] | 32 | |
| **712** | **10**-[3] *Ashdod Lakish* | **10** *(failed alliance)* | **14** | [26] | 1 **Chabataka** | *Isa 36:1; 39:1* |
| **711** | **11**-[4] *taken* | 11 | 15 | [27] | 2 /**Taharqa** | *1* |
| **710** | **12**-[5] | 12 Sargon II | 16 | [28] | 3 | *2* |
| **709** | **13**-[6] | 1 | 17 | [29] | 4 | *3* |
| **708** | **14**-[7] | 2 | 18 | [30] | 5 | *4* |
| **707** | **15**-[8] | 3 | 19 | [31] | 6 | *5* |
| **706** | **16**-[9] | 4 | 20 | [32] | 7 | *6* |
| **705** | **17 Sennacherib** | 5 **Sennacherib** | 21 | [33] | 8 | *7* |
| **704** | 1 | 1 | 22 | [34] | 9 | *8* |
| **703** | 2 | 2 Bêl-ibni | 23 | [35] | 10 | *9* |
| **702** | 3 | 1 | 24 | [36] | 11 | 10 |
| **701** | 4 | 2 | 25 | [37] | 12 | *11* |
| **700** | 5 | 3 Aššur-nâdin-šumi II | 26 | [38] | 13 | *12* |
| **699** | 6 | 1 | 27 | [39] | 14 | *13* |
| **698** | 7 [1] Arda-Mulissu | 2 | 28 | [40] | 15 | *14* |
| **697** | 8 [2] | 3 | **29** | [41] | 16 | ***15*** |
| **696** | 9 [3] | 4 | **1 Manasseh** | [42] | 17 | *2Ki 21:1* |
| **695** | 10 [4] | 5 | 2 | [43] | 18 | |
| **694** | 11 [5] | 6 | 3 | [44] | 19 | |
| **693** | 12 [6] | 1 Nergal-ušezib | 4 | [45] | 20 | |
| **692** | 13 [7] | 1 Mušezib-Marduk | 5 | [46] | 21 | |
| **691** | 14 [8] | 2 | 6 | [47] | 22 | |
| **690** | 15 [9] | 3 | 7 | [48] | 23 | |
| **689** | 16 [10] | 4 | 8 | [49] | 1 **Taharqa** | |
| **688** | 17 [11] | 1 Sennacherib | 9 | [50] | 2 | |
| **687** | 18 [12] | 2 | 10 | [51] | 3 | |
| **686** | 19 [13] | 3 | 11 | [52] | 4 | |
| **685** | 20 [14] | 4 | 12 | [53] | 5 | |
| **684** | 21 [15] | 5 | 13 | [54] | 6 | |
| **683** | 22 [1] Esarhaddon | 6 | 14 | [55] | 7 | |
| **682** | 23 [2] | 7 | 15 | [56] | 8 | |
| **681** | 24 [3] | 8 | 16 | [57] | 9 | |
| **680** | **1 Esarhaddon** | 1 Esarhaddon | 17 | [58] | 10 | |
| **679** | 2 | 2 | 18 | [59] | 11 | |
| **678** | 3 | 3 | 19 | [60] | 12 | |
| **677** | 4 | 4 | 20 | [61] | 13 | |
| **676** | 5 | 5 | 21 | [62] | 14 | |
| **675** | 6 | 6 | 22 | [63] | 15 | |
| **674** | 7 | 7 | 23 | [64] | 16 | |
| **673** | **8** *(Manasseh deported)* | 8 *(2Chr 33:11)* | 24 | [65] | 17 *Ezr 4:2* | *Isa 7:8,9* |
| **672** | 9 [1] Aššurbanipal | 9 | 25 | | 18 *(Israel deported into Assyria)* | |
| **671** | 10 [2] | 10 | 26 | | 19 | |
| **670** | 11 [3] | 11 | 27 | | 20 | |
| **669** | 12 [4] | 12 | 28 | | 21 | |
| **668** | **1 Aššurbanipal** | 1 Aššurbanipal | 29 | | 22 | |
| **667** | 2 | 1 Šamašumaukin | 30 | | 23 | |
| **666** | 3 | 2 *(lunar eclipse April 10)* | 31 | | 24 | *BM 45640* |
| **665** | 4 | 3 | 32 | | 25 | |
| **664** | **5** *(Thebes sacked* | 4 *(Nah 3:8)* | **33** | | 26 | *Ezr 4:10* |
| **663** | 6 | 5 | 34 | **1 Psammetichus I** | | |
| **662** | 7 | 6 | 35 | 2 | | |
| **661** | 8 | 7 | 36 | 3 | | |
| **660** | 9 | 8 | 37 | 4 | | |
| **659** | 10 | 9 | 38 | 5 | | |
| **658** | 11 | 10 | 39 | 6 | | |
| **657** | 12 | 11 | 40 | 7 | | |
| **656** | 13 | 12 | 41 | 8 | | |
| **655** | 14 | 13 | 42 | 9 | | |
| **654** | 15 | 14 | 43 | 10 | | |
| **653** | 16 | 15 | 44 | 11 | | |
| **652** | 17 | 16 | 45 | 12 | | |
| **651** | 18 | 17 | 46 | 13 | | |
| **650** | 19 | 18 | 47 | 14 | | |
| **649** | 20 | 19 | 48 | 15 | | |
| **648** | 21 | 20 **Kandalanu** | 49 | 16 | | |
| **647** | 22 | 1 | 50 | 17 | | |
| **646** | 23 | 2 | 51 | 18 | | |
| **645** | 24 | 3 | 52 | 19 | | |
| **644** | 25 | 4 | 53 | 20 | | |
| **643** | 26 | 5 | 54 | 21 | | |
| **642** | 27 | 6 | **55 Amon** | 22 | | *2Ki 21:1* |
| **641** | 28 | 7 | 1 | 23 | | *2Ki 21:19* |
| **640** | 29 | 8 | **2 Josias** | 24 | | *2Ki 22:1* |
| **639** | 30 | 9 | 1 | 25 | | |
| **638** | 31 | 10 | 2 | 26 | | |
| **637** | 32 | 11 | 3 | 27 | | |
| **636** | 33 | 12 | 4 | 28 | | |
| **635** | 34 | 13 | 5 | 29 | | |

| | | | | | |
|---|---|---|---|---|---|
| **634** | 35 | 14 | 6 | 30 | |
| **633** | 36 | 15 | 7 | 31 | |
| **632** | 37 | 16 | 8 | 32 | |
| **631** | 38 | 17 | 9 | 33 | |
| **630** | 39 | 18 | 10 | 34 | |
| **629** | 1 **Aššur-etel-ilâni** | 19 | 11 | 35 | |
| **628** | 2 [41] | 20 | 12 | 36 | |
| **627** | 3 [42] | 21 | 13 | 37 | |
| **626** | 4 **Sin-šar-iškun** | 22 **Nabopolassar** | 14 | 38 | |
| **625** | 1 | 1 | 15 | 39 | |
| **624** | 2 | 2 | 16 | 40 | |
| **623** | 3 | 3 | 17 | 41 | |
| **622** | 4 | 4 | 18 | 42 | |
| **621** | 5 | 5 *Lunar eclipse (22 April)* | 19 | 43 | *Almagest V,14* |
| **620** | 6 | 6 | 20 | 44 | |
| **619** | 7 | 7 | 21 | 45 | |
| **618** | 8 | 8 | 22 | 46 | |
| **617** | 9 | 9 | 23 | 47 | |
| **616** | 10 | 10 | 24 | 48 | |
| **615** | 11 | 11 | 25 | 49 | |
| **614** | 12 | 12 | 26 | 50 | |
| **613** | 13 | 13 | 27 | 51 | |
| **612** | 14 **Aššur-uballit II** | 14 *Nineveh destroyed* | 28 | 52 | *Nah 3:15-19* |
| **611** | 1 | 15 | 29 | 53 | |
| **610** | 2 | 16 | 30 | 54 | |
| **609** | 3 *Battle of Harran* | 17 *BM 21901* | **31 Joiaqim** | 1 **Necho II** | *2Ki 22:1;23:36* |
| **608** | *End of Assyria* | 18 | 1 | 2 | |

| | BABYLONIA | | | JUDEA | | | | EGYPT | | *reference* |
|---|---|---|---|---|---|---|---|---|---|---|
| **610** | | **16** | | | **30** | | | 54 | | |
| **609** | 3 *Battle of Harran* | **17** | [0] | **Joiaqim** | **31** | *0* | | 1 | **Necho II** | *2Ki 22:1;23:36* |
| **608** | | **18** | [1] | | **1** | *1* | | 2 | | |
| **607** | | **19** | [2] | | **2** | *2* | | 3 | | |
| **606** | | **20** | [3] | | **3** | *3* | | 4 | | |
| **605** | **Nebuchadnezzar** | **21** | 1 | | **4** | 4 | | 5 | *Battle of Karkemish* | *Jer 25:1;46:2* |
| **604** | | **1** | 2 | | **5** | *5* | | 6 | | |
| **603** | | **2** | 3 | | **6** | *6* | | 7 | | |
| **602** | | **3** | 4 | | **7** | *7* | | 8 | | |
| **601** | *Birth of Darius the* | **4** | 5 | | **8** | *8* | **0** | 9 | *Joiaqim vassal of* | *Dan 5:31* |
| **600** | *Mede (Harpagus)* | **5** | 6 | | **9** | *9* | **1** | 10 | *Nebuchadnezzar* | *2Ki 24:1* |
| **599** | | **6** | 7 | | **10** | *10* | **2** | 11 | | |
| **598** | *BM 21946* | **7** | 8 | **Zedekiah** | **11** | *11* | **3** | 12 | *Exile of Joiakîn* | *Jer 52:28* |
| **597** | | **8** | 9 | | **1** | *12* | *1* | 13 | *2Chr 36:9,10* | *2Ki 24:12* |
| **596** | | **9** | 10 | | **2** | *13* | *2* | 14 | | |
| **595** | | **10** | 11 | | **3** | *14* | *3* | 15 | | |
| **594** | | **11** | 12 | | **4** | *15* | *4* | 16 | | |
| **593** | | **12** | 13 | | **5** | *16* | *5* | 1 | Psammetichus II | |
| **592** | | **13** | 14 | | **6** | *17* | *6* | 2 | | |
| **591** | | **14** | 15 | | **7** | *18* | *7* | 3 | | |
| **590** | | **15** | 16 | | **8** | *19* | *8* | 4 | | |
| **589** | | **16** | 17 | *Jer 52:4* | **9** | *20* | *9* | 5 | *Siege of Jerusalem* | *Ezk 24:1* |
| **588** | *Jubilee violated* | **17** | 18 | *50* | **10** | *21* | *10* | **6/1** | **Apries** | *Jer 32:1* |
| **587** | | **18** | 19 | *1 Temple destroyed* | **11** | *22* | *11* | 2 | *Exile of the people* | *Jer 52:1,12,29* |
| **586** | *Ezk 26:1-12* | **19** | 20 | *2* | *1* | *23* | *12* | 3 | *1st year of exile* | *Ezk 33:21* |
| **585** | | **20** | 21 | *3 Dan 2:1* | *2* | *24* | *13* | 4 | | |
| **584** | | **21** | 22 | *4* | *3* | *25* | *14* | 5 | | |
| **583** | *Dan 4:29* | **22** | 23 | *5* | *4* | *26* | *15* | 6 | *Last exile* | *Jer 52:30* |
| **582** | | **23** | 24 | 6 | *5* | *27* | *16* | 7 | | |
| **581** | | **24** | 25 | *7* | *6* | *28* | *17* | 8 | | |
| **580** | | **25** | 26 | 8 | 7 | 29 | 18 | 9 | | |
| **579** | | **26** | 27 | 9 | 8 | 30 | 19 | 10 | | |
| **578** | | **27** | 28 | 10 | 9 | 31 | 20 | 11 | | |
| **577** | *(7 years of madness)* | **28** | 29 | 11 | 10 | 32 | 21 | 12 | | |
| **576** | | **29** | 30 | 12 | 11 | 33 | 22 | 13 | | |
| **575** | | **30** | 31 | 13 | 12 | 34 | 23 | 14 | | |
| **574** | *(Tyre, siege of 13 years)* | **31** | 32 | 14 | 13 | 35 | 24 | 15 | | |
| **573** | *Against Apion 1:156* | 32 | 33 | 15 | 14 | 36 | 25 | 16 | | *Ezk 40:1* |
| **572** | | 33 | 34 | 16 | 15 | 37 | 26 | 17 | | |
| **571** | | 34 | 35 | 17 | 16 | 38 | 27 | 18 | | *Ezk 29:12-20* |
| **570** | | 35 | 36 | 18 | 17 | 39 | 28 | 19 | | |
| **569** | | 36 | 37 | 19 | 18 | 40 | 29 | **1** | 20 **Amasis** | |
| **568** | *VT 4956 (eclipse)* | **37** | 38 | 20 | 19 | 41 | 30 | 2 | 21 | *Jer 43:10,13* |
| **567** | *(Egypt attacked)* | **38** | 39 | 21 | 20 | 42 | 31 | **3** | **22** *Death of Apries* | |
| **566** | | 39 | 40 | 22 | 21 | 43 | 32 | 4 | *1 Egypt desolated 40 years* | *Jer 44:30* |
| **565** | | 40 | 41 | 23 | 22 | 44 | 33 | 5 | 2 | |
| **564** | | 41 | 42 | 24 | 23 | 45 | 34 | 6 | 3 | |
| **563** | | 42 | 43 | 25 | 24 | 46 | 35 | 7 | 4 | |
| **562** | **Amêl Marduk** | 43 | | 26 | 25 | 47 | 36 | 8 | 5 | |
| **561** | | 1 | | 27 | 26 | 48 | 37 | 9 | *6 Jehoiachin liberated* | *Jer 52:31* |
| **560** | **Neriglissar** | 2 | | 28 | 27 | 49 | | 10 | 7 | |
| **559** | *(Cyrus II Persian king)* | 1 | | 29 | 28 | 50 | | 11 | 8 | |
| **558** | | 2 | | 30 | 29 | 51 | | 12 | 9 | |
| **557** | | 3 | | 31 | 30 | 52 | | 13 | 10 | |
| **556** | **Labashi-Marduk** | 4 | | 32 | 31 | 53 | | 14 | 11 | |
| **555** | **Nabonidus** | 1 | | 33 | 32 | 54 | | 15 | 12 | |
| **554** | | 2 | | 34 | 33 | 55 | | 16 | 13 | |
| **553** | **Belshazzar** | 3 | 0 | 35 | 34 | 56 | | 17 | 14 | |
| **552** | | 4 | 1 | 36 | 35 | 57 | | 18 | 15 | *Dan 7:1* |
| **551** | | 5 | 2 | 37 | 36 | 58 | | 19 | 16 | |
| **550** | *Harpagus Median king* | 6 | 3 | 38 | 37 | 59 | | 20 | 17 | *Dan 8:1,20-21* |
| **549** | *vassal of Cyrus II* | 7 | 4 | 39 | 38 | 60 | | 21 | 18 | |
| **548** | | 8 | 5 | 40 | 39 | 61 | | 22 | 19 | |
| **547** | | 9 | 6 | 41 | 40 | 62 | | 23 | 20 | |
| **546** | | 10 | 7 | 42 | 41 | 63 | | 24 | 21 | |
| **545** | | 11 | 8 | 43 | 42 | 64 | | 25 | 22 | |
| **544** | | 12 | 9 | 44 | 43 | 65 | | 26 | 23 | |
| **543** | | 13 | 10 | 45 | 44 | 66 | | 27 | 24 | |
| **542** | | 14 | 11 | 46 | 45 | 67 | | 28 | 25 | |
| **541** | | 15 | 12 | 47 | 46 | 68 | | 29 | 26 | |
| **540** | | 16 | 13 | 48 | 47 | 69 | | 30 | 27 | |
| **539** | *Fall of Babylon* | 17 | 14 | **Cyrus II** | 48 | 70 | | 31 | 28 | *Jer 25:11,12* |
| **538** | *Freedom year 1* | 1 | | 50 | 49 | | | 32 | 29 | *Is 43:1,3; 45:1* |
| **537** | | 2 | | | 50 | | | 33 | 30 | |
| **536** | | 3 | | | 51 | | | 34 | 31 | *Dan 10:1* |
| **535** | | 4 | | | 52 | | | 35 | 32 | |
| **534** | | 5 | | | 53 | | | 36 | 33 | |
| **533** | | 6 | | | 54 | | | 37 | 34 | |
| **532** | | 7 | | | 55 | | | 38 | 35 | |
| **531** | | 8 | | | 56 | | | 39 | 36 | |
| **530** | | 9 | 0 | **Cambyses II** | 57 | | | 40 | 37 | |
| **529** | | | 1 | | 58 | | | 41 | 38 | |
| **528** | | | 2 | | 59 | | | 42 | 39 | |
| **527** | | | 3 | | 60 | | | 43 | 40 | *Ezk 29:12-16* |
| **526** | | | 4 | | 61 | | 1 | **44** | **Psammetichus III** | |
| **525** | | | 5 | | 62 | | 2 | | | *Fall of Egypt* |
| **524** | | | 6 | | 63 | | | | | |
| **523** | *Lunar eclipse 16 July* | | 7 | | 64 | | | | | *BM 33066* |
| **522** | | 0 | 8 | **Darius I** | 65 | | | | | |
| **521** | | 1 | | | 66 | | | | | |
| **520** | | 2 | | | *67* | | | | | |
| **519** | | 3 | | | *68* | | | | | |
| **518** | | 4 | | | *69* | *50* | | | *End of Temple's desolation* | *Zek 7:1-5* |
| **517** | | 5 | | | *70* | *1* | | | *New jubilee cycle* | *Dan 9:2* |
| **516** | | 6 | | | | *2* | | | | |
| **515** | | 7 | | | | *3* | | | | |
| **514** | | 8 | | | | *4* | | | | |
| **513** | | 9 | | | | *5* | | | | |
| **512** | | 10 | | | | *6* | | | | |
| **511** | | 11 | | | | *7* | | | | |
| **510** | | 12 | | | | *8* | | | | |
| **509** | | 13 | | | | *9* | | | | |
| **508** | | 14 | | | | *10* | | | | |
| **507** | | 15 | | | | *11* | | | | |
| **506** | | 16 | | | | *12* | | | | |
| **505** | | 17 | | | | *13* | | | | |
| **504** | | 18 | | | | *14* | | | | |
| **503** | | 19 | | | | *15* | | | | |
| **502** | *Lunar eclipse 19 Nov.* | 20 | | | | *16* | | | | *Almagest IV:9* |
| **501** | | 21 | | | | *17* | | | | |
| **500** | | 22 | | | | *18* | | | | |
| **499** | | 23 | | | | *19* | | | | |
| **498** | | 24 | | | | *20* | | | | |
| **497** | | 25 | | | | *21* | | | | |
| **496** | | 26 | **0** | **Xerxes I** | | *22* | | | | |
| **495** | | 27 | **1** | | | *23* | | | | |
| **494** | | 28 | **2** | | | *24* | | | | |
| **493** | | 29 | **3** | | | *25* | | | *Vashti repudiated* | *Est 1:3* |
| **492** | | 30 | **4** | | | *26* | | | | |
| **491** | *Lunar eclipse 25 Apr.* | 31 | **5** | | | *27* | | | | *Almagest IV:1* |
| **490** | | 32 | **6** | | | *28* | | | | |
| **489** | | 33 | **7** | | | *29* | | | *Wedding of Xerxes* | *Est 2:16-17* |
| **488** | | 34 | **8** | | | *30* | | | | |
| **487** | | 35 | **9** | | | *31* | | | | |
| **486** | | 36 | **10** | | | *32* | | | | |
| **485** | *Babylonian revolt* | | **11** | | | *33* | | | | *Ezr 4:6* |
| **484** | *Est 2:21-3:7* | | **12** | | | *34* | | | | *Est 3:7-10* |
| **483** | | | **13** | | | *35* | | | | |
| **482** | | | **14** | | | *36* | | | | |
| **481** | | | **15** | | | *37* | | | | |
| **480** | | | **16** | | | *38* | | | | |
| **479** | | | **17** | | | *39* | | | | |
| **478** | | | **18** | | | *40* | | | | |
| **477** | | | **19** | | | *41* | | | | |
| **476** | | | **20** | | | *42* | | | | |
| **475** | *Lunar eclipse Jun. 26* | 0 | **21** | **Artaxerxes I** | | *43* | | | | *BM 32234* |
| **474** | *Lunar eclipse Dec. 20* | 1 | | | | *44* | | | | |
| **473** | | 2 | | | | *45* | | | | |
| **472** | | 3 | | | | *46* | | | | |
| **471** | | 4 | | | | *47* | | | | |
| **470** | | 5 | | | | *48* | | | | |
| **469** | | 6 | | | | *49* | | | | |
| **468** | | 7 | | | | *50* | | | *1st jubilee celebrated* | *Ezr 7:1-8,24* |
| **467** | | 8 | | | | *1* | | | | |
| **466** | | 6 | | | | *2* | | | | |
| **465** | | 10 | | | | *3* | | | | |
| **464** | | 11 | | | | *4* | | | | |
| **463** | | 12 | | | | *5* | | | | |
| **462** | | 13 | | | | *6* | | | | |
| **461** | | 14 | | | | *7* | | | | |
| **460** | | **15** | | | | *8* | | | | |
| **459** | | **16** | | | | *9* | | | | |
| **458** | | **17** | | | | *10* | | | | |
| **457** | | **18** | | | | *11* | | | | |
| **456** | | **19** | | | | *12* | | | | |
| **455** | | **20** | | | | *13* | *1* | | *Beginning of 483 years* | *Dan 9:24-27* |
| **454** | | **21** | | | | *14* | *2* | | *(483 = 69x7)* | *Neh 2:1-9* |
| **453** | | **22** | | | | *15* | *3* | | | |
| **452** | | **23** | | | | *16* | *4* | | | |
| **451** | | **24** | | | | *17* | *5* | | | |
| **450** | | **25** | | | | *18* | *6* | | | |
| **449** | | **26** | | | | *19* | *7* | | | |
| **448** | | **27** | | | | *20* | *8* | | | |
| **447** | | **28** | | | | *21* | *9* | | | |
| **446** | | **29** | | | | *22* | *10* | | | |
| **445** | | **30** | | | | *23* | *11* | | | |
| **444** | | **31** | | | | *24* | *12* | | | |
| **443** | | **32** | | | | *25* | *13* | | *Inspection of Nehemiah* | *Neh 5:14* |
| **442** | | **33** | | | | *26* | *14* | | | |
| **441** | | **34** | | | | *27* | *15* | | | |
| **440** | | **35** | | | | *28* | *16* | | | |
| **439** | | **36** | | | | *29* | *17* | | | |
| **438** | | **37** | | | | *20* | *18* | | | |
| **437** | | **38** | | | | *31* | *19* | | | |
| **436** | | **39** | | | | *32* | *20* | | | |
| **435** | | **40** | | | | *33* | *21* | | | |
| **434** | | **41** | **0** | **Darius B** | | *34* | *22* | | | |
| **433** | | **42** | **1** | | | *35* | *23* | | | |
| **432** | | [] | **2** | | | *36* | *24* | | | |
| **431** | | [] | **3** | | | *37* | *25* | | | |
| **430** | | [] | **4** | | | *38* | *26* | | | |
| **429** | | [] | **5** | | | *39* | *27* | | | |
| **428** | | [] | **6** | | | *40* | *28* | | | |
| **427** | | [] | **7** | | | *41* | *29* | | | |
| **426** | | [] | **8** | | | *42* | *30* | | | |
| **425** | | **50** | **0** | **Xerxes II** | | *43* | *31* | | | |
| **424** | | **0** | 51 | **Darius II** | | *44* | *32* | | | |
| **423** | | **1** | | | | *45* | *33* | | | |
| **422** | | **2** | | | | *46* | *34* | | | |
| **421** | | **3** | | | | *47* | *35* | | | |
| **420** | | **4** | | | | *48* | *36* | | | |
| **419** | | **5** | | | | *49* | *37* | | | |
| **418** | | **6** | | | | *50* | *38* | | | |
| **417** | | **7** | | | | *1* | *39* | | | |
| **416** | | **8** | | | | *2* | *40* | | | |
| **415** | | **9** | | | | *3* | *41* | | | |
| **414** | | **10** | | | | *4* | *42* | | | |
| **413** | | **11** | | | | *5* | *43* | | | |
| **412** | | **12** | | | | *6* | *44* | | | |
| **411** | | **13** | | | | *7* | *45* | | | |
| **410** | | **14** | | | | *8* | *46* | | | |
| **409** | | **15** | | | | *9* | *47* | | | |
| **408** | | **16** | | | | *10* | *48* | | | |
| **407** | | **17** | | | | *11* | *49* | | *(49 = 7x7)* | *(Dan 9:25)* |
| **406** | | **18** | | | | *12* | *50* | | *Jerusalem city achieved* | *Neh 12:22-43* |
| **405** | | **19** | **0** | **Artaxerxes II** | | *13* | *51* | | *(inauguration)* | |
| **404** | | | **1** | | | *14* | *52* | | | |
| **403** | | | **2** | | | *15* | *53* | | | |
| **402** | | | **3** | | | *16* | *54* | | | |
| **401** | | | **4** | | | *17* | *55* | | | |

## MESOPOTAMIAN CHRONOLOGY ON THE PERIOD 1133-609

The Mesopotamian chronology of this period is reconstructed using the number of Assyrian eponyms (1 a year) and the length of Babylonian reigns (#) combined with the set of synchronisms among Assyrian and Babylonian kings in Annals:

| ASSYRIAN KING | # | Reign | BABYLONIAN KING | # | Reign | JUDEAN RULER | # | Reign |
|---|---|---|---|---|---|---|---|---|
| Aššur-rêš-iši I | 18 | **1133**-1115 | Ninurta-nâdin-šumi | 6 | **1133**-1127 | Eli (*Philistines*) | 40 | 1162-1122 |
| Tiglath-pileser I | 39 | 1115 - | Nebuchadnezzar I | 22 | 1127-1105 | Samson | 20 | 1122-1102 |
| | | | Enlil-nâdin-apli | 4 | 1105-1101 | Sons of Samuel | 5 | 1102-1097 |
| | | -1076 | Marduk-nâdin-ahhê | 18 | 1101-1083 | Saul | 40 | 1097 - |
| Ašared-apil-Ekur | 2 | 1076-1074 | Marduk-šapik-zêri | 13 | 1083-1070 | | | -1057 |
| Aššur-bêl-kala | 18 | 1074-1056 | Adad-apla-iddina | 22 | 1070-1048 | David | 40 | 1057 - |
| Erîba-Adad II | 2 | 1056-1054 | Marduk-ahhê-erîba | 1 | 1048-1047 | | | |
| Šamšî-Adad IV | 4 | 1054-1050 | Marduk-zêr-[…] | 12 | 1047-1035 | | | |
| Aššurnaṣirpal I | 19 | 1050-1031 | Nabû-šum-libur | 8 | 1035-1027 | | | -1017 |
| Shalmaneser II | 12 | 1031-1019 | Simbar-šipak | 18 | 1027-1009 | Solomon | 40 | 1017 - |
| Aššur-nêrârî IV | 6 | 1019-1013 | Ea-mukîn-zêri | 1 | 1009-1008 | | | |
| Aššur-rabi II | 41 | 1013 - | Kaššu-nâdin-ahi | 2 | 1008-1006 | | | |
| | | | Eulmaš-šakin-šumi | 17 | 1006 -989 | | | |
| | | | Ninurta-kudurri-uṣur I | 3 | 989-986 | | | |
| | | | Širiki-šuqamuna | 1 | 986-985 | | | |
| | | -972 | Mâr-bîti-apla-uṣur | 5 | 985-980 | | | -977 |
| Aššur-rêš-iši II | 5 | 972-967 | Nabû-mukîn-apli | 36 | 980-944 | Roboam | 17 | 977-960 |
| Tiglath-pileser II | 32 | 967-935 | Ninurta-kudurri-uṣur II | 3 | 944-943 | Abiyam | 3 | 960-957 |
| Aššur-dân II | 23 | 935-912 | Mâr-bîti-ahhê-iddin | | 943- ? | Asa | 41 | 957-916 |
| Adad-nêrârî II | 21 | **912**-891 | Šamaš-mudammiq | | ? -900 | Jehosaphat | 25 | 916-891 |
| Tukultî-Ninurta II | 7 | 891-884 | Nabû-šum-ukîn I | 12 | 900-888 | Jehoram | 8 | 893-885 |
| Aššurnaṣirpal II | 26 | 884-859 | Nabû-apla-iddina | 33 | 888-855 | [Athaliah] | 6 | 885-879 |
| Shalmaneser III | 35 | 859-824 | Marduk-zâkir-šumi I | 36 | 855-819 | Joas | 40 | 879-839 |
| Šamšî-Adad V | 13 | 824-811 | Marduk-balâssu-iqbi | 6 | 819-813 | Amasiah | 29 | 839-810 |
| Adad-nêrârî III | 28 | 811-783 | Bâba-ah-iddina | | 813- ? | Uziah | 52 | 810 - |
| Shalmaneser IV | 10 | 783 - | 5 unknown kings | | ? | | | |
| | | | Ninurta-apla-[…] | | ? | | | |
| | | -773 | Marduk-apla-uṣur | | ? -770 | | | |
| Aššur-dân III | 18 | **773**-755 | Erîba-Marduk | 9 | 770-761 | | | -758 |
| Aššur-nêrârî V | 10 | 755-745 | Nabû-šum-iškun | 13 | 761-748 | Jotham | 16 | 758-742 |
| Tiglath-pileser III | 18 | 745 - | Nabû-naṣir | 14 | 748-734 | Ahaz | 16 | 742 - |
| | | | Nabû-nâdin-zêri | 2 | 734-732 | | | |
| | | | Nabû-šum-ukîn II | 1 | 732-731 | | | |
| | | | Nabû-mukîn-zêri | 2 | 731-729 | | | |
| | | -727 | Tiglath-pileser III Pûlu | 2 | 729-727 | | | -726 |
| Shalmaneser V | 5 | 727-722 | Shalmaneser V Ulûlaiu | 5 | 727-722 | Ezechias | 29 | 726 - |
| Sargon II | 17 | 722 - | Merodachbaladan II | 12 | 722-710 | | | |
| | | -705 | Sargon II | 5 | 710-705 | | | |
| Sennacherib | 24 | 705 - | Sennacherib | 2 | 705-703 | | | |
| | | | Marduk-zakir-šumi II | 0 | 703 | | | |
| | | | Bêl-ibni | 3 | 703-700 | | | -697 |
| | | | Aššur-nâdin-šumi | 6 | 700-694 | Manasseh | 55 | 697 - |
| | | | Nergal-ušezib | 1 | 694-693 | | | |
| | | | Mušezib-Marduk | 4 | 693-689 | | | |
| | | -681 | Sennacherib | 8 | 689-681 | | | |
| Esarhaddon | 12 | 681-669 | Esarhaddon | 12 | 681-669 | | | |
| Aššurbanipal | 42 | 669-627 | Šamaš-šum-ukîn | 40 | 668-648 | | | -642 |
| Aššur-etel-ilâni | 4 | 630 - | Kandalanu | 21 | 648 - | Amon | 2 | 642-640 |
| | | | Sin-šum-lišir | | -627 | Josias | 31 | 640 - |
| | | -626 | Sin-šar-iškun | 1 | 627-626 | | | |
| Sin-šar-iškun | 14 | 626-612 | Nabopolassar | 21 | 626 - | | | |
| Aššur-uballiṭ II | 3 | 612-**609** | | | **-605** | | | **-609** |

## MESOPOTAMIAN CHRONOLOGY ON THE PERIOD 1799-1133

Fifteen royal chronicles[40] make it possible to partially reconstruct[41] the Babylonian and Kassite chronologies[42]:

| n° | KASSITE KING | Reign | # | King Lists | Highest date | BABYLONIAN KING | Reign | # |
|---|---|---|---|---|---|---|---|---|
| | | | | | | Sûmû-abum | 1799-1785 | 14 |
| | | | | | | Sûmû-la-Il | 1785-1749 | 36 |
| | | | | | | Sâbium | 1749-1735 | 14 |
| | | | | | | Apil-Sîn | 1735-1717 | 18 |
| | | | | | | Sîn-muballiṭ | 1717-1697 | 20 |
| | | | | | | Hammurabi | **1697**-1654 | 43 |
| 1 | Gandaš | *1661-1635* | 26 | **[2]6** | | Samsu-iluna | 1654 - | 38 |
| 2 | Agum I | *1635-1613* | 22 | **22** | | | -1616 | |
| 3 | Kaštiliaš I | *1613-1591* | 22 | **22** | | Abi-ešuḫ | 1616-1588 | 28 |
| 4 | Ušši | *1591-1583* | 8 | **8** | | Ammiditana | 1588 - | 37 |
| 5 | Abirattaš | *1583-1567* | *[16]* | | | | | |
| 6 | Kaštiliaš II | *1567-1551* | *[16]* | | | | -1551 | |
| 7 | Urzigurumaš | *1551-1535* | *[16]* | | | Ammiṣaduqa | **1551-1530** | 21 |
| 8 | Harbašihu | *1535-1519* | *[16]* | | | Samsuditana | 1530 - | 31 |
| 9 | Tiptakzi | *1519-1503* | *[16]* | | | | -**1499** | |
| 10 | Agum II | *1503-1487* | *[16]* | | | *"Babylon's restoration"* | **1498** - | 41 |
| 11 | Burna-Buriaš I | *1487-1471* | *[16]* | | | | | |
| 12 | Kaštiliaš III | *1471-1455* | *[16]* | | | | -1457 | |
| 13 | Ulam-Buriaš | *1455-1439* | *[16]* | | | | **1457** - | |
| 14 | Agum III | *1439-1423* | *[16]* | | | | | |
| 15 | Kadašman-Harbe I | *1423-1407* | *[16]* | | | | | |
| 16 | Kara-indaš | *1407-1391* | *[16]* | | | | | |
| 17 | Kurigalzu Ier | *1391-1375* | *[16]* | | | | | |
| 18 | Kadašman-Enlil I | **1375**-1360 | 15 | | **15** | | | |
| 19 | Burna-Buriaš II | 1360-1333 | 27 | | **27** | | | |
| 20 | Kara-ḫardaš | 1333 | 0 | | **0** | | | |
| 21 | Nazi-Bugaš | 1333 | 0 | | **0** | | | |
| 22 | Kurigalzu II | 1333-1308 | 25 | [**25**] | 24 | | | |
| 23 | Nazi-Maruttaš | 1308-1282 | 26 | **26** | 24 | | | |
| 24 | Kadašman-Turgu | 1282-1264 | 18 | **18** | 17 | | | |
| 25 | Kadašman-Enlil II | 1264-1255 | 9 | [10 (+x)] | 8 [**9**?] | | | |
| 26 | Kudur-Enlil | 1255-1246 | 9 | [6] | **9** | | | |
| 27 | Šagarakti-šuriaš | 1246-1233 | 13 | 13 | **13** | | | |
| 28 | Kaštiliašu IV | 1233-1225 | 8 | **8** | 8 | | | |
| | [Tukulti-Ninurta I] | [1225-1224] | [1] | | **[1]** | | | |
| 29 | Enlil-nâdin-šumi | 1225-1224 | 1 | 1,5 | **1** | | | |
| 30 | Kadašman-Harbe II | 1224-1223 | 1 | 1,5 | **1** | | | |
| 31 | Adad-šuma-iddina | 1223-1217 | 6 | **6** | 0 | | | |
| 32 | Adad-šuma-uṣur | 1217-1187 | 30 | **30** | 13 | | | |
| 33 | Meli-Šipak | 1187-1172 | 15 | **15** | 12 | | | |
| 34 | Marduk-apla-iddina | 1172-1159 | 13 | **13** | 6 | | | |
| 35 | Zababa-šuma-iddina | 1159-1158 | 1 | **1** | | | | |
| 36 | Enlil-nâdin-ahi | 1158-**1155** | 3 | **3** | | | | |
| | | | | | | Marduk-kabit-aḫḫešu | **1155**-1141 | 14 |
| | | | | | | Itti-Marduk-balaṭu | 1141-**1133** | 8 |

40 J.J. GLASSNER – Chroniques mésopotamiennes (n°22)
Paris 1993 Éd. Belles Lettres pp. 137-179.

41 F. JOANNÈS - Dictionnaire de la civilisation mésopotamienne
Paris 2001 Éd. Robert Laffont pp. 164,522,565,758.
F. JOANNÈS - La Mésopotamie au 1er millénaire avant J.C.
2000 Paris Ed. Armand Colin pp. 186-187.

42 J.A. BRINKMAN – Materials and Studies for Kassite History Vol. I
Chicago 1976 Ed. The Oriental Institute of the University of Chicago pp. 6-34.

After the fall of the city, the Babylonian dynasty was replaced by the Kassites in the north, and by the Sealand dynasty in the south, whose beginning is not well known[43]. It begins with Gandaš, a contemporary of Samsu-iluna, and it is precisely dated from Kadašman-Enlil I. Synchronisms[44] with Kassites kings (highlighted reigns) are as follows:

- The reigns of Kassites kings are mentioned in a Babylonian king list[45] very incomplete (Babylonian King List A) which assigns to these 36 kings a total of 576 years, implying an average duration of 16 years by reign. The first four reigns have respective durations of: [1]6/[2]6[46], 22, 22 and 8 years.
- The appearance of Ulam-Buriaš coincides with the end of the restoration of Babylon (year 41 of "recovery").
- A tablet (VAT 1429) describes the Kassite king Agum II as *bukašu* "Duke" of Babylon, who is actually the first Kassite king of Babylon[47].
- Alliance between Assyrian Puzur-Aššur III (1491-1467) and Kassite Burna-Buriaš I.
- Alliance between Assyrian Aššur-bêl-nišešu (1411-1403) and Kassite Kara-indaš.
- A Babylonian chronicle mentions, in the year 9 of Samsu-iluna, an assault by Kassite troops, likely led by Gandaš, the first Kassite king[48].

The reign of Kassite King Gandaš, obtained from average durations, is dated 1661-1635 and coincides with the reign of the Assyrian king Samsu-iluna (1654-1616). The reign of Kassite King Agum II (1503-1487) is consistent with a fall of Babylon in -1499. If one accepts the total duration of 576 years, the reign of Kassite King Gandaš had to start around 1731 (= 1155 + 576) instead of 1651, which would support the Middle Chronology, but in this case the average length of reigns would increase from 16 to 32 years (from Abirattaš to Agum II), but no known reign has reached such a length, in addition, this contradicts the average value of 16 years indicated by the Chronicle (this value is consistent with those of the known reigns). Finally, according to the Middle Chronology, interruption of Babylonian kingdoms would have been complete for a century (from 1595 to 1495), which is very unlikely:

| MIDDLE CHRONOLOGY | | | | | |
|---|---|---|---|---|---|
| KASSITE KING | Reign | Length | BABYLONIAN KING | Reign | Length |
| Gandaš | *1760-1734* | [2]6 | Samsu-iluna | **1750** - | 38 |
| Agum I | *1734-1712* | 22 | | -1712 | |
| Kaštiliaš I | *1712-1690* | 22 | Abi-ešuḫ | 1712 - | 28 |
| Ušši | *1690-1682* | 8 | | -1684 | |
| Abirattaš | *1682-1650* | *[32]** | Ammiditana | 1684-1647 | 37 |
| Kaštiliaš II | *1650-1618* | *[32]** | Ammiṣaduqa | 1647-1626 | 21 |
| Urzigurumaš | *1618-1586* | *[32]** | Samsuditana | 1626-**1595** | 31 |
| Harbašihu | *1586-1554* | *[32]** | | | |
| Tiptakzi | *1554-1522* | *[32]** | | * | - |
| Agum II | *1522-1490* | *[32]** | *"Babylon's restoration"* | **1495** - | 41 |
| Burna-Buriaš I | *1490-1470* | *[20]* | | | |
| Kaštiliaš III | *1470-1454* | *[16]* | | -1454 | |
| Ulam-Buriaš | *1454-1438* | *[16]* | | **1454** - | |
| Agum III | *1438-1422* | *[16]* | | | |

[43] F. JOANNÈS - Dictionnaire de la civilisation mésopotamienne
Paris 2001 Éd. Robert Laffont pp. 164.
[44] K. GRAYSON – Texts from Cuneiform Sources Volume V Assyrian and Babylonian Chronicles (ABC 20, 21)
Winona Lake 2000 Ed. Eisenbrauns pp. 157-170.
[45] J.B. PRITCHARD - Ancient Near Eastern Texts
Princeton 1969 Ed. Princeton University Press p. 272).
[46] J.A. BRINKMAN – Materials and Studies for Kassite History Vol. I
Chicago 1976 Ed. The Oriental Institute of the University of Chicago p. 128.
[47] J. FREU, M. MAZOYER – Des origines à la fin de l'Ancien royaume hittite
Paris 2007Éd. L'Harmattan p. 114.
[48] A. GOETZE – The Kassites and near Eastern Chronology
in: *Journal of Cuneiform Studies* 18:4 (1964) p. 97.

According to a Babylonian Chronicle[49]: *he did battle against him [...] their corpses, the sea [...] he changed and Samsu-iluna [...] Iliman attacked and [brought about] the defeat of [his] army. Albishi, son of Samsu-iluna, set out to conquer Iliman. He decided to dam the Tigris. He dammed the Tigris but did not capture Iliman. At the time of Samsu-ditana the Hittites marched against Akkad. Ea-gamil, king of the Sealand, f[led] to Elam. After he had gone, Ulamburiash, brother of Kashtiliash (III), the Kassite, mustered his army and conquered the Sealand. He was master of the land. Agum (III), son of Kashtiliash (III), mustered his army and marched to the Sealand.* Thus, after the fall of Babylon *[in 1499 BCE]*, Agum II a Kassite King of Chaldean origin began to dominate northern Babylonia (land of Akkad) and the Sealand Kings, perhaps of Sumerian origin, began to dominate southern Babylonia (land of Sumer) up to Ulam-Buriaš who defeated them *[around 1450 BCE]*. A Synchronistic King List[50] and a tablet (KAV 216, Assur 14616c) very difficult to read give the following list: *Samsud[itana reigned 31 years]. 11 kings, the cycle of [Babylon; they reigned 300 years]. The cycle of Babylon [changed, her kingship went to Sealand]. At Urukuga, 60 [years] Ilimalum, king; 56 [years] Itti-ili-nibi; 36 [years] Damqi-ilišu; 15 [years] Iškibal; 26 [years] Šušši, brother; 55 [years] Gulkišar; 50 [years] Pešgaldarameš; 28 [years] Ayadaragalama, his son, same; 26 [years] Akurulana; 7 [years] Melamkukurra; 9 [years] Ea-gam[il]; 368 [years] 11 kings, dynasty of Urukuga.* If one adds the sum of these abnormally long reigns (368 years), Iliman, the first king, should have reigned from 1818 to 1758, which is not possible according to the synchronisms (highlighted) . In fact, co-regencies were common among the kings of Sealand dynasty since the year 7 of Pešgaldarameš is also dated to Ayadaragalama and his year 29 is followed by the accession of Ayadaragalama[51]. Kassite and Sealand dynasties being close in time and space, it is more likely to have an average of 16 years of reign.

| AKKADIAN | | | | | | SUMERIAN | | |
|---|---|---|---|---|---|---|---|---|
| | | | BABYLONIAN | Reign | # | ISINIAN | Reign | # |
| | | | Sâbium | 1749 - -1735 | 14 | Iter-piša | 1740-1736 | [4] |
| | | | | | | Ur-dukuga | 1736-1732 | [4] |
| | | | Apil-Sîn | 1735-1717 | 18 | Sîn-mâgir | 1732-1721 | 11 |
| | | | Sîn-muballiṭ | 1717-1697 | 20 | Damiq-ilîšu I | 1721-1698 | 23 |
| KASSITE | Reign | # | Hammurabi | 1697-1654 | 43 | SEALAND | Reign | # |
| Gandaš | *1661-1635* | [2]6 | Samsu-iluna | 1654 - -1616 | 38 | Ilum-maz-ilî | *1654 - -1594* | 60 |
| Agum I | *1635-1613* | 22 | | | | | | |
| Kaštiliaš I | *1613-1591* | 22 | Abi-ešuḫ | 1616-1588 | 28 | | | |
| Ušši | *1591-1583* | 8 | Ammiditana | 1588 - -1551 | 37 | Itti-ili-nîbî | *1594-1578* | *[16]* |
| Abirattaš | *1583-1567* | *[16]* | | | | Damqi-ilišu II | *1578-1562* | *[16]* |
| Kaštiliaš II | *1567-1551* | *[16]* | | | | Iškibal | *1562-1546* | *[16]* |
| Urzigurumaš | *1551-1535* | *[16]* | Ammiṣaduqa | 1551-1530 | 21 | Šušši | *1546-1530* | *[16]* |
| Harbašihu | *1535-1519* | *[16]* | Samsuditana | 1530 - -**1499** | 31 | Gulkišar | *1530-1514* | *[16]* |
| Tiptakzi | *1519-1503* | *[16]* | | | | Pešgaldarameš | *1514-1498* | *[16]* |
| Agum II | *1503-1487* | *[16]* | | | 300 | Ayadaragalama | *1498-1482* | *[16]* |
| Burna-Buriaš I | *1487-1471* | *[16]* | | | | Akurulana | *1482-1466* | *[16]* |
| Kaštiliaš III | *1471 - -1455* | *[16]* | | | | Melamkukurra | *1466-1459* | 7 |
| | | | | | | Ea-gam[il] | *1459 - -1450* | 9 |
| Ulam-Buriaš | *1455-1439* | *[16]* | | | | | | |
| Agum III | *1439-1423* | *[16]* | | | | | | |
| Kadašman-Harbe I | *1423-1407* | *[16]* | | | | | | |
| Kara-indaš | *1407-1391* | *[16]* | | | | | | |
| Kurigalzu Ier | *1391-1375* | *[16]* | | | | | | |
| Kadašman-Enlil I | **1375**-1360 | 15 | | | | | | |

---

49 K. GRAYSON – Texts from Cuneiform Sources Volume V Assyrian and Babylonian Chronicles Winona Lake 2000 Ed. Eisenbrauns pp. 156.

50 J.B. PRITCHARD - Ancient Near Eastern Texts Princeton 1969 Ed. Princeton University Press pp. 271-272.

51 S. DALLEY – Babylonian Tablets from the First Sealand Dynasty in the Schøyen Collection in: *Cornell University Studies in Assyriology and Sumerology* Vol. 9 (CDL Press, 2009) pp. 4-10.

Several synchronisms with Assyrian reigns on the period 1400-1200 confirm the previous chronology, taking into account that before Aššur-dân I the Assyrians years are lunar (354 days) and no longer solar (365 days), making it necessary to reduce Assyrian reigns of 1 year every 33 years [1 year = 33x(365 - 354 days)][52]. The following synchronisms highlight this discrepancy:

- The Assyrian king Tukulti-Ninurta I replaced the Babylonian king Kaštiliašu IV (1233-1225) by Enlil-nâdin-šumi (1225-1224) during the 19th eponym[53] Ina-Aššur-šumi-aṣbat[54]. The first eponymous being the year of accession, the 19th eponym refers to year 18.
- The disappearance of the Mitannian empire (Hanigalbat) is dated in year 6/7 of Shalmaneser I since there are at least 5 eponyms before this victory[55] and 7 at most[56]. But peace and alliance concluded by Hattusili III in the year 21 of Ramses II and the tightening of links between Hattusili III and Kadašman-Turgu (1282-1264) were responses to the threat on the Eastern border of Hatti following the disappearance of Hanigalbat[57], which implies the following synchronism: the year 21 of Ramses II fits the year 7 of Shalmaneser I. Thus the accession of Kadašman-Enlil II (1264-1255) matches the year 19 of Ramses II[58] (1283 = 1264 + 19). Gasche proposed to advance the Babylonian chronology of 5 years to calibrate the Egyptian chronology with an accession of Ramses II in 1279[59], instead of 1283. This solution is not acceptable, because if the reign of Enlil-nâdin-ahi is shifted by 5 years (1153-1150 instead of 1158-1155), this contradicts the accuracy of the Assyrian eponyms and Babylonian chronology on this period. Consistent with this reliability, the most logical choice is to anchor the Egyptian chronology on Babylonian chronology and not vice versa.
- The letter EA 16[60] d'Aššur-uballiṭ I is addressed to the pharaoh Aÿ (1327-1323).

| n° | ASSYRIAN KING | | | | | gap |
|---|---|---|---|---|---|---|
| 72 | Erîba-Adad I | 1391-1364 | 27 | EGYPTIAN KING | | |
| 73 | Aššur-uballiṭ I | 1364-1328 | 36 | | | |
| | | | | Aÿ | 1327-1364 | **3<** |
| 74 | Enlil-nêrârî | 1328-1318 | 10 | | | |
| 75 | Arik-dên-ili | 1318-1306 | 12 | | | |
| 76 | Adad-nêrârî I | 1306-1274 | 32 | BABYLONIAN KING | | |
| 77 | Salmanazar I | 1274-1267 | 30 | | | |
| | (year 7) | **1267**-1244 | | Kadašman-Enlil II | **1264**-1255 | **3** |
| 78 | Tukultî-Ninurta I | 1244-1226 | 37 | | | |
| | (year 18) | **1226**-1207 | | Enlil-nâdin-šumi | **1225**-**1224** | **2** |
| 79 | Aššur-nâdin-apli | 1207-1203 | 4 | | | |
| 80 | Aššur-nêrârî III | 1203-1197 | 6 | | | |
| 81 | Enlil-kudurri-uṣur | 1197-1192 | 5 | | | |
| 82 | Ninurta-apil-Ekur | 1192-1179 | 13 | | | |
| 83 | Aššur-dân Ier | 1179-**1133** | 46 | | | |

[52] There is an exact difference of 6 days = 33x(29,530588x12) - 32x(365,24219). In fact, (Nx1,0306889) lunar years = N solar years.
[53] W. RÖLLING – Eponymen in den Mittelassyrischen Dokumenten aus Tall Seh Hamad/Dur-Katlimmu
in: *Zeitschrift für Assyrologie und vorderasiatische Archäologie* 94:1 (2004) pp 18-51.
[54] E.C. CANCIK-KIRSCHBAUM – Mittelassyrischen aus Tall Seh Hamad
in: *Berichte der Ausgrabung Tall Seh Hamad* 4:1 (1996) pp.
[55] A. HARRAK – Assyria and Hanigalbat
Hildesheim 1987 Georg Olms Verlag pp. 117-120, 157-162.
[56] Y. BLOCH – The Order of Eponyms in the Reign of Shalmaneser I
in: *Ugarit-Forschungen* Band 40/2008 (Münster 2009) pp. 143-178.
[57] J. FREU – De la confrontation à l'entente cordiale: Les relations assyro-hittites
in: Hittite Studies in Honor of Harry A. Hoffner Jr. (2003) Ede. Eisenbrauns pp. 102,103.
[58] W.A. WARD - The Present Status of Egyptian Chronology
in: *Bulletin of the American Schools of Oriental Research* 288 (1991) pp. 55,56.
[59] H. GASCHE, J.A. ARMSTRONG, S.W. COLE – Dating the Fall of Babylon
in: Mesopotamian History and Environment (1998) Chicago p. 65.
[60] W.L. MORAN - Les lettres d'El Amarna
in: *LIPO* n°13 Paris 1987 Éd. Cerf pp. 106-109.

## ASSYRIAN CHRONOLOGY ON THE PERIOD 1799-1133

The exact period of the adoption of the lunisolar calendar (Babylonian-inspired) by the Assyrians is difficult to determine because of the small number of documents. According to the *Assyrian Royal List*, eponyms appeared at the time of Sulili (around -1950) and were recorded from Êrišu I (1873-1834). The Assyrian dating system is based on the principle: 1 eponym = 1 year, the paleo-Assyrian calendar had to appear at that time. The names of 12 months of the year were influenced by other neighboring calendars (Sumerian and Akkadian)[61] and stabilized only in the eponymy of Ḫabil-kenum[62] (around -1650):

| | SUMERIAN | AKKADIAN | JULIAN |
|---|---|---|---|
| I | BÀRA-ZAG-GAR | Nisannu | March/April |
| II | GUD-SI-SÀ$_2$ | Ayyaru | April/May |
| III | SIG$_4$-GA | Simanu | May/June |
| IV | ŠU-NUMUN-NA | Du'ùzu | June/July |
| V | NE-IZI-GAR | Abu | July/August |
| VI | KIN-$^{d}$INNINA | Ulûlû | August/Sept. |
| VII | DU$_6$-KÙ$_3$ | Tašrîtu | September/Oct. |
| VIII | APIN-DU$_8$-A | Araḫsamna | October/Nov. |
| IX | GAN-GAN-E | Kisilimu | November/Dec. |
| X | AB-BA-È$_3$ | Tebêtu | December/Jan. |
| XI | ZIZ$_2$-AM | Šabâtu | Jannuary/Feb. |
| XII | ŠE-KIN-KU$_5$ | Addâru | February/March |

| | PALEO-ASSYRIAN | ASSYRIAN |
|---|---|---|
| *i* | *Ṣip'im* | Ṣippu |
| *ii* | *Qarrâtim* | Qarrâtu |
| *iii* | *Kanwarta* | Kalmartu |
| *iv* | *Te'inâtim* | $^{d}$Sin |
| *v* | *Kuzallu* | Kuzallu |
| *vi* | *Allanâtim* | Allanâtu |
| *vii* | *Bêltî-ekallim* | Belêt-ekalli |
| *viii* | *Ša sarratim* | Ša sarrâte |
| *ix* | *Narmak Aššur ša kênâtim* | Ša kênâte |
| *x* | *Maḫḫurili* | Muḫḫur ilâni |
| *xi* | *Ab šarrâni* | Abû šarrâni |
| *xii* | *Ḫubur* | Ḫubur |

Under the increasing influence of international relations, caused by new political and trade links, the Assyrian calendar is gradually replaced by the standard Mesopotamian calendar established by the Babylonian king Samsu-iluna (1654-1616). However two major issues stand out these calendars: 1) the Babylonian contracts are frequently dated: "day D, month M, year Y of King-so" while the Assyrian contracts are sometimes (1% of cases)[63] dated: "day D, month M, eponym so and so", 2) the Assyrians, unlike the Babylonians, never mention astronomical observations, which implies a lack of synchronization between the solar year (365 days) with 12 lunar months (354 days) through intercalary months. This fundamental difference can be detected by 1) military campaigns that took place (almost) always outside of the rainy season, between the spring equinox (month I of the Babylonian calendar) and the autumnal equinox (month VII) and 2) a statement of contracts in some seasons[64] (period 1800-1700): spring (vernal equinox on April; 5) winter (winter solstice on January 5;) harvest (summer solstice on July 8); the end of harvest and the beginning of plowing in autumn (autumn equinox on October 7).

Contracts of Assyrian merchants sometimes associate a month to a season. The one sent to Ilî-âLUM (dated to -1800) parallels the month of Assur Narmak with spring, another addressed to Šu-Hubur (from the same period) juxtaposes the term of the year with the harvest[65] (July). Another contract, whose Iddin-Suen is the guarantor, put the month of Ṣip'im at harvest time[66]. A second set of contracts and letters, when Šamšî-Adad

[61] Paleo-Assyrian months are preceded by the Akkadian word *waraḫ* "lunation" and Assyrian months by the Sumerian word ITI "month".
[62] M.E. COHEN - The Cultic Calendars of the Ancient Near East
Maryland 1993 Ed. CDL Press pp. 237-247, 297-305.
[63] C. MICHEL – Correspondance des marchands de Kaniš au début du II$^{e}$ millénaire avant J.-C.
in: *Littératures Anciennes du Proche-Orient* 19 (Cerf, 2001) pp. 547-548.
[64] K.R. VEENHOF, J. EIDEM – Mesopotamia. The Old Assyrian Period
in: *Orbis Biblicus et Orientalis* 160/5 (2008) pp. 234-245.
[65] R. LABAT – Unusual Eponymy-datings from Mari and Assyria
in: *Revue d'Assyriologie et d'archéologie orientale* 74:1 (1974) pp. 15-20.
[66] J. G. DERCKSEN – The beginning of the Old Assyrian year
in: *Nouvelles Assyriologiques Brèves et Utilitaires* 2011 N°4 pp. 90-91.

annexed Mari (1697-1680), shows that there was a lag of 5 months between the Amorite calendar of Šamšî-Adad (who died in early Ṣip'im)[67] and the one of Mari[68]. Some of these texts[69] put in connection the month of Ṣip'im with the harvest (in July) and the month of Te'inâtim with the harvest of late figs (in September / October):

| | N° | MARI | N° | ŠAMŠÎ-ADAD | (*)[70] | N° | PALEO-ASSYRIAN | JULIAN | # |
|---|---|---|---|---|---|---|---|---|---|
| *VI* | *i°* | *Ḫubur (ḫilib)* | *i** | *Niqmum* | *(vii)* | *i* | *Ṣip'im* | July (*harvest*) | 8 |
| *VII* | *ii°* | *Kinûnum* | *ii** | *Kinûnum* | *(viii)* | *ii* | *Qarrâtim* | August | 9 |
| *VIII* | *iii°* | *Dagan* | *iii** | *Tamḫîrum* | *(ix)* | *iii* | *Kanwarta* | September | 10 |
| *IX* | *iv°* | *Lîliatum* | *iv** | *Nabrûm* | *(x)* | *iv* | *Te'inâtim* | October (*figs*) | 11 |
| *X* | *v°* | *Bêlet-bîrî* | *v** | *Mammîtum* | *(xi)* | *v* | *Kuzallu* | November | 12 |
| *XI* | *vi°* | *Kiskissum* | *vi** | *Mana* | *(xii)* | *vi* | *Allanâtim* | December | 1 |
| *XII* | *vii°* | *Ebûrum* | *vii** | *Ayyarum* | *(i)* | *vii* | *Bêltî-ekallim* | January (*winter*) | 2 |
| *I* | *viii°* | *Urâḫum* | *viii** | *Niggalum* | *(ii)* | *viii* | *Ša sarratim* | February | 3 |
| *II* | *ix°* | *Malkânum* | *ix** | *Maqrânum* | *(iii)* | *ix* | *Narmak Aššur ša kînâtim* | March | 4 |
| *III* | *x°* | *Laḫḫum* | *x** | *Du'uzum* | *(iv)* | *x* | *Maḫḫurili* | April (*spring*) | 5 |
| *IV* | *xi°* | *Abum* | *xi** | *Abum* | *(v)* | *xi* | *Ab šarrâni* | May | 6 |
| *V* | *xii°* | *Ḫibirtum* | *xii** | *Tîrum* | *(vi)* | *xii* | *Ḫubur* | June | 7 |

These equivalences show that the paleo-Assyrian calendar was not synchronized with the spring equinox as the Babylonian calendar was. Ṣip'im marks the beginning of the Assyrian year, since a multi-year contract is completed in conjunction with this month (July at that time) and another (TPAK 1, 98) reports that it is the revival (*edâš*) [of the year]. Since no Assyrian contract is completely dated, it is not possible to establish an exact correspondence between months, in addition, the two series (in -1800 and -1700), being separated by a period of about 100 years, the coincidence with the seasons is fortuitous since 98 solar years = 101 lunar years.

The following inscription[71] of Tiglath-pileser I (1115-1076) containing a double date can be used to synchronize the Assyrian calendar: *I crossed the Euphrates 28 times, 2 times in one year, in pursuit of the Arameans aḫlamû (...) I captured the palaces of Babylon which belonged to Marduk-nadîn-ahhê king of Karduniash*[72]*, and I burned them. In the eponym of Aššur-šumu-ereš (and) in the eponym of Ninuaya, 2 times, I drove a battle of chariots online against Marduk-nadîn-ahhê king of Karduniash, and I defeated him (...) Month of Hibur, equivalent of the (Babylonian) month of Kislev, 18*$^{th}$ *day [eponym] of Taklak-ana-Aššur.* Another inscription says: *I crossed the Euphrates [27?] times, 2 times in one year, in pursuit of the Arameans aḫlamû (...) Month of Kuzallu, 13*$^{th}$ *day, eponym of Ninuaya son of Aššur-aplu-lišir.* Depending on the date at the end of the first inscription, the Babylonian calendar had become the reference. Assyrian kings performing a traditional military campaign each year, the mention of 28 crossings of the Euphrates, including 2 in one year, implies to date this inscription shortly after the year 1088 (= 1115 - 27). Thus, at that time (in -1088), the 12$^{th}$ month of the Assyrian calendar (Ḫubur) matched the 9$^{th}$ month of the Babylonian calendar (Kislev), which confirms their desynchronization. The Babylonian year began on 1$^{st}$ Nisan, or April 12 in -1088, when the Assyrian year began on 1$^{st}$ Ṣippu or January 13 in -1088.

[67] C. MICHEL – Correspondance des marchands de Kaniš au début du IIe millénaire avant J.-C.
in: *Littératures Anciennes du Proche-Orient* 19 (Cerf, 2001) pp. 309-310, 376-377.
[68] D. CHARPIN, N. ZIEGLER – Florilegium marianum V. Mari et le Proche-Orient à l'époque amorrite
in: *Mémoires de N.A.B.U.* 6 (2003) pp. 155-156.
[69] J. G. DERCKSEN – Weeks, Months and Years in Old Assyrian Chronology
in: *Bibliotheca Orientalis* LXVII 3/4 (2011) pp. 234-243.
[70] Months inside brackets refer to the old numbering assuming a beginning of the Assyrian year at the winter solstice. The calendar of Mari begins at *Urâḫum* (= *(w)arḫum* "month") and ends at *Ebûrum* ("Harvest"), to the autumn equinox (October).
[71] A.K. GRAYSON – Assyrian Royal Inscriptions part 2
Wiesbaden 1976 Ed. Otto Harrassowitz pp. 24-29.
[72] A synchronic list places the event in the 2nd year of Marduk-nadin-ahhê (1101-1083).

The presence of a double date in the reign Tiglath-pileser I shows that the new calendar adopted by the Assyrians (the Babylonian calendar) was not yet familiar. This change, which occurred shortly before the reign of Tiglath-pileser I, implies a desynchronization of eponyms since the beginning of the Assyrian year began on 1st Ṣippu while the Babylonian year began on 1st Nisan. The eponym marking each new Assyrian year was therefore chosen from the month of Nisan and not from the month of Ṣippu, for practical reasons. Indeed, the equivalence: 1 year = 1 eponym = 1 campaign, is generally verified but, for reasons of stewardship (the army on campaign had to be fed, in addition, the movements should be done on practicable grounds) military campaigns took place outside the rainy season, between the spring equinox (month I) and the autumnal equinox (month VII). The completion of two campaigns in one year is indeed exceptional. The number of years (Nb) is equal to the number of campaigns minus one:

| | month | | | | Nb | **year** | **King** / *eponym* |
|---|---|---|---|---|---|---|---|
| **-1089** | 1 | X | *i* | Ṣippu | 26 | [25] | **Tiglath-pileser I** |
| | 2 | XI | *ii* | Qarrâtu | | | |
| | 3 | XII | *iii* | Kalmartu | | | |
| | 4 | I | *iv* | $^{d}$Sin | | **[26]** | *Ninuaya son of Aššur-aplu-lišir* |
| | 5 | II | *v* | Kuzallu | | | |
| | 6 | III | *vi* | Allanâtu | | | |
| | 7 | IV | *vii* | Belêt-ekalli | | | |
| | 8 | V | *viii* | Ša sarrâte | | | |
| | 9 | VI | *ix* | Ša kênâte | | | |
| | 10 | VII | *x* | Muḫḫur ilâni | | | |
| | 11 | VIII | *xi* | Abû šarrâni | | | |
| | 12 | IX | *xii* | Ḫubur | | | |
| **-1088** | 1 | X | *i* | Ṣippu | 27 | | |
| | 2 | XI | *ii* | Qarrâtu | | | |
| | 3 | XII | *iii* | Kalmartu | | | |
| | 4 | I | *iv* | $^{d}$Sin | | [27] | *Taklak-ana-Aššur* |
| | 5 | II | *v* | Kuzallu | | | |
| | 6 | III | *vi* | Allanâtu | | | |
| | 7 | IV | *vii* | Belêt-ekalli | | | |
| | 8 | V | *viii* | Ša sarrâte | | | |
| | 9 | VI | *ix* | Ša kênâte | | | |
| | 10 | VII | *x* | Muḫḫur ilâni | | | |
| | 11 | VIII | *xi* | Abû šarrâni | | | |
| | 12 | IX | *xii* | Ḫubur | | | *(inscription at the end of the former Assyrian year)* |
| **-1087** | 1 | X | *i* | Ṣippu | 28 | | |
| | 2 | XI | *ii* | Qarrâtu | | | |
| | 3 | XII | *iii* | Kalmartu | | | |
| | 4 | I | *iv* | $^{d}$Sin | | [28] | *(Eponym)* |
| | 5 | II | *v* | Kuzallu | | | |
| | 6 | III | *vi* | Allanâtu | | | |
| | 7 | IV | *vii* | Belêt-ekalli | | | |
| | 8 | V | *viii* | Ša sarrâte | | | |
| | 9 | VI | *ix* | Ša kênâte | | | |
| | 10 | VII | *x* | Muḫḫur ilâni | | | |
| | 11 | VIII | *xi* | Abû šarrâni | | | |
| | 12 | IX | *xii* | Ḫubur | | | |

The beginning of regnal years is different depending on dating systems (in 1088 BCE): 1st Nisan (12 April) with accession for Babylonians and for Judeans, 1st Ṣippu (13 January) with accession for Assyrians, 1st Thot (22 May) without accession for Egyptians, 1st Tishri (5 October) without accession for Israelites (the accession year is the length time between the accession and the first year of reign, "with accession" means that the accession year is reckoned as "year 0" and "without accession" means that the accession year is reckoned as "year 1"). Thus, according to the Assyrian calendar of this period, year 1 of Tiglath-pileser I, based on eponyms, not 1st Ṣippu, began on 1st Nisan (April -1114, and accession year in -1115).

It is noted that during the reign of Aššur-dân I (1179-1133) eponyms still begin 1st Nisan, instead of 1 Ṣippu, and that Assyrian lunar years without intercalation remain the norm. However, as the Babylonian year began on 1st Nisan (shortly after the spring equinox) Assyrian years thus coincide with Babylonian lunar years (with intercalation). The period between Aššur-dân I and Tiglath-pileser I is therefore transitional.

The previous system of dating is still used during the reign of Aššur-dân I. Indeed, the 46th year of Aššur-dân I began to the eponym Pišqîya (April -1133) then Ninurta-tukultî-Aššur reigned from months Ša kênâte to Abu šarrâni (from February to April -1132), then Mutakkil-Nusku briefly (few days), then year 1 of Aššur-reš-iši I which began to the eponym Sîn-šêya. There is a gap[73] between the eponyms that start on 1st Nisan and Assyrian year beginning on 1st Ṣippu, June 16 in -1132:

| | month | | | Assyrian | **year** | **King** | *eponym* |
|---|---|---|---|---|---|---|---|
| **-1133** | 1 | X | *viii* | Ša sarrâte | **45** | **Aššur-dân I** | |
| | 2 | XI | *ix* | Ša kênâte | | | |
| | 3 | XII | *x* | Muḫḫur ilâni | | | |
| | 4 | I | *xi* | Abû šarrâni | **46** | | *Pišqîya* |
| | 5 | II | *xii* | Ḫubur | | | |
| | 6 | III | *i* | Ṣippu | | | |
| | 7 | IV | *ii* | Qarrâtu | | | |
| | 8 | V | *iii* | Kalmartu | | | |
| | 9 | VI | *iv* | $^{d}$Sin | | | |
| | 10 | VII | *v* | Kuzallu | | | |
| | 11 | VIII | *vi* | Allanâtu | | | |
| | 12 | IX | *vii* | Belêt-ekalli | | | |
| **-1132** | 1 | X | *viii* | Ša sarrâte | | | |
| | 2 | XI | *ix* | Ša kênâte | **0** | **Ninurta-tukultî-Aššur** | |
| | 3 | XII | *x* | Muḫḫur ilâni | | | |
| | 4 | I | *xi* | Abû šarrâni | | **Mutakkil-Nusku** | *Sîn-šêya* |
| | 5 | II | *xii* | Ḫubur | **1** | **Aššur-reš-iši I** | |
| | 6 | III | *i* | Ṣippu | | | |
| | 7 | IV | *ii* | Qarrâtu | | | |
| | 8 | V | *iii* | Kalmartu | | | |
| | 9 | VI | *iv* | $^{d}$Sin | | | |
| | 10 | VII | *v* | Kuzallu | | | |
| | 11 | VIII | *vi* | Allanâtu | | | |
| | 12 | IX | *vii* | Belêt-ekalli | | | |
| **-1131** | 1 | X | *viii* | Ša sarrâte | | | |
| | 2 | XI | *ix* | Ša kênâte | | | |
| | 3 | XII | *x* | Muḫḫur ilâni | | | |
| | 4 | I | *xi* | Abû šarrâni | **2** | | |
| | 5 | II | *xii* | Ḫubur | | | |
| | 6 | III | *i* | Ṣippu | | | |
| | 7 | IV | *ii* | Qarrâtu | | | |
| | 8 | V | *iii* | Kalmartu | | | |
| | 9 | VI | *iv* | $^{d}$Sin | | | |

The following synchronism shows that before King Aššur-dân I, Assyrian eponyms started on 1st Ṣippu, not on 1st Nisan. Actually, the capture of Babylon and the replacement of its king (Kaštiliašu IV) are dated to (Ina)-Aššur-šuma-aṣbat[74], the 19th eponym[75] of Tukultî-Ninurta I, which corresponds to the year 8 of Kaštiliašu IV (1233-1225) dated 1225 BCE[76]. The order of eponyms from the capture of Babylon is uncertain[77], but the sequence of eponyms in this period seems to be: Ina-Aššur-šuma-aṣbat (No. 18), Ninu'aju (No. 19), Bêr-nâdin-apli (No. 20), Abi-ili son of Katiri (No. 21), Šalmanu-šuma-uṣur (No. 22).

[73] Y. BLOCH – Solving the Problems of the Assyrian King List: Toward a Precise Reconstruction of the Middle Assyrian Chronology in: *Journal of Ancient Civilizations* Vol. 25 (2010, Northeast Normal University), pp. 1-87.
[74] E.C. CANCIK-KIRSCHBAUM – Zu den Eponymenfolgen des 13.Jahrhunderts v. Chr. in Dûr-Katlimmu in: *Berichte der Ausgrabung Tall Seh Hamad* 4 (1996) pp. 9-18.
[75] H. FREYDANK – Zu den Eponymenfolgen des 13.Jahrhunderts v. Chr. in Dûr-Katlimmu in: *Altorientalische Forschungen* 32 (2005) 1 pp. 45-56.
[76] The capture of Babylon is shifted by 10 years (in -1215) if it is calibrated on the current Egyptian chronology.
[77] Y. BLOCH – The Order of Eponyms in the Reign of Tukultî-Ninurta I in: *Orientalia* 79:1 (2010) pp. 1-35.

| | ASSYRIAN EPONYM | SON OF | RANK | YEAR | BABYLONIAN KING | YEAR |
|---|---|---|---|---|---|---|
| **-1242** | **Tukultî-Ninurta Ier** | | 1st | [29]/[0] | Šagarakti-šuriaš | 4 |
| **-1241** | Qibi-Aššur | Ibašši-ili | 2nd | [1] | | 5 |
| **-1240** | Mušallim-Adad | Šalmanu-qarrâd | 3nd | [2] | | 6 |
| **-1239** | Adad-bêl-gabbe | King | 4th | [3] | | 7 |
| **-1238** | Šunu-qardû | | 5th | [4] | | 8 |
| **-1237** | Libur-zanin-Aššur | | 6th | [5] | | 9 |
| **-1236** | Aššur-nâdin-apli | King | 7th | [6] | | 10 |
| **-1235** | Urad-ilani (?) | | 8th | [7] | | 11 |
| **-1234** | Adad-uma''i | | 9th | [8] | | 12 |
| **-1233** | Abattu | Adad-šamši | 10th | [9] | Kaštiliašu IV | 13/0 |
| **-1232** | Abattu | Adad-šumu-lêšir | 11th | [10] | | 1 |
| **-1231** | Aššur-da''an | | 12th | [11] | | 2 |
| **-1230** | Etel-pî-Aššur | Kurbânu | 13th | [12] | | 3 |
| **-1229** | Uṣur-namkûr-šarri | | 14th | [13] | | 4 |
| **-1228** | Aššur-bêl-ilâni | | 15th | [14] | | 5 |
| **-1227** | Aššur-zera-iddina | | 16th | [15] | | 6 |
| **-1226** | Aššur-mušabši (?) | | 17th | [16] | | 7 |
| **-1225** | Ina-Aššur-šuma-aṣbat | Aššur-nâdin-šume | 18th | [17] | Enlil-nâdin-šumi | 8/0 |
| **-1224** | Ninu'aju | Aššur-iddin | 19th | [18]/1 | Kadašman-Harbe II | 1/0 |
| **-1223** | Bêr-nâdin-apli | | 20th | [19]/2 | Adad-šuma-iddina | 1/0 |
| **-1222** | Abi-ili | Katiri | 21th | [20]/3 | | 1 |
| **-1221** | Šalmanu-šuma-uṣur | | 22th | [21]/4 | | 2 |
| **-1220** | Ellil-nâdin-apli (?) | | 23th | [22]/5 | | 3 |
| **-1219** | Kaštiliašu (?) | | 24th | [23]/6 | | 4 |
| **-1218** | Bêr-išmanni (?) | | 25th | [24]/7 | | 5 |
| **-1217** | Ilî-padâ (?) | Aššur-iddin | 26th | [25] | Adad-šuma-uṣur | 6/0 |
| **-1216** | Qarrad-Aššur (?) | Aššur-iddin | 27th | [26] | | 1 |
| **-1215** | Sarniqu (?) | | 28th | [27] | | 2 |
| **-1214** | Ninurta-nâdin-apli (?) | Bukruni | 29th | [28] | | 3 |
| **-1213** | Urad-Kube (?) | Aššur-bel-ilani | 30th | [29] | | 4 |
| **-1212** | Mudammiq-Nusku (?) | Ibašši-ili | 31th | [30] | | 5 |
| **-1211** | Kidin-Aššur (?) | | 32th | [31] | | 6 |
| **-1210** | Sin-uballiṭ (?) | | 33th | [32] | | 7 |
| **-1209** | Nabu-bela-uṣur (?) | | 34th | [33] | | 8 |
| **-1208** | Riš-Aššur (?) | | 35th | [34] | | 9 |
| **-1207** | Aššur-nirari (?) | Šarri | 36th | [35] | | 10 |
| | Samedu (?) | Aššur-zera-iddina | 37th | | | |
| **-1206** | **Aššur-nâdin-apli** | | 1st | [36]/[0] | | 11 |
| **-1205** | | | 2nd | [1] | | 12 |

Counting reigns by Babylonian scribes seems incorrect since Tukultî-Ninurta I regimented Babylonia (not reigned) for 7 years, through three successive Viceroys (whose first two were killed by the King of Elam) or 1.5 years, 1.5 years and 6 years, giving a total of 9 years[78]. In fact, the system used is the cause of these differences. The 7 years of Tukultî-Ninurta I match the 7 eponyms and the 3 years (= 1.5 + 1.5) of the vassal kings match the 3 eponyms or 2 years reign, because 1.5 year was no sense in the Babylonian system (the Assyrian year started on 1st Ṣippu, or March 27 in -1225)[79].

[78] J.M. MUNN RANKIN – Assyrian Military Power, 1300-1200 B.C.
in: *The Cambridge Ancient History* Vol. 2 Part 2 (2000, Cambridge University Press), pp. 287-291.
[79] N = 1225, (N – 1088)x365,24219 = (141)x12x29,530588 + 72 => 72 + 13 = 85th day of the year = 27 March.

| | Month | | | Assyrian | [A] | [B] | **King** *eponym* |
|---|---|---|---|---|---|---|---|
| **-1225** | 1 | X | *x* | Muḫḫur ilâni | 17 | **7** | [A] **Tukultî-Ninurta I** Assyrian<br>[B] **Kaštiliašu IV** Babylonian |
| | 2 | XI | *xi* | Abû šarrâni | | | |
| | 3 | XII | *xii* | Ḫubur | | | |
| | 4 | I | *i* | Ṣippu | 18 | **8** | *Ina-Aššur-šuma-aṣbat* |
| | 5 | II | *ii* | Qarrâtu | | | |
| | 6 | III | *iii* | Kalmartu | | | |
| | 7 | IV | *iv* | $^{d}$Sin | | **0** | (*Babylon taken*)<br>[B] **Enlil-nâdin-šumi** (Babylonian Viceroy) |
| | 8 | V | *v* | Kuzallu | | | |
| | 9 | VI | *vi* | Allanâtu | | | |
| | 10 | VII | *vii* | Belêt-ekalli | | | |
| | 11 | VIII | *viii* | Ša sarrâte | | | |
| | 12 | IX | *ix* | Ša kênâte | | | |
| **-1224** | 1 | X | *x* | Muḫḫur ilâni | | | |
| | 2 | XI | *xi* | Abû šarrâni | | | |
| | 3 | XII | *xii* | Ḫubur | | | |
| | 4 | I | *i* | Ṣippu | 19 | **1** | *Ninu'aju son of Aššur-iddin* |
| | 5 | II | *ii* | Qarrâtu | | | |
| | 6 | III | *iii* | Kalmartu | | | |
| | 7 | IV | *iv* | $^{d}$Sin | | | |
| | 8 | V | *v* | Kuzallu | | | |
| | 9 | VI | *vi* | Allanâtu | | | |
| | 10 | VII | *vii* | Belêt-ekalli | | | |
| | 11 | VIII | *viii* | Ša sarrâte | | | |
| | 12 | IX | *ix* | Ša kênâte | | | |
| **-1223** | 1 | X | *x* | Muḫḫur ilâni | | | |
| | 2 | XI | *xi* | Abû šarrâni | | **0** | [B] **Kadašman-Harbe II** (Babylonian Viceroy) |
| | 3 | XII | *xii* | Ḫubur | | | |
| | 4 | I | *i* | Ṣippu | 20 | **1** | *Bêr-nâdin-apli* |
| | 5 | II | *ii* | Qarrâtu | | | |
| | 6 | III | *iii* | Kalmartu | | | |

Tukultî-Ninurta I ruled over Babylonia for 7 years from the 19th to the 26th eponym. Enlil-nâdin-šumi and Kadašman-Harbe II each of them ruled Babylonia for 1.5 years from the 18th to the 20th eponym. The third pro-Assyrian vassal king, Adad-šuma-iddina, was subsequently reversed by Babylonian officers at the 26th eponym. The Assyrians would have liked to impose their candidate Enlil-kudur-uṣur (?), but the Babylonians settled Adad-šuma-uṣur, freeing themselves from the Assyrian suzerainty.

It is then possible, through the fall of Mitanni, to determine under which Assyrian king took place the calendar change (without intercalation then with), because it is precisely dated. This remarkable event (between April and October) is dated in year 7 of Shalmaneser I and coincides with the accession of Kadašman-Enlil II (-1264), now between these two events there are 152 eponyms (= 23 + 37 + 4 + 6 + 5 + 13 + 46 +18), or 152 "years" instead of the 149 solar years (= 1264 - 1115).

| ASSYRIAN KING | Reign with intercalation | | Reign without intercalation | | gap |
|---|---|---|---|---|---|
| Shalmaneser I | 1274-**1267** | 7 | 1271-**1264** | (-1) | 3 |
| | 1267-1244 | 23 | 1264-1242 | | |
| Tukultî-Ninurta I | 1244-12**27** | 17 | 1242-12**25** | (-1) | 2 |
| | 1227-1207 | 20 | 1225-1206 | | |
| Aššur-nâdin-apli | 1207-1203 | 4 | 1206-1203 | (-1) | 1 |
| Aššur-nêrârî III | 1203-1197 | 6 | 1203-1197 | | 0 |
| Enlil-kudurri-uṣur | 1197-1192 | 5 | 1197-1192 | | 0 |
| Ninurta-apil-Ekur | 1192-1179 | 13 | 1192-1179 | | 0 |
| Aššur-dân I | **1179**-1133 | 46 | | | |
| Ninurta-tukultî-Aššur | 1133 | 0 | | | |
| Mutakkil-Nusku | 1133 | 0 | | | |
| Aššur-rêš-iši I | 1133-1115 | 18 | | | |
| Tiglath-pileser I | **1115**-1076 | 39 | | | |

If it were lunar years of 354.36706 days (instead of the solar year of 365.24219 days), the collapse of Mitanni would fall November -1262. As the period between the two events is 149 years, the eponyms thus began 1st Nisan from the reign of Aššur-dân I (1179-1133) instead of 1 Ṣippu marking the beginning of Assyrian year.

The collapse of Mitanni (called Hanigalbat by the Assyrians) is dated in year 7 of Shalmaneser I, or during his 8th eponym[80] among a set of 30, the latter being Ubru (correspondence between Babylonian and Assyrian years is approximate)[81].

| | ASSYRIAN EPONYM | SON OF | RANK | regnal year | BABYLONIAN KING | regnal year |
|---|---|---|---|---|---|---|
| **-1271** | **Salmanazar Ier** | | 1 | [0] | Kadašman-Turgu | 11 |
| **-1270** | Mušabši'u-Sibitti (?) | | 2 | [1] | | 12 |
| **-1269** | Šerrîya | | 3 | [2] | | 13 |
| **-1268** | Aššur-kâšid | | 4 | [3] | | 14 |
| **-1267** | Aššur-mušabši | Iddin-Mêr | 5 | [4] | | 15 |
| **-1266** | Aššur-mušabši | Anu-mušallim | 6 | [5] | | 16 |
| **-1265** | Qibi-Aššur | Šamaš-aḫa-iddina | 7 | [6] | | 17 |
| **-1264** | Aššur-nâdin-šumâte | *(collapse of Mitanni)* | 8 | [7] | Kadašman-Enlil II | 18/0 |
| **-1263** | Abî-ilî | Aššur-šumu-lêšer | 9 | [8] | | 1 |
| **-1262** | Aššur-âlik-pâni | | 10 | [9] | | 2 |
| **-1261** | Mušallim-Aššur | | 11 | [10] | | 3 |
| **-1260** | Ilî-qarrad (?) | | 12 | [11] | | 4 |
| **-1259** | Qibi-Aššur | Ṣilli-Marduk | 13 | [12] | | 5 |
| **-1258** | Ina-pî-Aššur-lišlim (?) | | 14 | [13] | | 6 |
| **-1257** | Adad-šamši | Adad-šumu-lêšer | 15 | [14] | | 7 |
| **-1256** | Kidin-Sîn | Adad-têya | 16 | [15] | | 8 |
| **-1255** | Bêr-šumu-lêšir | | 17 | [16] | Kudur-Enlil | 9/0 |
| **-1254** | Aššur-dammeq | Abî-ilî | 18 | [17] | | 1 |
| **-1253** | Bêr-bêl-lîte | | 19 | [18] | | 2 |
| **-1252** | Ištar-êriš | Šulmanu-qarrâd | 20 | [19] | | 3 |
| **-1251** | Lullâyu | Adad-šumu-iddina | 21 | [20] | | 4 |
| **-1250** | Aššur-kettî-îde | | 22 | [21] | | 5 |
| **-1249** | Ekaltâyu | | 23 | [22] | | 6 |
| **-1248** | Aššur-da'issunu | Ululayu | 24 | [23] | | 7 |
| **-1247** | Riš-Adad (?) | | 25 | [24] | | 8 |
| **-1246** | Nabû-bêla-uṣur | | 26 | [25] | Šagarakti-šuriaš | 9/0 |
| **-1245** | Usât-Marduk | | 27 | [26] | | 1 |
| **-1244** | Ellil-ašared | | 28 | [27] | | 2 |
| **-1243** | Ittabši-dên-Aššur | | 29 | [28] | | 3 |
| | Ubru | | **30** | | | |
| **-1242** | **Tukultî-Ninurta Ier** | | 1 | [29]/[0] | | 4 |
| **-1241** | Qibi-Aššur | Ibašši-ili | 2 | [1] | | 5 |

The reign of Shalmaneser I can be restored through the prosopography of 30 eponyms attributed to him, with the following uncertainties:

- The place of some eponyms (those with?) is not certain.
- It is possible that some eponyms are not canonical, but this possibility is very low. Indeed, when a eponym died in the year of his eponymy, he was replaced by another eponym who became canonical. However, among the 84 Assyrian dated reigns (No. 33 to No. 117), nine have a duration of 0 year (overrepresentation due to assassinations), one lasted 1 year and one lasted 2 years, giving an average of two dead during his 1st year of reign out of 84 cases recorded (1 out 40).

[80] Y. BLOCH – The Order of Eponyms in the Reign of Shalmaneser I in: *Ugarit-Forschungen* Band 40/2008 (Münster 2009) pp. 143-178.

[81] The 30 eponyms of Shalmaneser I correspond to 30 lunar years and equivalent to around 29 solar years. Thus the accession of Shalmaneser I fell in -1271 but the accession of Tukultî-Ninurta I fell in -1242.

- The contract referenced MI 82970 indicating that: *a reception of wool, recorded by the scribe Nabû-Mudammeq, dated 26th day of the month Ša-sarrâte, eponymous year of Aššur-nâdin-šumâte*, only specifies that the transaction took place: *the day when the king went to Hanigalbat and that the country of Habriuri revolted*. The collapse of Mitanni had therefore take place (just little?) before this date.

Years of Ramses II's reign started in June (from his accession: 27th day of month XI°) and Assyrian year began[82] on June 5 in -1264.

| | month | | | | [A] | [B] | [C] | [D] | [E] | King |
|---|---|---|---|---|---|---|---|---|---|---|
| -1264 | 1 | VII° | X | *viii* | 19 | 17 | [11?] | [6] | [3] | [A] **Ramses II** Egyptian |
| | 2 | VIII° | XI | *ix* | | | | | | [B] **Kadašman-Turgu** Babylonian |
| | 3 | IX° | XII | *x* | | | | | | [C] **Šattuara II** Mitannian |
| | 4 | X° | I | *xi* | | 18 | | | [4] | [D] **Salmanazar I** Assyrian |
| | 5 | XI° | II | *xii* | | | | | | [E] **Ḫattušili III** Hittite |
| | 6 | XII° | III | *i* | 20 | 0 | | [7] | | |
| | 7 | I° | IV | *ii* | | | | | | |
| | 8 | II° | V | *iii* | | | | | | |
| | 9 | III° | VI | *iv* | | | | | | *Collapse of Mitanni (Hanigalbat) ?* |
| | 10 | IV° | VII | *v* | | | | | | |
| | 11 | V° | VIII | *vi* | | | | | | |
| | 12 | VI° | IX | *vii* | | | | | | |
| -1263 | 1 | VII° | X | *viii* | | | *** | *** | | *The collapse of Mitanni just happened* |
| | 2 | VIII° | XI | *ix* | | | | | | |
| | 3 | IX° | XII | *x* | | | | | | |
| | 4 | X° | I | *xi* | | 1 | | | [5] | [B] **Kadašman-Enlil II** |
| | 5 | XI° | II | *xii* | | | | | | |
| | 6 | XII° | III | *i* | 21 | | | [8] | | |
| | 7 | I° | IV | *ii* | | | | | | |
| | 8 | II° | V | *iii* | | | | | | |
| | 9 | III° | VI | *iv* | | | | | | |
| | 10 | IV° | VII | *v* | | | | | | |
| | 11 | V° | VIII | *vi* | *** | | | | *** | *Peace treaty between Ramses II and Ḫattušili III* |
| | 12 | VI° | IX | *vii* | | | | | | |
| -1262 | 1 | VII° | X | *viii* | | | | | | |
| | 2 | VIII° | XI | *ix* | | | | | | |
| | 3 | IX° | XII | *x* | | | | | | |
| | 4 | X° | I | *xi* | | 2 | | | [6] | |
| | 5 | XI° | II | *xii* | | | | | | |
| | 6 | XII° | III | *i* | 22 | | | [9] | | |
| | 7 | I° | IV | *ii* | | | | | | |

The system of eponyms is an institution typically Assyrian (*bît limmi* of Aššur), these calendars dated by eponyms seem to have had the same type of functioning (no intercalary month). As the eponyms of Assyria appear in some texts from Hattuša, the Hittite calendar should look like the lunar calendar of Assyrian type, but religious festivals occurring at regular intervals (monthly, annually or on a longer cycle) particularly in spring, beginning of the Hittite year (as the festival *purulli* marking the New Year[83]) and in autumn, it is unclear whether Hittite kings counted their reigns in solar years (as the Babylonians) or lunar (as the Assyrians). In his annals, the first 10 years of Hittite king Muršili (1322-1295) are punctuated by seasonal religious festivals[84] (and therefore solar).

[82] L'année assyrienne lunaire (AL) débute le 13 janvier en -1088, or comme elle se décale de 10,875 jours (= AS – AL) chaque année solaire, l'année N (>1088) est décalée de (N – 1088 = AS)x10,875 jours par rapport à celle de -1088. Calcul de ce décalage: (AS)x365,24219 = (AL)x12x29,530588 + J, le nombre J + 13 donne la position du jour dans l'année julienne.
Exemple, si N = 1264, (N – 1088)x365,24219 = (181)x12x29,530588 + 142 => 142 + 13 =155e jour de l'année julienne = 5 juin.
Years of reign in brackets are calculated from the estimated duration of these reigns.
[83] H. OTTEN – Ein Text zum Neujahrsfest aus Boğazköy
in: *Orientalistische Literaturzeitung* 51 (1956), pp. 101-105.
P. H. J. HOUWINK VAN TEN CATE -Brief Comments on the Hittite Cult Calendar: The Outline of the AN.TAḪ.ŠUM Festival
in: *Kaniššuwar, A Tribute to Hans Güterbock* (dir. H. A. Hoffner Jr. et G. Beckman), Chicago, 1986.
[84] T. BRYCE – The Kingdom of the Hittites
Oxford 2005 Ed. Oxford University Press pp. 190-220.

Chronology of Assyrian reigns can therefore be fully reconstructed starting from Aššur-uballiṭ II (612-609) up to Êrišu I (No. 33), since all the years of reign between these two kings are known, being aware that Assyrian years are solar up to Aššur-dân I (1179-1133) and lunar prior to this king. The durations of four reigns are missing (No. 65, 66, 37, and 38), but they can be calculated through synchronisms from Assyrian annals that indicate the exact length between the reconstruction of some famous temples[85].

- Shalmaneser I (No. 77), for example, states in his inscriptions that the temple of Assur (Ehursagkurkurra) was built by Ušpiya and rebuilt by Erišu I, then 159 years later by Šamšî-Adad I and 580 years later by himself. Shalmaneser I do not specify the point used to determine these durations, but Esarhaddon gives a figure of 126 years for the duration between Erišu I and Šamšî-Adad I, proving that Shalmaneser I included the 33-year reign of Šamšî-Adad I in his calculation (159 = 126 +33). The 159 years have therefore to start at the end of the reign of Erišu I to the end of the reign of Šamšî-Adad I and 580 years are completed at the beginning of the reign of Shalmaneser I (in 1271 BCE). So there are 421 lunar years ago (421 = 580 - 159) between the reigns of Šamšî-Adad I and Shalmaneser I, or a duration of 409 solar years[86], which sets the end of the reign of Šamšî-Adad I in 1680 (= 1271 + 409)
- Tiglath-pileser I (No. 87) states in his annals having rebuilt the temple called Anu-Adad at the beginning of his reign (in 1115), which was built 641 years earlier by Šamšî-Adad I. These 641 years (= 68 solar + 573 lunar) correspond to 624 (= 68 + 556) solar years, which dates back the reign of Šamšî-Adad I in 1739 (= 1115 + 624) instead of 1712, but the scribe has probably used a King list with a reign of 40 years instead of 11 for Išme-Dagan I (since Šamšî-Adad I died in the year 17 of Hammurabi and Išme-Dagan I died in the year 28 this king)[87], which reduces the 641 years to 612 (= 641 - 29), or a duration of 596 solar years, which fixes the beginning of the reign of Samsi-Adad I in 1711 (= 1115 + 596), in good agreement with the previous date of 1712 (= 1680 + 33 - 1).
- Esarhaddon (No. 112) also claims to have rebuilt the temple of Aššur. In an inscription (Assur A) dated eponym Issi-Adad-anînu (679), at the beginning of his reign, he claims that 129 years elapsed between the reconstruction of Erišu I and the one of Šamšî-Adad I, and 434 years later Shalmaneser I has rebuilt again the temple, then 580 years later Esarhaddon finally rebuilt the temple. The information of Esarhaddon seems accurate. Indeed, the first term is correct, because it actually falls in the reign of Shalmaneser I (679 + 580 = 1259). The duration between the beginning of the reign of Esarhaddon and the end of the reign of Šamšî-Adad I is 1014 years (= 580 solar + 434 lunar), which corresponds to 1001 solar years, that sets the end of the reign of Šamšî-Adad I in 1680 (= 679 + 1001). The reign of this king can therefore be set from 1712 to 1680. His death in the year 17 of Hammurabi allows to anchor it to the Babylonian chronology[88]. After his death the documents dated in different calendars allows some synchronisms[89].
- The paleo-Assyrian dynasty begins after the fall of Ur[90] with king Puzur-Aššur I (No. 30), which enables us to date the fall of this city around -1912 (the average length of an Assyrian reign is 14 years over all the period).

---

85 H. GASCHE, J.A. ARMSTRONG, S.W. COLE – Dating the Fall of Babylon in: Mesopotamian History and Environment (1998) Chicago pp. 57-80.

86 1 solar year = 1 lunar year x 1.0306889 (= 365.24219/12x29,530588).

87 H. GASCHE, J.A. ARMSTRONG, S.W. COLE – Dating the Fall of Babylon in: Mesopotamian History and Environment (1998) Chicago p. 52.

88 H. GASCHE, J.A. ARMSTRONG, S.W. COLE & V.G. GURZADYAN – A correction to Dating the Fall of Babylon in: *Akkadica* 108 (1998) pp.1-4.

89 D.A. BARREYRA FRACAROLI – The Chronology of Zimri-Lim's Reign A Report in: *Aula Orientalis* XXIX:2 (2011) pp.185-198.

90 F. JOANNES - Dictionnaire de la civilisation mésopotamienne Paris 2001 Éd. Robert Laffont pp. 617-621,823.

Overview of the period 1954-609:

| N° | ASSYRIAN KING | Reign | length | # | | synchronisms |
|---|---|---|---|---|---|---|
| 27 | Sulili (Zariqum) | 1954-1940 | *[14]* | | | *First lists of eponyms (lost)* |
| 28 | Kikkia | 1940-1927 | *[14]* | *(-1)* | | |
| 29 | Akia | 1927-1913 | *[14]* | | | *Fall of Ur* (in **-1912**) |
| 30 | Puzur-Aššur I | **1913**-1900 | *[14]* | *(-1)* | | *Beginning of the Paleo-Assyrian period* |
| 31 | Šalim-ahum | 1900-1886 | *[14]* | | | |
| 32 | Ilu-šumma | 1886-1873 | *[14]* | *(-1)* | | |
| 33 | Êrišu I | 1873-**1834** | 40 | (-1) | 40 | *First Chronicles* |
| 34 | Ikunum | 1834-1821 | 14 | (-1) | 159 | *(eponym starting on 1$^{st}$ Sippu)* |
| 35 | Sargon I | 1821-1782 | 40 | (-1) | | |
| 36 | Puzur-Aššur II | 1782-1774 | 8 | | | |
| 37 | Naram-Sîn | 1774-1722 | [54] | (-2) | | |
| 38 | Êrišu II | 1722-1712 | [10] | | | |
| 39 | Šamšî-Adad I | 1712-**1680** | 33 | (-1) | | year 33 of Šamšî-Adad I = |
| 40 | Išme-Dagan I | 1680-1670 | 11 | (-1) | 434 | year 17 of Hammurabi |
| 41 | Aššur-dugul | 1670-1664 | 6 | | | 41* Mut-Aškur/ Rimu-x/ Asîsum |
| 42 | Aššur-apla-idi | 1664 | 0 | | | |
| 43 | Nâṣir-Sîn | 1664 | 0 | | | |
| 44 | Sîn-namir | 1664 | 0 | | | |
| 45 | Ipqi-Ištar | 1664 | 0 | | | |
| 46 | Adad-ṣalûlu | 1664 | 0 | | | |
| 47 | Adasi | 1664 | 0 | | | |
| 48 | Bêlu-bâni | 1664-1654 | 10 | | | |
| 49 | Libbaya | 1654-1638 | 17 | (-1) | | |
| 50 | Šarma-Adad I | 1638-1626 | 12 | | | |
| 51 | Puzur-Sîn | 1626-1615 | 12 | (-1) | | |
| 52 | Bazaya | 1615-1588 | 28 | (-1) | | |
| 53 | Lullaya | 1588-1582 | 6 | | | |
| 54 | Šû-Ninûa | 1582-1568 | 14 | | | |
| 55 | Šarma-Adad II | 1568-1565 | 3 | | | |
| 56 | Êrišu III | 1565-1553 | 13 | (-1) | | |
| 57 | Šamšî-Adad II | 1553-1547 | 6 | | | |
| 58 | Išme-Dagan II | 1547-1531 | 16 | | | |
| 59 | Šamšî-Adad III | 1531-1516 | 16 | (-1) | | |
| 60 | Aššur-nêrârî I | 1516-1491 | 26 | (-1) | | |
| 61 | Puzur-Aššur III | 1491-1467 | 24 | | | |
| 62 | Enlil-nâṣir I | 1467-1455 | 13 | (-1) | | |
| 63 | Nûr-ili | 1455-1443 | 12 | | | |
| 64 | Aššur-šadûni | 1443-1443 | 0 | | | |
| 65 | Aššur-rabi I$^{er}$ | 1443-1433 | [10*] | | | |
| 66 | Aššur-nâdin-aḫḫe I | 1433-1424 | [10*] | (-1) | | |
| 67 | Enlil-naṣir II | 1424-1418 | 6 | | | |
| 68 | Aššur-nêrârî II | 1418-1411 | 7 | | | |
| 69 | Aššur-bêl-nišešu | 1411-1403 | 9 | (-1) | | |
| 70 | Aššur-rê'im-nišešu | 1403-1395 | 8 | | | |
| 71 | Aššur-nâdin-aḫḫe II | 1395-1385 | 10 | | | |
| 72 | Erîba-Adad I | 1385-1358 | 27 | - | | |
| 73 | Aššur-uballiṭ I | 1358-1323 | 36 | (-1) | | |
| 74 | Enlil-nêrârî | 1323-1313 | 10 | - | | |
| 75 | Arik-dên-ili | 1313-1302 | 12 | (-1) | | |
| 76 | Adad-nêrârî I | 1302-1271 | 32 | (-1) | | |
| 77 | Salmanazar I | 1271-**1259** | 12 | | | 434 *eponyms from Išme-Dagan I* |
| | | **1259**-1242 | 18 | (-1) | 580 | 452 = 434 + 12 |
| 78 | Tukultî-Ninurta I | 1242-1206 | 37 | (-1) | | |
| 79 | Aššur-nâdin-apli | 1206-1203 | 4 | (-1) | | |
| 80 | Aššur-nêrârî III | 1203-1197 | 6 | - | | |
| 81 | Enlil-kudurri-uṣur | 1197-1192 | 5 | - | | |
| 82 | Ninurta-apil-Ekur | 1192-1179 | 13 | - | | |
| 83 | Aššur-dân I | 1179-**1133** | 46 | - | | *(eponym starting on 1st Nisan:* |
| 84 | Ninurta-tukultî-Aššur | 1133 | 0 | - | | *as Babylonian intercalation)* |
| 85 | Mutakkil-Nusku | 1133 | 0 | - | | |
| 86 | Aššur-rêš-iši I | 1133-1115 | 18 | - | | |
| 87 | Tiglath-phalazar I | 1115-1076 | 39 | - | | |
| 88 | Ašared-apil-Ekur | 1076-1074 | 2 | - | | *(Babylonian calendar used)* |
| 89 | Aššur-bêl-kala | 1074-1056 | 18 | - | | |
| 90 | Erîba-Adad II | 1056-1054 | 2 | - | | |
| 91 | Šamšî-Adad IV | 1054-1050 | 4 | - | | |
| 92 | Aššurnaṣirpal I | 1050-1031 | 19 | - | | |
| 93 | Salmanazar II | 1031-1019 | 12 | - | | |
| 94 | Aššur-nêrârî IV | 1019-1013 | 6 | - | | |
| 95 | Aššur-rabi II | 1013-972 | 41 | - | | |
| 96 | Aššur-rêš-iši II | 972-967 | 5 | - | | |
| 97 | Tiglath-phalazar II | 967-935 | 32 | - | | |
| 98 | Aššur-dân II | 935-912 | 23 | - | | |
| 99 | Adad-nêrârî II | 912-891 | 21 | - | | |
| 100 | Tukultî-Ninurta II | 891-884 | 7 | - | | |
| 101 | Aššurnaṣirpal II | 884-859 | 25 | - | | |
| 102 | Salmanazar III | 859-824 | 35 | - | | |
| 103 | Šamšî-Adad V | 824-811 | 13 | - | | |
| 104 | Adad-nêrârî III | 811-783 | 28 | - | | |
| 105 | Salmanazar IV | 783-773 | 10 | - | | |
| 106 | Aššur-dân III | 773-755 | 18 | - | | |
| 107 | Aššur-nêrârî V | 755-745 | 10 | - | | |
| 108 | Tiglath-phalazar III | 745-727 | 18 | - | | |
| 109 | Salmanazar V | 727-722 | 5 | - | | |
| 110 | Sargon II | 722-705 | 17 | - | | |
| 111 | Sennacherib | 705-681 | 24 | - | | |
| 112 | Assarhaddon | **681-669** | 12 | (72) | - | |
| 113 | Aššurbanipal | 669-627 | 42 | | - | |
| | [Aššur-etel-ilâni] | [630-627] | [3] | | - | |
| 114 | Aššur-etel-ilâni | 627-626 | 1 | | - | |
| 115 | Sin-šar-iškun | 626-612 | 14 | | - | |
| 116 | Aššur-uballiṭ II | 612-**609** | 3 | | - | |

This chronology obtained from Assyrian king lists is confirmed on the period from Êrišu I (No. 33) to Aššur-dugul (No. 40) thanks to lists of eponyms[91], in addition, some comments associated with eponyms allow to fix several synchronisms, especially the start and the duration of certain reigns.

| No. | ASSYRIAN KING | year | Comments from chronicles | Number of eponyms |
|---|---|---|---|---|
| 33 | Êrišu I | 40 | year 1 eponym Šu-Ištar son of Abila (No. 1) | 40 |
| 34 | Ikunum | 14 | year 1 eponym Iddin-Suen brother of Šuli (No. 41) | 14 |
| 35 | Sargon I | 40 | year 1 eponym Aššur-malik son of Agatum (No. 55) | 40 |
| 36 | Puzur-Aššur II | 8 | year 1 eponym Aššur-nada son of Puzur-Ana (No. 95) | 8 |
| 37 | Naram-Sîn | [54] | year 1 eponym Šu-Su'en son of Babilum (No. 103) | 54* |
| 38 | Êrišu II | [10] | Šamšî-Adad I conquers Assyria, eponym Ibni-Ištar (No. 157) | 10* |
| 39 | Šamšî-Adad I | 33 | Death of Šamšî-Adad I, eponym Ṭab-ṣilla-Aššur (No. 199) | 33* |
| 40 | Aššur-dugul | 6 | Usurper | ? |

Overview of the period 1873-1664 year by year:

91 K.R. VEENHOF – Some displaced Tablets from Kârum Kanesh (Kültepe)
C. GÜNBATTI – An Eponym List (KEL G) from Kültepe
G. KRYSZAT – Herrscher, Kult und Kulttradition in Anatolien nach den Quellen aus den altassyrischen Handelskolonien. Teil 3/1 in: *Altorientalische Forschungen* Band 38 (2008) 1 pp. 10-27, 103-132, 156-171.
G. KRYSZAT – Herrscher, Kult und Kulttradition in Anatolien nach den Quellen aus den altassyrischen Handelskolonien. Teil 3/2 in: *Altorientalische Forschungen* Band 38 (2008) 2 pp. 195-219.

| | N° | EPONYM | SON OF | COMMENTS IN CHRONICLES | | |
|---|---|---|---|---|---|---|
| **-1873** | | | | accession of **Êrišu I**er | 0 | |
| **-1872** | 1 | Šu-Ištar | Abila | (beginning of the list KEL A) | 1 | |
| **-1871** | 2 | Šu-Kuttum | Iššuhum | | 2 | |
| **-1870** | 3 | Iddin-ili | Kurub-Ištar | | 3 | |
| **-1869** | 4 | Šu-Anum | Isaliya | | 4 | |
| **-1868** | 5 | Anah-ili | Kiki | | 5 | |
| **-1867** | 6 | Su'etaya | Iribum | | 6 | |
| **-1866** | 7 | Daya | Iššuhum | | 7 | |
| **-1865** | 8 | Ili-ellat | | | 8 | |
| **-1864** | 9 | Šamaš-ṭab | | | 9 | |
| **-1863** | 10 | Akusa | | | 10 | |
| **-1862** | 11 | Idnaya | Šudaya | | 11 | |
| **-1861** | 12 | Quqadum | Buzu | | 12 | |
| **-1860** | 13 | Puzur-Ištar | Bedaki | | 13 | |
| **-1859** | 14 | Laqipum | Babidi | | 14 | |
| **-1858** | 15 | Šu-Laban | Kurub-Ištar | | 15 | |
| **-1857** | 16 | Šu-Belum | Iššuhum | | 16 | |
| **-1856** | 17 | Nab-Su'en | Šu-Ištar | | 17 | |
| **-1855** | 18 | Hadaya | Elali | | 18 | |
| **-1854** | 19 | Ennum-Aššur | Begaya | | 19 | |
| **-1853** | 20 | Ikunum | Šudaya | | 20 | |
| **-1852** | 21 | AŠ.DINGIR | Iddida | | 21 | |
| **-1851** | 22 | Buzutaya | Iššuhum | | 22 | |
| **-1850** | 23 | Šu-Ištar | Ammaya | | 23 | |
| **-1849** | 24 | Iddin-Aššur | GUDU | | 24 | |
| **-1848** | 25 | Puzur-Aššur | I.NUN | | 25 | |
| **-1847** | 26 | Quqadum | Buzu | | 26 | |
| **-1846** | 27 | Ibni-Adad | Susaya | | 27 | |
| **-1845** | 28 | Irišum | Adad-rabi | | 28 | |
| **-1844** | 29 | Menanum | Begaya | | 29 | |
| **-1843** | 30 | Iddin-Su'en | Šalim-ahum | | 30 | |
| **-1842** | 31 | Puzur-Aššur | Idnaya | | 31 | |
| **-1841** | 32 | Šûli | Uphakum | | 32 | |
| | 33 | Laqip | Zukua | | 33 | |
| **-1840** | 34 | Puzur-Ištar | Erisu'a | | 34 | |
| **-1839** | 35 | Aguwa | Adad-rabi | | 35 | |
| **-1838** | 36 | Šû-Su'en | Ṣilliya | | 36 | |
| **-1837** | 37 | Ennum-Aššur | Begaya | | 37 | |
| **-1836** | 38 | Enna-Su'en | Pussanum | | 38 | |
| **-1835** | 39 | Ennanum | Uphakum | | 39 | |
| **-1834** | 40 | Buzi | Adad-rabi | accession of **Ikunum** | **40** | |
| **-1833** | 41 | Iddin-Suen | brother of Šuli | Šuli son of Šalmah | 1 | |
| **-1832** | 42 | Ikunum | Šudaya | | 2 | |
| **-1831** | 43 | Dan-Wer | Ahu-ahi | | 3 | |
| **-1830** | 44 | Šu-Anum | Nerabtim | | 4 | |
| **-1829** | 45 | Il-massu | Aššur-ṭab | | 5 | |
| **-1828** | 46 | Šu-Hubur | Šuli | | 6 | |
| **-1827** | 47 | Idua | Ṣulili | | 7 | |
| **-1826** | 48 | Laqip | Puzur-Laba | | 8 | |
| **-1825** | 49 | Šu-Anum | du hapirum | | 9 | |
| **-1824** | 50 | Uku | Bila | | 10 | |
| **-1823** | 51 | Aššur-malik | Panaka | | 11 | |
| **-1822** | 52 | Dan-Aššur | Puzur-Wer | | 12 | |
| **-1821** | 53 | Šu-Kubum | Ahu-ahi | | 13 | |
| **-1820** | 54 | Irišum | Idi-Aššur | accession of **Sargon I** | **14** | |
| **-1819** | 55 | Aššur-malik | Agatum | | 1 | |
| **-1818** | 56 | Aššur-malik | Enania | | 2 | |
| **-1817** | 57 | Ibisua | Suen-nada | | 3 | |
| **-1816** | 58 | Bazia | Bal-Tutu | | 4 | |
| **-1815** | 59 | Puzur-Ištar | Sabasia | | 5 | |
| **-1814** | 60 | Pišaḫ-Ili | Adin | | 6 | |
| **-1813** | 61 | Asqudum | Lapiqum | | 7 | |
| **-1812** | 62 | Ili-pilaḫ | Damqum | | 8 | |
| **-1811** | 63 | Qulali | [-] | | 9 | |
| **-1810** | 64 | Susaya | [-] | | 10 | |
| **-1809** | 65 | Amaya | of Armourer | | 11 | |
| | 66 | Ipḫurum | Ili-ellat | | 12 | |
| **-1808** | 67 | Kudanum | Laqip | | 13 | |
| **-1807** | 68 | Ili-bani | Ikunum | | 14 | |
| **-1806** | 69 | Šu-Kubum | Susaya | | 15 | |
| **-1805** | 70 | Quqidi | Amur-Aššur | | 16 | |
| **-1804** | 71 | Abia | Nur-Suen | | 17 | |
| **-1803** | 72 | Šu-Ištar | Šukutum | | 18 | |
| **-1802** | 73 | Bazia | Šepa-lim | | 19 | |
| **-1801** | 74 | Šu-Ištar | Ikunum, the Star | | 20 | |
| **-1800** | 75 | Abia | Šu-Dagan | **Babylonian king** | 21 | |
| **-1799** | 76 | Salia | Šabakuranum | **Sûmû-abum** | 22 | **0** |
| **-1798** | 77 | Ibni-Adad | Baqqunum | | 23 | **1** |
| **-1797** | 78 | Aḫmarši | Malkum-išar | | 24 | **2** |
| **-1796** | 79 | Sukkalia | Minanum | | 25 | **3** |
| **-1795** | 80 | Iddin-Aššur | Kubidi | | 26 | **4** |
| **-1794** | 81 | Šudaya | Ennanum | | 27 | **5** |
| **-1793** | 82 | Al-ṭab | Pilaḫ-Aššur | | 28 | **6** |
| **-1792** | 83 | Aššur-dammiq | Abarsisum | | 29 | **7** |
| **-1791** | 84 | Puzur-Niraḫ | Puzur-Suen | | 30 | **8** |
| **-1790** | 85 | Amur-Aššur | Karria | | 31 | **9** |
| **-1789** | 86 | Buzuzu | Ibbi-Suen | | 32 | **10** |
| **-1788** | 87 | Šu-Ḫubur | Elali | | 33 | **11** |
| **-1787** | 88 | Ilšu-rabi | Bazia | | 34 | **12** |
| **-1786** | 89 | Alaḫum | Inaḫ-ili | | 35 | **13** |
| **-1785** | 90 | Ṭab-Aššur | Suḫarum | **Sûmû-a-il** | 36 | **14** |
| **-1784** | 91 | Elali | Ikunum | | 37 | **1** |
| **-1783** | 92 | Iddin-abum | Narbitum | | 38 | **2** |
| | 93 | Adad-bani | Iddin-Aššur | | 39 | |
| **-1782** | 94 | Aššur-iddin | Šuli | accession of **Puzur-Aššur II** | **40** | **3** |
| **-1781** | 95 | Aššur-nada | Puzur-Ana | | 1 | **4** |
| **-1780** | 96 | Kubia | Karria | | 2 | **5** |
| **-1779** | 97 | Ili-dan | Elali | | 3 | **6** |
| **-1778** | 98 | Ṣilulu | Uku | | 4 | **7** |
| **-1777** | 99 | Aššur-nada | Ili-binanni | | 5 | **8** |
| **-1776** | 100 | Ikuppi-Ištar | Ikua | | 6 | **9** |
| **-1775** | 101 | Buzutaya | Šuli | | 7 | **10** |
| **-1774** | 102 | Innaya | Amuraya | accession of **Naram-Sîn** | **8** | **11** |
| **-1773** | 103 | Šu-Su'en | Babilum | (beginning of the list MEC A) | 1 | **12** |
| **-1772** | 104 | Aššur-malik | Al-ahum | | 2 | **13** |
| **-1771** | 105 | Aššur-imitti | Ili-bani | | 3 | **14** |
| **-1770** | 106 | Enna-Su'en | Šu-Aššur | | 4 | **15** |
| **-1769** | 107 | Akutum | Al-ahum | | 5 | **16** |
| **-1768** | 108 | Maṣi-ili | Irišum | | 6 | **17** |
| **-1767** | 109 | Iddin-ahum | Kudanum | | 7 | **18** |
| **-1766** | 110 | Samaya | Šu-Belum | (beginning of the list KEL G) | 8 | **19** |
| **-1765** | 111 | Ili-alum | Sukkalliya | | 9 | **20** |
| **-1764** | 112 | Ennamanum | Aššur-malik | | 10 | **21** |
| **-1763** | 113 | Ennum-Aššur | Dunea | | 11 | **22** |
| **-1762** | 114 | Enna-Su'en | Šu-Ištar | | 12 | **23** |
| **-1761** | 115 | Hana-Narum | [-] | | 13 | **24** |
| **-1760** | 116 | Dadiya | [-] | | 14 | |
| | 117 | Kapatiya | [-] | | 15 | **25** |
| **-1759** | 118 | Išme-Aššur | Ea-dan | | 16 | **26** |
| **-1758** | 119 | Aššur-muttabbil | Azizum | | 17 | **27** |
| **-1757** | 120 | Šu-Nirah | Azuzaya | | 18 | **28** |
| **-1756** | 121 | Iddin-Abum | [-] | | 19 | **29** |
| **-1755** | 122 | Ili-dan | Azua | | 20 | **30** |
| **-1754** | 123 | Aššur-imitti | Iddin-Ištar | | 21 | **31** |
| **-1753** | 124 | Buṣiya | Abiya | | 22 | **32** |
| **-1752** | 125 | Dadiya | Šu-Ilabrat | Birth of Šamšî-Adad Ier | 23 | **33** |
| **-1751** | 126 | Puzur-Ištar | Nur-ilišu | Assombrissement du soleil | 24 | **34** |
| **-1750** | 127 | Isaya | Dagan-malkum | | 25 | **35** |
| **-1749** | 128 | Abu-šalim | Ilu-alum | **Sabium** | 26 | **36** |
| **-1748** | 129 | Aššur-re'i | Ili-emuqi | (end of the list KEL A) | 27 | **1** |
| **-1747** | 130 | Ṭab-Aššur | Uzua | | 28 | **2** |
| **-1746** | 131 | Šu-Rama | Uzua | | 29 | **3** |
| **-1745** | 132 | Sin-išme'anni | | (end of the list MEC A) | 30 | **4** |
| **-1744** | 133 | Aššur-malik | Šu-Haniš | (beginning of the list MEC B) | 31 | **5** |
| **-1743** | 134 | Dan-Ea | Abu-qar | | 32 | **6** |
| **-1742** | 135 | Enna-Sîn | Iddin-abum | | 33 | **7** |
| **-1741** | 136 | Aššur-balaṭi | | | 34 | **8** |
| **-1740** | 137 | Enna-Sîn | | | 35 | **9** |
| **-1739** | 138 | Iṭur-Aššur | | | 36 | **10** |
| **-1738** | 139 | Šu-Belum | | | 37 | **11** |
| **-1737** | 140 | Šarrum-Adad | | | 38 | **12** |
| **-1736** | 141 | Šu-Laban | | | 39 | **13** |
| **-1735** | 142 | Aššur-imitti | | **Apil-Sîn** | 40 | **14** |
| **-1734** | 143 | Dadaya | | | 41 | **1** |
| | 144 | Dadaya | | | 42 | |
| **-1733** | 145 | Ah-šalim | | | 43 | **2** |
| **-1732** | 146 | Uṣur-ši-Ištar | | | 44 | **3** |
| **-1731** | 147 | Kataya | | | 45 | **4** |
| **-1730** | 148 | Šu-Su'en | | | 46 | **5** |
| **-1729** | 149 | Abu-šalim | | | 47 | **6** |
| **-1728** | 150 | Šudaya | | | 48 | **7** |
| **-1727** | 151 | Šu-Dadum | | | 49 | **8** |
| **-1726** | 152 | Aššur-dugul | | | 50 | **9** |
| **-1725** | 153 | Puzur-Ištar | | | 51 | **10** |
| **-1724** | 154 | Atanah | | Šamšî-Adad Ier conquered Ekallatum | 52 | **11** |
| **-1723** | 155 | Irišum | | Ekallatum | 53 | **12** |
| **-1722** | 156 | Aššur-ennam | | Ekallatum (**Êrišu II**) | **54** | **13** |
| **-1721** | 157 | Ibni-Ištar | Sîn-išme'anni | Šamšî-Adad Ier conquered Aššur | 1 | **14** |
| **-1720** | 158 | Aššur-bel-malki | | | 2 | **15** |
| **-1719** | 159 | Belanum | | | 3 | **16** |
| **-1718** | 160 | Sukkallum | | Êrišu II | 4 | **17** |
| **-1717** | 161 | Amur-Aššur | | Êrišu II **Sîn-muballit** | 5 | **18** |
| **-1716** | 162 | Aššur-nišu | | Êrišu II | 6 | **1** |
| **-1715** | 163 | Manawirum | | | 7 | **2** |
| **-1714** | 164 | Idnaja | Aššur-imitti | | 8 | **3** |
| **-1713** | 165 | Dadaya | | (end of the list MEC B) | 9 | **4** |
| | 166 | Puzur-Nirah | | (beginning of the list MEC D) | **10** | |
| **-1712** | 167 | Abiya | | (Assyrian reign of **Šamšî-Adad Ier**) | 1 | **5** |
| **-1711** | 168 | Edinum | | | 2 | **6** |
| **-1710** | 169 | Aššur-taklatu | | (end of the list MEC D) | 3 | **7** |
| **-1709** | 170 | Isim-Su'en | | | 4 | **8** |
| **-1708** | 171 | Adad-bani | | | 5 | **9** |
| **-1707** | 172 | Abi-šagiš | | | 6 | **10** |
| **-1706** | 173 | Ṭab-ṣilla-Aššur | | | 7 | **11** |
| **-1705** | 174 | Iddin-Aššur | | | 8 | **12** |
| **-1704** | 175 | Namiya | | | 9 | **13** |
| **-1703** | 176 | Attal-šarri | Ili-alum | | 10 | **14** |
| **-1702** | 177 | Dadaya | | | 11 | **15** |
| **-1701** | 178 | Ani-m[alik] | | | 12 | **16** |
| **-1700** | 179 | Idna-Aššur* | | (illegible part of the list KEL G) | 13 | **17** |
| **-1699** | 180 | Atanum* | | | 14 | **18** |
| **-1698** | 181 | Aššur-taklatu* | - | (eponymous chronicle of Mari)[92] | 15 | **19** |
| **-1697** | 182 | Haya-malik* | Dudanum | (beginning of the list MEC E) **Hammurabi** | 16 | **20** |
| **-1696** | 183 | Šalim-Aššur* | Šalim-Anum | | 17 | **1** |
| **-1695** | 184 | Šalim-Aššur | Uṣranu | | 18 | **2** |
| **-1694** | 185 | Ennam-Aššur | | (position uncertain) | 19 | **3** |
| **-1693** | 186 | Sîn-muballiṭ | | | 20 | **4** |
| **-1692** | 187 | Riš-Šamaš | | | 21 | **5** |
| **-1691** | 188 | Ibni-Adad | | | 22 | **6** |
| **-1690** | 189 | Aššur-imitti | | | 23 | **7** |
| **-1689** | 190 | Ahiyaya | Takigi | (non canical eponym?) | 24 | **8** |
| **-1688** | 191 | Ili-ellat | Aššur-nišu | | 25 | **9** |
| **-1687** | 192 | Rigmanum | | | 26 | **10** |
| **-1686** | 193 | Ikuppiya | (*alliance with Qatna*) | (*Yasmah-Addu became vice-roy of Mari*) | 27 | **11** |
| **-1685** | 194 | Asqudum | | | 28 | **12** |
| **-1684** | 195 | Aššur-malik | | (*total lunar eclipse dated 7/12/1684*) | 29 | **13** |
| **-1683** | 196 | Awiliya | | (end of the list MEC E) | 30 | **14** |
| **-1682** | 197 | Nimar-Su'en | | | 31 | **15** |
| **-1681** | 198 | Adad-bani | | | 32 | **16** |
| **-1680** | 199 | Ṭab-ṣilla-Aššur | | death of **Šamšî-Adad I** | 33 | **17** |
| | 200 | Ennam-Aššur | | (accession of **Išme-Dagan I**) | 1 | |
| **-1679** | 201 | Aššur-emuqi | | | 2 | **18** |
| **-1678** | 202 | Abu-šalim | | | 3 | **19** |
| **-1677** | 203 | Pussanum | Adad-rabi | | 4 | **20** |
| **-1676** | 204 | Ikuppi-Ištar | | | 5 | **21** |
| **-1675** | 205 | Ahiya | Takiki | | 6 | **22** |
| **-1674** | 206 | Beliya | Enna-Su'en | | 7 | **23** |
| **-1673** | 207 | Ili-bani | | | 8 | **24** |
| **-1672** | 208 | Aššur-taklaku | | | 9 | **25** |
| **-1671** | 208 | Sassapum | Aššur-malik | | 10 | **26** |
| **-1670** | 209 | Ahu-waqar | | (accession of **Aššur-dugul**) | **11** | **27** |
| **-1669** | 210 | Kizurum | | | 1 | **28** |
| **-1668** | 211 | Dadiya | | | 2 | **29** |
| **-1667** | 212 | Yam-NE? | | | 3 | **30** |
| **-1666** | 213 | Adad-bani | | | 4 | **31** |
| **-1665** | 214 | Ennam-Aššur | Aššur-taklaku | | 5 | **32** |
| **-1664** | 215 | Ataya | Šamaya | | 6 | **33** |

This list of eponyms used for reconstituting Assyrian reigns (from several partial lists)[93], contains the following difficulties:

- The Assyrian king list compiled under Šamšî-Adad I states that the eponyms from Sulili (Zariqum) to Il-šumma (Kings No. 27 to 32) were lost, suggesting a beginning of Assyrian eponyms only from Sulili (-1954) and a compilation from Êrišu I (-1873).
- After the accession of King Ikunum, a list gives Šuli son of Šalmah as eponym instead of Iddin-Suen brother of Šuli (eponym No. 41). Rather than assume an oversight in the lists and thus keep these two eponyms, the presence of a canonical eponym replacing an noncanonical eponym (died during the year of his eponymy) is more likely.

[92] J.-J. GLASSNER – Chroniques mésopotamiennes
Paris 2004 Éd. Les Belles Lettres pp. 157-160.
D. CHARPIN, N. ZIEGLER – Florilegium marianum V. Mari et le Proche-Orient à l'époque amorrite
in: *Mémoires de N.A.B.U.* 6 (2003) pp. 156-157.

[93] A complete list of eponyms should contain about 150 names (size of KEL G list). At the time of Esarhaddon, for example, the reigns of Êrišu I (King No. 33) and Sennacherib (King No. 111) were separated by 1213 eponyms, which could be inscribed on about 8 tablets of 150 names.

- The darkening of the sun mentioned during the Puzur-Ištar eponym (No. 126), the year just after the birth of Šamšî-Adad I, has been interpreted by some as a solar eclipse[94]. However, there was no total solar eclipse visible in Assyria (between Ashur and Nineveh) over this period (1800-1700), but only two partial eclipses slightly visible[95]. Moreover, the term used *[n]a-ah-du-ur*, means an eclipse in a metaphorical way and is different from the usual *antalûm* used in Mari[96]. These two comments have been added later in the list of eponyms, because Šamšî-Adad I was initially an Amorite king who became part of the Assyrian dynasty only at the end of his glorious reign. Thus for the Assyrian copyist of that time, the birth of Šamšî-Adad I actually marked the end (the eclipse) of the authentic Assyrian dynasty.
- Neither death nor the accession of Êrišu II are detailed in the lists, but this reign can be framed by two dates: the 1st year of Naram-Sîn (in -1773) during the eponymy Šu-Su'en in the beginning of the list MEC A, and the death of Šamšî-Adad during the eponymy of Ṭab-ṣilla-Aššur (in -1680), after 33 years of reign. Thus the death of Êrišu II must go back to 1713 (= 1680 + 33), beginning of the list MEC D. The eponyms of the list KEL G being completely unreadable at least 11 lines, most likely 16 lines (eponyms No. 179 to 194), they were supplemented by the list MEC E whose recovery remains uncertain[97]. Since the accession of Naram-Sîn is in -1774 and that the death of Êrišu II is in -1712, then the two kings ruled a total of 62 solar years (= 1774 - 1712), or 64 lunar years (or eponyms). The reign of Naram-Sîn was over 27 years since the list KEL A includes 27 eponyms after his accession. However, according to Assyrian king lists, his reign is [-]4 years, implying a duration of either 34, 44 or 54 years, the last two being the most likely possibilities[98]. Indeed, during the eponymy Ibni-Ištar (eponym No. 157) it is stated that "Šamšî-Adad I conquered Assyria" which seems to correspond to the 1st year of Êrišu II, his father Naram-Sin being died the previous year (beginning of the list MEC D). This would mean that the Amorite king Šamšî-Adad I conquered Assyria only gradually, starting with the city of Ekallatum at the end of the reign of Naram-Sîn. So when Êrišu II ascended the throne he reigned no longer than over a small part of Assyria and his death, after 10 years of reign, what was left of Assyria was absorbed by Šamšî-Adad I.
- The alliance with Qatna under eponymy of Ikuppiya coincides with the installation of Yasmah-Addu[99] (1687-1680) as king of Mari, by Šamšî-Adad I.

This reconstruction of the list of eponyms confirms the reliability of Assyrian king lists. Assyrian scribes could easily date a past event by equivalence: 1 eponym = 1 year. However the eponymous year was lunar (354.37 days) before Aššur-dân I, then was solar (365.24 days) from his reign (but Babylonian calendar with intercalation being adopted only from the reign of Tiglath-pileser I). The paleo-Assyrian calendar (or Amorite) was lunar while the calendar of Mari was lunisolar[100] like the one of Babylon. Synchronization among various calendars of the past is made difficult by these changing paradigms (unreported). For instance, on the death of Šamšî-Adad I it is possible to get the following synchronisms among months of several different calendars[101] (at least five):

---

94 C. MICHEL, P. ROCHER – La chronologie du IIe millénaire revue à l'ombre d'une éclipse de soleil in: *Jaarbericht (...) Ex Oriente Lux* N° 35/36 (1997-2000) Chicago pp. 111-126.

95 On October 10, 1737 BCE (of magnitude 0.92) and that on September 8, 1791 BCE (of magnitude 0.92)

96 As the sentence: *on the 26th day of the month Sivan, in the 7th year [of Simbar-šipak], the day turned to night*, did not describe a solar eclipse.

97 D. CHARPIN, N. ZIEGLER – Florilegium marianum V. Mari et le Proche-Orient à l'époque amorrite in: *Mémoires de N.A.B.U.* 6 (2003) pp. 72-73, 134-169.

98 K.R. VEENHOF – The Old Assyrian List of Year Eponyms from Karum Kanish and its Chronological Implications Ankara 2002 Ed Turkish Historical Society pp. 1-78.

99 D. CHARPIN – Rapport sur les conférences 1995-1996 in: Livret 11 1995-1996 (École Pratique des Hautes Études, 1997) pp. 15-16.

100 However the day 30 could be 29 or 1 (J.M. SASSON -Zimri-Lim Takes the Grand Tour in: *Biblical Archaeologist* 47, 1984, pp. 246-252).

101 D. CHARPIN, N. ZIEGLER – Florilegium marianum V. Mari et le Proche-Orient à l'époque amorrite in: *Mémoires de N.A.B.U.* 6 (2003) pp. 134-176, 260-262.

| | BABYLONIAN | | JULIAN | | MARIOTE | | AMORRITE | PALEO-ASSYRIAN |
|---|---|---|---|---|---|---|---|---|
| X | Tebêtu | 1 | January (*winter*) | *xi*° | *Abum (IV)* | *xi** | *Abum* | *Ab šarrâni (v*)* |
| XI | Šabâtu | 2 | February | *xii*° | *Ḫibirtum (V)* | *xii** | *Tîrum* | *Ḫubur (vi*)* |
| XII | Addâru | 3 | March | *i*° | *Ḫubur (Ḫilib)* | *i** | *Niqmum* | *Ṣip'im (vii*)* |
| I | Nisannu | 4 | April (*spring*) | *ii*° | *Kinûnum (VII)* | *ii** | *Kinûnum* | *Qarrâtim (viii*)* |
| II | Ayyaru | 5 | May | *iii*° | *Dagan (VIII)* | *iii** | *Tamḫîrum* | *Kanwarta (ix*)* |
| III | Simanu | 6 | June | *iv*° | *Lîliatum (IX)* | *iv** | *Nabrûm* | *Te'inâtim (x*)* |
| IV | Du'ùzu | 7 | July (*summer*) | *v*° | *Bêlet-bîrî (X)* | *v** | *Mammîtum* | *Kuzallu (xi*)* |
| V | Abu | 8 | August | *vi*° | *Kiskissum (XI)* | *vi** | *Mana* | *Allanâtim (xii*)* |
| VI | Ulûlû | 9 | September | *vii*° | *Ebûrum (XII)* | *vii** | *Ayyarum* | *Bêltî-ekallim (i*)* |
| VII | Tašrîtu | 10 | October (*autumn*) | *viii*° | *Urâḫum (I)* | *viii** | *Niggalum* | *Ša sarratim (ii*)* |
| VIII | Araḫsamna | 11 | November | *ix*° | *Malkânum (II)* | *ix** | *Maqrânum* | *Narmak Aššur (iii*)* |
| IX | Kisilimu | 12 | December | *x*° | *Laḫḫum (III* | *x** | *Du'uzum* | *Maḫḫurili (iv*)* |

The end of Šamšî-Adad I's reign is dated on February 20, -1679[102] because this king died on 14/*xii*°/**33**. The month *VI* in Mari coincides with the Assyrian month *i** (months *VI* to *XII* are dated "after the eponym Ṭab-ṣilla-Aššur"). The fall of Larsa is dated [1-6]/XII/**30** of Hammurabi and matches the [1-6]/*VI*/**60** of Rîm-Sîn I, because Zimrî-Lîm congratulated Hammurabi for his taking Larsa in his letter dated 7/*VI*/**12** (*ARM* XXV 9).

| | month | | | | [A] | [B] | [C] | [D] | **King** / *eponym* |
|---|---|---|---|---|---|---|---|---|---|
| **-1680** | 1 | *IV* | *xi*° | X | **6** | **32** | **16** | **46** | |
| | 2 | *V* | *xii*° | XI | | | | | |
| | 3 | *VI* | *i*° | XII | | **33** | | | *Ṭab-ṣilla-Aššur* |
| | 4 | *VII* | *ii*° | I | | | **17** | **47** | |
| | 5 | *VIII* | *iii*° | II | | | | | |
| | 6 | *IX* | *iv*° | III | | | | | |
| | 7 | *X* | *v*° | IV | | | | | |
| | 8 | *XI* | *vi*° | V | | | | | |
| | 9 | *XII* | *vii*° | VI | | | | | |
| | 10 | *I* | *viii*° | VII | **7** | | | | [A] **Yasmah-Addu** king of Mari |
| | 11 | *II* | *ix*° | VIII | | | | | [B] **Šamšî-Adad I** king of Assyria |
| | 12 | *III* | *x*° | IX | | | | | [C] **Hammurabi** king of Babylon |
| **-1679** | 1 | *IV* | *xi*° | X | | | | | [D] **Rîm-Sîn I** king of Larsa |
| | 2 | *V* | *xii*° | XI | | **0** | | | |
| | 3 | *VI* | *i*° | XII | **0** | **1** | | | *after Ṭab-ṣilla-Aššur* |
| | 4 | *VII* | *ii*° | I | | | **18** | **48** | |
| | 5 | *VIII* | *iii*° | II | | | | | [A] **Zimrî-Lîm** king of Mari |
| | 6 | *IX* | *iv*° | III | | | | | [B] **Išme-Dagan I** king of Assyria |
| | 7 | *X* | *v*° | IV | | | | | |
| | 8 | *XI* | *vi*° | V | | | | | |
| | 9 | *XII* | *vii*° | VI | | | | | |
| | 10 | *I* | *viii*° | VII | **1** | | | | *Ennam-Aššur* |
| | 11 | *II* | *ix*° | VIII | | | | | |
| | 12 | *III* | *x*° | IX | | | | | |
| **-1678** | 1 | *IV* | *xi*° | X | | | | | *(Feast of Ištar in month xi° Ab Šarrani)* |
| | 2 | *V* | *xii*° | XI | | | | | |
| | 3 | *VI* | *i*° | XII | | **2** | | | *Aššur-emuqi* |
| | 4 | *VII* | *ii*° | I | | | **19** | **49** | |
| | 5 | *VIII* | *iii*° | II | | | | | |
| | 6 | *IX* | *iv*° | III | | | | | |
| | 7 | *X* | *v*° | IV | | | | | |
| | 8 | *XI* | *vi*° | V | | | | | |
| | 9 | *XII* | *vii*° | VI | | | | | |
| | | *XIIb* | *viii*° | VII | | | | | |
| | 10 | *I* | *ix*° | VIII | **2** | | | | |
| | 11 | *II* | *x*° | IX | | | | | |
| | 12 | *III* | *xi*° | X | | | | | |

The presence or the absence of intercalation further complicates synchronizations among calendars. For instance, the year 1 of Zimrî-Lîm has an intercalary month (*xii°b*) but other years are strangely irregular (2:*xii°b*; 5:*ii°b*, *iii°b*, *v°b*; 8:*i°b*; 10:*v°b*; 11:*v°b*)[103]. On the other hand the feast of Ištar seems to be celebrated without intercalation[104].

[102] An exorcist priest (*wašipum*) is consulted on 11/*xii*°/**33** and the oil for the offering king's burial came on 16/*xii*°/**33**. In 1679 BCE, 1st Nisan is dated April 5, 1st Tishri on September 30 and 1st Ṣip'im March 7. It is interesting to notice that the year 33 of Šamši-Adad I started with a total lunar eclipse (bad omen) http://eclipse.gsfc.nasa.gov/5MCLEmap/-1699--1600/LE-1678-03-21T.gif

Mesopotamian chronologies are anchored by numerous synchronisms (highlighted in light blue) and dated by astronomical phenomena (boxed Julian years). Synchronisms with Elamite, Egyptian and Israelite chronologies are given only for information:

| | UR III | | ELAM | | | | ASSYRIA | | EGYPT | |
|---|---|---|---|---|---|---|---|---|---|---|
| **1967** | Šulgi | 35 | AWAN | | SIMAŠKI | | Amînum | | Amenemhat I | 8 |
| **1966** | | 36 | Kutir-lagamar | 24 | Girname | | | | | 9 |
| **1965** | | 37 | | 25 | | | | | | 10 |
| **1964** | | 38 | | 26 | | | | | | 11 |
| **1963** | | 39 | | 27 | | | | | | 12 |
| **1962** | | 40 | | 28 | | | | | | 13 |
| **1961** | | 41 | | 29 | | | | | | 14 |
| **1960** | | 42 | | 30 | Ebarat I | | | | | 15 |
| **1959** | | 43 | | 31 | | | | | | 16 |
| **1958** | | 44 | | 32 | | | | | | 17 |
| **1957** | | 45 | | 33 | | | | | | 18 |
| **1956** | | 46 | | 34 | | | | | | 19 |
| **1955** | | 47 | | 35 | | | | | | 20 |
| **1954** | | 48 | | 36 | | | Sulili (Zariqum) | | | 21 |
| **1953** | Amar-Sîn | 1 | | | | | | | | 22 |
| **1952** | | 2 | | | | | | | | 23 |
| **1951** | | 3 | | | | | | | | 24 |
| **1950** | | 4 | | | Tazatta I | | | | | 25 |
| **1949** | | 5 | | | | | | | | 26 |
| **1948** | | 6 | | | | | | | | 27 |
| **1947** | | 7 | | | | | | | | 28 |
| **1946** | | 8 | | | | | | | | 29 |
| **1945** | | 9 | | | | | | | Sesostris I | 1 |
| **1944** | Šû-Sîn | 1 | | | | | | | | 2 |
| **1943** | | 2 | | | | | | | | 3 |
| **1942** | | 3 | | | | | | | | 4 |
| **1941** | | 4 | | | | | | | | 5 |
| **1940** | | 5 | | | | | Kikkia | | | 6 |
| **1939** | | 6 | | | | | | | | 7 |
| **1938** | | 7 | | | | | | | | 8 |
| **1937** | | 8 | | | | | | | | 9 |
| **1936** | | 9 | | | | | | | | 10 |
| **1935** | Ibbi-Sîn | 1 | | | Tazitta II | | | | | 11 |
| **1934** | | 2 | | | | | | | | 12 |
| **1933** | | 3 | | | | | | | | 13 |
| **1932** | | 4 | | | **LARSA** | | | | | 14 |
| **1931** | | 5 | | | | | | | | 15 |
| **1930** | | 6 | | | Naplânum | 1 | | | | 16 |
| **1929** | | 7 | | | | 2 | | | | 17 |
| **1928** | | 8 | | | | 3 | | | | 18 |
| **1927** | | 9 | | | | 4 | | | | 19 |
| **1926** | | 10 | | | | 5 | Akia | | | 20 |
| **1925** | | 11 | | | | 6 | | | | 21 |
| **1924** | | 12 | **ISIN** | | | 7 | | | | 22 |
| **1923** | | 13 | | | | 8 | | | | 23 |
| **1922** | | 14 | Išbi-Erra | 1 | | 9 | | | | 24 |
| **1921** | | 15 | | 2 | | 10 | | | | 25 |
| **1920** | | 16 | | 3 | | 11 | | | | 26 |

[103] W. HEIMPEL – Letters to the King of Mari: A New Translation, With Historical Introduction, Notes, and Commentary Leiden 2003 Ed. Eisenbrauns pp. 54-56.
[104] For example the feast of Ištar is celebrated month *xi* in 1 year of Zimrî-Lîm, month *ix* in years 6-8 and month *viii* in year 12, which implies a lag of about 3 months on 12 years, indicating a lack of intercalation (at least in one of the two calendars).

| | | | | | | | | | | |
|---|---|---|---|---|---|---|---|---|---|---|
| **1919** | | 17 | | 4 | | 12 | | | | 27 |
| **1918** | | 18 | | 5 | | 13 | | | | 28 |
| **1917** | | 19 | | 6 | | 14 | | | | 29 |
| **1916** | | 20 | | 7 | | 15 | | | | 30 |
| **1915** | | 21 | | 8 | | 16 | | | | 31 |
| **1914** | | 22 | | 9 | | 17 | | | | 32 |
| **1913** | | 23 | | 10 | | 18 | | | | 33 |
| **1912** | | 24 | | 11 | | 19 | Puzur-Aššur I | | | 34 |
| **1911** | | | | 12 | | 20 | | | | 35 |
| **1910** | | | | 13 | | 21 | | | | 36 |
| **1909** | | | | 14 | Iemṣium | 1 | | | | 37 |
| **1908** | | | | 15 | | 2 | | | | 38 |
| **1907** | | | | 16 | | 3 | | | | 39 |
| **1906** | | | | 17 | | 4 | | | | 40 |
| **1905** | | | | 18 | | 5 | | | | 41 |
| **1904** | | | | 19 | | 6 | | | | 42 |
| **1903** | | | | 20 | | 7 | | | | 43 |
| **1902** | | | | 21 | | 8 | | | | 44 |
| **1901** | | | | 22 | | 9 | | | | 45 |
| **1900** | | | | 23 | | 10 | Šalim-ahum | | Amenemhat II | 1 |
| **1899** | | | | 24 | | 11 | | | | 2 |
| **1898** | | | | 25 | | 12 | | | | 3 |
| **1897** | | | | 26 | | 13 | | | | 4 |
| **1896** | | | | 27 | | 14 | | | | 5 |
| **1895** | | | | 28 | | 15 | | | | 6 |
| **1894** | | | | 29 | | 16 | | | | 7 |
| **1893** | | | | 30 | | 17 | | | | 8 |
| **1892** | | | | 31 | | 18 | | | | 9 |
| **1891** | | | | 32 | | 19 | | | | 10 |
| **1890** | | | | 33 | | 20 | | | | 11 |
| **1889** | | | Šû-ilîšu | 1 | | 21 | | | | 12 |
| **1888** | | | | 2 | | 22 | | | | 13 |
| **1887** | | | | 3 | | 23 | | | | 14 |
| **1886** | | | | 4 | | 24 | Ilu-šumma | | | 15 |
| **1885** | | | | 5 | | 25 | | | | 16 |
| **1884** | | | | 6 | | 26 | | | | 17 |
| **1883** | | | | 7 | | 27 | | | | 18 |
| **1882** | | | | 8 | | 28 | | | | 19 |
| **1881** | | | | 9 | Sâmium | 1 | | | | 20 |
| **1880** | | | | 10 | | 2 | | | | 21 |
| **1879** | | | Iddin-Dagân | 1 | | 3 | | | | 22 |
| **1878** | | | | 2 | | 4 | | | | 23 |
| **1877** | | | | 3 | | 5 | | | | 24 |
| **1876** | | | | 4 | | 6 | | | | 25 |
| **1875** | | | | 5 | | 7 | | | | 26 |
| **1874** | | | | 6 | | 8 | | | | 27 |
| **1873** | | | | 7 | | 9 | | | | 28 |
| **1872** | | | | 8 | | 10 | Êrišu I | 1 | | 29 |
| **1871** | | | | 9 | | 11 | | 2 | | 30 |
| **1870** | | | | 10 | | 12 | | 3 | | 31 |
| | | | | | | | | 4 | | |
| **1869** | | | | 11 | | 13 | | 5 | | 32 |
| **1868** | | | | 12 | | 14 | | 6 | | 33 |
| **1867** | | | | 13 | | 15 | | 7 | | 34 |
| **1866** | | | | 14 | | 16 | | 8 | | 35 |
| **1865** | | | | 15 | | 17 | | 9 | | 36 |
| **1864** | | | | 16 | | 18 | | 10 | | 37 |
| **1863** | | | | 17 | | 19 | | 11 | | 38 |
| **1862** | | | | 18 | | 20 | | 12 | Sesostris II | 1 |
| **1861** | | | | 19 | | 21 | | 13 | | 2 |
| **1860** | | | | 20 | | 22 | | 14 | | 3 |
| **1859** | | | | 21 | | 23 | | 15 | | 4 |
| **1858** | | | Išme-Dagân | 1 | | 24 | | 16 | | 5 |
| **1857** | | | | 2 | | 25 | | 17 | | 6 |
| **1856** | | | | 3 | | 26 | | 18 | | 7 |
| **1855** | | | | 4 | | 27 | | 19 | | 8 |
| **1854** | | | | 5 | | 28 | | 20 | Sesostris III | 1 |
| **1853** | | | | 6 | | 29 | | 21 | | 2 |
| **1852** | | | | 7 | | 30 | | 22 | | 3 |
| **1851** | | | | 8 | | 31 | | 23 | | 4 |
| **1850** | | | | 9 | | 32 | | 24 | | 5 |
| **1849** | | | | 10 | | 33 | | 25 | | 6 |
| **1848** | | | | 11 | | 34 | | 26 | | 7 |
| **1847** | | | | 12 | | 35 | | 27 | | 8 |
| **1846** | | | | 13 | Zabâia | 1 | | 28 | | 9 |
| **1845** | | | | 14 | | 2 | | 29 | | 10 |
| **1844** | | | | 15 | | 3 | | 30 | | 11 |
| **1843** | | | | 16 | | 4 | | 31 | | 12 |
| **1842** | | | | 17 | | 5 | | 32 | | 13 |
| **1841** | | | | 18 | | 6 | | 33 | | 14 |
| **1840** | | | | 19 | | 7 | | 34 | | 15 |
| **1839** | | | | 20 | | 8 | | 35 | | 16 |
| **1838** | | | Lipit-Eštar | 1 | | 9 | | 36 | | 17 |
| **1837** | | | | 2 | Gungunum | 1 | | 37 | | 18 |
| **1836** | | | | 3 | | 2 | | 38 | | 19 |
| **1835** | | | | 4 | | 3 | | 39 | Amenemhat III | 1 |
| **1834** | | | | 5 | | 4 | | 40 | | 2 |
| **1833** | | | | 6 | | 5 | Ikunum | 1 | | 3 |
| **1832** | | | | 7 | | 6 | | 2 | | 4 |
| **1831** | | | | 8 | | 7 | | 3 | | 5 |
| **1830** | | | | 9 | | 8 | | 4 | | 6 |
| | | | | | | | | 5 | | |
| **1829** | | | | 10 | | 9 | | 6 | | 7 |
| **1828** | | | | 11 | | 10 | | 7 | | 8 |
| **1827** | | | Ur-Ninurta | 1 | | 11 | | 8 | | 9 |
| **1826** | | | | 2 | | 12 | | 9 | | 10 |
| **1825** | | | | 3 | | 13 | | 10 | | 11 |
| **1824** | | | | 4 | | 14 | | 11 | | 12 |
| **1823** | | | | 5 | | 15 | | 12 | | 13 |
| **1822** | | | | 6 | | 16 | | 13 | | 14 |
| **1821** | | | | 7 | | 17 | | 14 | | 15 |
| **1820** | | | | 8 | | 18 | Sargon I | 1 | | 16 |
| **1819** | | | | 9 | | 19 | | 2 | | 17 |
| **1818** | | | | 10 | | 20 | | 3 | | 18 |
| **1817** | | | | 11 | | 21 | | 4 | | 19 |
| **1816** | | | | 12 | | 22 | | 5 | | 20 |
| **1815** | | | | 13 | | 23 | | 6 | | 21 |
| **1814** | | | | 14 | | 24 | | 7 | | 22 |
| **1813** | | | | 15 | | 25 | | 8 | | 23 |
| **1812** | | | | 16 | | 26 | | 9 | | 24 |
| **1811** | | | | 17 | | 27 | | 10 | | 25 |
| **1810** | | | | 18 | Abî-sarê | 1 | | 11 | | 26 |
| **1809** | | | | 19 | | 2 | | 12 | | 27 |
| **1808** | | | | 20 | | 3 | | 13 | | 28 |
| **1807** | | | | 21 | | 4 | | 14 | | 29 |
| **1806** | | | | 22 | | 5 | | 15 | | 30 |
| **1805** | | | | 23 | | 6 | | 16 | | 31 |
| **1804** | | | | 24 | | 7 | | 17 | | 32 |
| **1803** | | | | 25 | | 8 | | 18 | | 33 |
| **1802** | | | | 26 | | 9 | | 19 | | 34 |
| **1801** | | | | 27 | | 10 | | 20 | | 35 |
| | | | | | | | | 21 | | |
| **1800** | **BABYLON** | | | 28 | | 11 | | 22 | | 36 |
| **1799** | | 0 | Bûr-Sîn | 1 | Sûmû-El | 1 | | 23 | | 37 |
| **1798** | Sûmû-abum | 1 | | 2 | | 2 | | 24 | | 38 |
| **1797** | | 2 | | 3 | | 3 | | 25 | | 39 |
| **1796** | | 3 | | 4 | | 4 | | 26 | | 40 |
| **1795** | | 4 | | 5 | | 5 | | 27 | | 41 |
| **1794** | | 5 | | 6 | | 6 | | 28 | | 42 |
| **1793** | | 6 | | 7 | | 7 | | 29 | | 43 |
| **1792** | | 7 | | 8 | | 8 | | 30 | | 44 |
| **1791** | | 8 | | 9 | | 9 | | 31 | | 45 |
| **1790** | | 9 | | 10 | | 10 | | 32 | Amenemhat IV | 1 |
| **1789** | | 10 | | 11 | | 11 | | 33 | | 2 |
| **1788** | | 11 | | 12 | | 12 | | 34 | | 3 |
| **1787** | | 12 | | 13 | | 13 | | 35 | | 4 |
| **1786** | | 13 | | 14 | | 14 | | 36 | | 5 |
| **1785** | | 14 | | 15 | | 15 | | 37 | | 6 |
| **1784** | Sûmû-la-Il | 1 | | 16 | | 16 | | 38 | | 7 |
| **1783** | | 2 | | 17 | | 17 | | 39 | | 8 |
| **1782** | | 3 | | 18 | | 18 | | 40 | | 9 |
| **1781** | | 4 | | 19 | | 19 | Puzur-Aššur II | 1 | Neferusebek | *1* |
| **1780** | | 5 | | 20 | | 20 | | 2 | | *2* |
| **1779** | | 6 | | 21 | | 21 | | 3 | | *3* |
| **1778** | | 7 | Lipit-Enlil | 1 | | 22 | | 4 | | *4* |
| **1777** | | 8 | | 2 | | 23 | | 5 | Ugaf | *1* |
| **1776** | | 9 | | 3 | | 24 | | 6 | | *2* |
| **1775** | | 10 | | 4 | | 25 | | 7 | Amenemhat V | *1* |
| **1774** | | 11 | | 5 | | 26 | | 8 | | *2* |
| **1773** | | 12 | Erra-imittî | 1 | | 27 | Naram-Sîn | 1 | | *3* |
| **1772** | | 13 | | 2 | | 28 | | 2 | | *4* |
| **1771** | | 14 | | 3 | | 29 | | 3 | (Ameny)Qemau | *1* |
| **1770** | | 15 | | 4 | Nûr-Adad | 1 | | 4 | | *2* |
| **1769** | | 16 | | 5 | | 2 | | 5 | | *3* |
| **1768** | | 17 | | 6 | | 3 | | 6 | | *4* |
| **1767** | | 18 | | 7 | | 4 | | 7 | | *5* |
| **1766** | | 19 | Enlil-Bâni | 1 | | 5 | | 8 | Sehetepibre | *1* |
| | | | | | | | | 9 | | |
| **1765** | | 20 | | 2 | | 6 | | 10 | | *2* |
| **1764** | | 21 | | 3 | | 7 | | 11 | | *3* |
| **1763** | | 22 | | 4 | | 8 | | 12 | | *4* |
| **1762** | | 23 | | 5 | | 9 | | 13 | | *5* |
| **1761** | | 24 | | 6 | | 10 | | 14 | Iufni | *1* |
| **1760** | | 25 | | 7 | | 11 | | 15 | Amenemhat VI | *1* |
| **1759** | | 26 | | 8 | | 12 | | 16 | | *2* |
| **1758** | | 27 | | 9 | | 13 | | 17 | | *3* |
| **1757** | | 28 | | 10 | | 14 | | 18 | | *4* |
| **1756** | | 29 | | 11 | | 15 | | 19 | | *5* |
| **1755** | | 30 | | 12 | | 16 | | 20 | Nebnun | *1* |
| **1754** | | 31 | | 13 | Sîn-iddinam | 1 | | 21 | Hornedjheritef | *1* |
| **1753** | | 32 | | 14 | | 2 | | 22 | | *2* |
| **1752** | | 33 | | 15 | | 3 | | 23 | | *3* |
| **1751** | | 34 | | 16 | | 4 | | 24 | | *4* |
| **1750** | | 35 | | 17 | | 5 | | 25 | | *5* |
| **1749** | | 36 | | 18 | | 6 | | 26 | Sewadjkare | *1* |
| **1748** | Sâbium | 1 | | 19 | | 7 | | 27 | | *2* |
| **1747** | | 2 | | 20 | Sîn-irîbam | 1 | | 28 | | *3* |
| **1746** | | 3 | | 21 | | 2 | | 29 | | *4* |
| **1745** | | 4 | | 22 | Sîn-iqišam | 1 | | 30 | | *5* |
| **1744** | | 5 | | 23 | | 2 | | 31 | Nedjemebre | *1* |
| **1743** | | 6 | | 24 | | 3 | | 32 | Sebekhotep I | *1* |
| **1742** | | 7 | Zambîya | 1 | | 4 | | 33 | | *2* |
| **1741** | | 8 | | 2 | | 5 | | 34 | | *3* |
| **1740** | | 9 | | 3 | Silî-Adad | 1 | | 35 | Rensebeb | *4* |
| **1739** | | 10 | Iter-piša | 1 | Warad-Sîn | 1 | | 36 | Hor I | *1* |
| **1738** | | 11 | | 2 | | 2 | | 37 | | *2* |
| **1737** | | 12 | | 3 | | 3 | | 38 | | *3* |
| **1736** | | 13 | | 4 | | 4 | | 39 | | *4* |
| **1735** | | 14 | Ur-dukuga | 1 | | 5 | | 40 | | *5* |
| **1734** | Apil-Sîn | 1 | | 2 | | 6 | | 41 | Amenemhat VII | *1* |
| | | | | | | | | 42 | | |
| **1733** | | 2 | | 3 | | 7 | | 43 | | *2* |
| **1732** | | 3 | | 4 | | 8 | | 44 | | *3* |
| **1731** | | 4 | Sîn-mâgir | 1 | | 9 | | 45 | | *4* |
| **1730** | | 5 | | 2 | | 10 | | 46 | | *5* |
| **1729** | | 6 | | 3 | | 11 | | 47 | | *6* |
| **1728** | | 7 | | 4 | | 12 | | 48 | | *7* |
| **1727** | | 8 | | 5 | | 13 | | 49 | Sebekhotep II | *1* |
| **1726** | | 9 | | 6 | Rîm-Sîn I | 1 | | 50 | | *2* |
| **1725** | | 10 | | 7 | | 2 | | 51 | | *3* |
| **1724** | | 11 | | 8 | | 3 | | 52 | | *4* |
| **1723** | | 12 | | 9 | | 4 | | 53 | | *5* |
| **1722** | | 13 | | 10 | | 5 | | 54 | | *6* |
| **1721** | | 14 | | 11 | | 6 | Êrišu II | 1 | | *7* |
| **1720** | | 15 | Damiq-ilîšu | 1 | | 7 | | 2 | Kendjer | *1* |
| **1719** | | 16 | | 2 | | 8 | | 3 | | *2* |
| **1718** | | 17 | | 3 | | 9 | | 4 | | *3* |
| **1717** | | 18 | | 4 | | 10 | | 5 | | *4* |
| **1716** | Sîn-muballiṭ | 1 | | 5 | | 11 | | 6 | | *5* |
| **1715** | | 2 | | 6 | | 12 | | 7 | Semenkhkare | *1* |
| **1714** | | 3 | | 7 | | 13 | | 8 | | *2* |
| **1713** | | 4 | | 8 | | 14 | | 9 | | *3* |
| **1712** | | 5 | | 9 | | 15 | | 10 | | *4* |
| **1711** | | 6 | | 10 | | 16 | Šamšî-Adad I | 1 | | *5* |
| **1710** | | 7 | | 11 | | 17 | | 2 | Antef IV | *1* |
| **1709** | | 8 | | 12 | | 18 | | 3 | | *2* |
| **1708** | | 9 | | 13 | | 19 | | 4 | | *3* |
| **1707** | | 10 | | 14 | | 20 | | 5 | | *4* |
| **1706** | | 11 | | 15 | | 21 | | 6 | | *5* |
| **1705** | | 12 | | 16 | | 22 | | 7 | Seth | *1* |
| **1704** | | 13 | | 17 | | 23 | | 8 | | *2* |
| **1703** | | 14 | | 18 | | 24 | | 9 | | *3* |
| **1702** | | 15 | | 19 | | 25 | | 10 | | *4* |
| | | | | | | | | 11 | | |
| **1701** | | 16 | | 20 | | 26 | | 12 | Sebekhotep III | *1* |
| **1700** | | 17 | | 21 | | 27 | | 13 | | *2* |
| **1699** | | 18 | | 22 | | 28 | | 14 | | *3* |
| **1698** | | 19 | | 23 | | 29 | | 15 | | *4* |
| **1697** | | 20 | | | | 30 | | 16 | Neferhotep I | *1* |
| **1696** | Hammurabi | 1 | | | | 31 | | 17 | | *2* |
| **1695** | | 2 | | | | 32 | | 18 | | *3* |
| **1694** | | 3 | | | | 33 | | 19 | | *4* |
| **1693** | | 4 | | | | 34 | | 20 | | *5* |
| **1692** | | 5 | | | | 35 | | 21 | | *6* |
| **1691** | | 6 | | | | 36 | | 22 | | *7* |
| **1690** | | 7 | | | | 37 | | 23 | | *8* |
| **1689** | | 8 | | | | 38 | | 24 | | *9* |
| **1688** | | 9 | | | | 39 | | 25 | | *10* |
| **1687** | | 10 | | | | 40 | | 26 | | *11* |
| **1686** | | 11 | | | | 41 | | 27 | Sahathor | *12* |
| **1685** | | **12** | | | | 42 | *Asqudum* | **28** | Sebekhotep IV | *1* |
| **1684** | | 13 | | | | 43 | *(lunar eclipse)* | 29 | | *2* |
| **1683** | | 14 | | | | 44 | | 30 | | *3* |
| **1682** | | 15 | | | | 45 | | 31 | | *4* |
| **1681** | | 16 | | | | 46 | | 32 | | *5* |
| **1680** | | 17 | | | | 47 | | 33 | | *6* |
| **1679** | | 18 | | | | 48 | Išme-Dagan I | 1 | | *7* |
| **1678** | | 19 | | | | 49 | | 2 | | *8* |
| **1677** | | 20 | | | | 50 | | 3 | | *9* |
| **1676** | | 21 | | | | 51 | | 4 | Sebekhotep V | *1* |
| **1675** | | 22 | | | | 52 | | 5 | | *2* |
| **1674** | | 23 | | | | 53 | | 6 | | *3* |
| **1673** | | 24 | | | | 54 | | 7 | | *4* |
| **1672** | | 25 | | | | 55 | | 8 | | *5* |
| **1671** | | 26 | | | | 56 | | 9 | Sebekhotep VI | *1* |
| **1670** | | 27 | | | | 57 | | 10 | | *2* |
| | | | | | | | | 11 | | |
| **1669** | | 28 | | | | 58 | Aššur-dugul | 1 | Ibia | *1* |
| **1668** | | 29 | | | | 59 | | 2 | | *2* |
| **1667** | | 30 | | | | 60 | | 3 | | *3* |
| **1666** | | 31 | | | | | | 4 | | *4* |
| **1665** | | 32 | | | | | | 5 | | *5* |
| **1664** | | 33 | | | | | | 6 | | *6* |
| **1663** | | 34 | | | | | Bêlu-bâni | 1 | | *7* |
| **1662** | | 35 | | | | | | 2 | | *8* |
| **1661** | | 36 | | | | | | 3 | | *9* |
| **1660** | | 37 | | | | | | 4 | | *10* |
| **1659** | | 38 | | | | | | 5 | | *11* |
| **1658** | | 39 | | | | | | 6 | Aÿ | *1* |
| **1657** | | 40 | | | | | | 7 | | *1* |
| **1656** | | 41 | | | | | | 8 | | *2* |
| **1655** | | 42 | | | | | | 9 | | *3* |
| **1654** | | 43 | | | | | | 10 | | *4* |
| **1653** | Samsu-iluna | 1 | | | | | Libbaya | 1 | | *5* |
| **1652** | | 2 | | | | | | 2 | | *6* |
| **1651** | | 3 | | | | | | 3 | | *7* |
| **1650** | | 4 | | | | | | 4 | | *8* |
| **1649** | | 5 | | | | | | 5 | | *9* |
| **1648** | | 6 | | | | | | 6 | | *10* |
| **1647** | | 7 | | | | | | 7 | | *11* |
| **1646** | | 8 | | | | | | 8 | | *12* |
| **1645** | | 9 | | | | | | 9 | | *13* |
| **1644** | | 10 | | | | | | 10 | | *14* |
| **1643** | | 11 | | | | | | 11 | | *15* |
| **1642** | | 12 | | | | | | 12 | | *16* |
| **1641** | | 13 | | | | | | 13 | | *17* |
| **1640** | | 14 | | | | | | 14 | | *18* |
| **1639** | | 15 | | | | | | 15 | | *19* |
| | | | | | | | | 16 | | |
| **1638** | | 16 | | | | | | 17 | | *20* |
| **1637** | | 17 | | | | | Šarma-Adad I | 1 | | *21* |
| **1636** | | 18 | | | | | | 2 | | *22* |
| **1635** | | 19 | | | | | | 3 | | *23* |
| **1634** | | 20 | | | | | | 4 | | *24* |
| **1633** | | 21 | | | | | | 5 | Ani | *1* |
| **1632** | | 22 | | | | | | 6 | | *2* |
| **1631** | | 23 | | | | | | 7 | Sewadjtu | *1* |
| **1630** | | 24 | | | | | | 8 | | *2* |
| **1629** | | 25 | | | | | | 9 | | *3* |
| **1628** | | 26 | | | | | | 10 | Neferhotep II | *1* |
| **1627** | | 27 | | | | | | 11 | | *2* |
| **1626** | | 28 | | | | | | 12 | | *3* |
| **1625** | | 29 | | | | | Puzur-Sîn | 1 | Hori | *1* |
| **1624** | | 30 | | | | | | 2 | | *2* |
| **1623** | | 31 | | | | | | 3 | | *3* |
| **1622** | | 32 | | | | | | 4 | | *4* |
| **1621** | | 33 | | | | | | 5 | | *5* |
| **1620** | | 34 | | | | | | 6 | Sebekhotep VII | *1* |
| **1619** | | 35 | | | | | | 7 | | *2* |
| **1618** | | 36 | | | | | | 8 | | |
| **1617** | | 37 | | | | | | 9 | | |
| **1616** | | 38 | | | | | | 10 | | |
| **1615** | Abi-ešuḫ | 1 | | | | | | 11 | | |
| **1614** | | 2 | | | **ISRAEL** | | | 12 | | |
| **1613** | | 3 | | | | 0 | Bazaya | 1 | | |
| **1612** | | 4 | | | Apopi | 1 | | 2 | | |
| **1611** | | 5 | | | (*birth*) | 2 | | 3 | | |
| **1610** | | 6 | | | | 3 | | 4 | | |
| **1609** | | 7 | | | | 4 | | 5 | | |
| **1608** | | 8 | | | | 5 | | 6 | | |
| **1607** | | 9 | | | | 6 | | 7 | | |
| **1606** | | 10 | | | | 7 | | 8 | | |
| | | | | | | | | 9 | | |
| **1605** | | 11 | | | | 8 | | 10 | | |
| **1604** | | 12 | | | | 9 | | 11 | | |
| **1603** | | 13 | | | | 10 | | 12 | | |
| **1602** | | 14 | | | | 11 | | 13 | | |
| **1601** | | 15 | | | | 12 | | 14 | | |
| **1600** | | 16 | | | | 13 | | 15 | | |
| **1599** | | 17 | | | | 14 | | 16 | | |
| **1598** | | 18 | | | | 15 | | 17 | | |
| **1597** | | 19 | | | | 16 | | 18 | | |
| **1596** | | 20 | | | | 17 | | 19 | | |
| **1595** | | 21 | | | | 18 | | 20 | | |
| **1594** | | 22 | | | | 19 | | 21 | | |
| **1593** | | 23 | | | | 20 | | 22 | | |
| **1592** | | 24 | | | | 21 | | 23 | | |
| **1591** | | 25 | | | | 22 | | 24 | | |
| **1590** | | 26 | | | | 23 | | 25 | | |
| **1589** | | 27 | | | | 24 | | 26 | | |
| **1588** | | 28 | | | | 25 | | 27 | | |
| **1587** | Ammiditana | 1 | | | | 26 | | 28 | | |
| **1586** | | 2 | | | | 27 | Lullaya | 1 | | |
| **1585** | | 3 | | | | 28 | | 2 | | |
| **1584** | | 4 | | | | 29 | | 3 | | |
| **1583** | | 5 | | | | 30 | | 4 | | |
| **1582** | | 6 | | | | 31 | | 5 | | |
| **1581** | | 7 | | | | 32 | | 6 | | |
| **1580** | | 8 | | | (*papyrus Rhind*) | **33** | Šû-Ninûa | 1 | | |
| **1579** | | 9 | | | | 34 | | 2 | | |
| **1578** | | 10 | | | | 35 | | 3 | | |
| **1577** | | 11 | | | | 36 | | 4 | | |
| **1576** | | 12 | | | | 37 | | 5 | | |
| **1575** | | 13 | | | | 38 | | 6 | | |
| | | | | | | | | 7 | | |
| **1574** | | 14 | | | | 39 | | 8 | | |
| **1573** | | 15 | | | (*Turin Canon*) | **40** | | 9 | **Dynasty XVII** | *0* |
| **1572** | | 16 | | | Moses (Apopi) | 1 | | 10 | Râhotep | *1* |
| **1571** | | 17 | | | (*Madian stay*) | 2 | | 11 | | *2* |
| **1570** | | 18 | | | | 3 | | 12 | | *3* |
| **1569** | | 19 | | | | 4 | | 13 | | *4* |
| **1568** | | 20 | | | | 5 | | 14 | Sobekemsaf I | *1* |
| **1567** | | 21 | | | | 6 | Šarma-Adad II | 1 | | *2* |
| **1566** | | 22 | | | | 7 | | 2 | Sobekemsaf II | *1* |
| **1565** | | 23 | | | | 8 | | 3 | | *2* |
| **1564** | | 24 | | | | 9 | Êrišu III | 1 | | *3* |
| **1563** | | 25 | | | | 10 | | 2 | | *4* |
| **1562** | | 26 | | | | 11 | | 3 | | *5* |
| **1561** | | 27 | | | | 12 | | 4 | | *6* |
| **1560** | | 28 | | | | 13 | | 5 | | *7* |
| **1559** | | 29 | | | | 14 | | 6 | | *8* |
| **1558** | | 30 | | | | 15 | | 7 | | *9* |
| **1557** | | 31 | | | | 16 | | 8 | | *10* |
| **1556** | | 32 | | | | 17 | | 9 | Antef VI | *1* |
| **1555** | | 33 | | | | 18 | | 10 | | *2* |
| **1554** | | 34 | | | | 19 | | 11 | Antef VII | *1* |
| | | | | | | | | 12 | | |
| **1553** | | 35 | | | | 20 | | 13 | | *2* |
| **1552** | | 36 | | | | 21 | Šamšî-Adad II | 1 | | *3* |
| **1551** | | 37 | | | | 22 | | 2 | | *4* |
| **1550** | Ammiṣaduqa | **1** | | | | 23 | | 3 | | *5* |
| **1549** | | **2** | | | | 24 | | 4 | | *6* |
| **1548** | | **3** | | | | 25 | | 5 | | *7* |
| **1547** | | **4** | | | | 26 | | 6 | | *8* |
| **1546** | | **5** | | | | 27 | Išme-Dagan II | 1 | | *9* |
| **1545** | | **6** | | | | 28 | | 2 | Antef VIII | *10* |
| **1544** | | **7** | | | | 29 | | 3 | Senakhtenrê | *1* |
| **1543** | | **8** | | | | 30 | | 4 | Seqenenrê Taa | 1 |
| **1542** | | **9** | | | | 31 | | 5 | | 2 |
| **1541** | | **10** | | | | 32 | | 6 | | 3 |
| **1540** | | **11** | | | | 33 | | 7 | | 4 |
| **1539** | | **12** | | | | 34 | | 8 | | 5 |
| **1538** | | **13** | | | | 35 | | 9 | | 6 |
| **1537** | | **14** | | | | 36 | | 10 | | 7 |
| **1536** | | **15** | | | | 37 | | 11 | | 8 |
| **1535** | | **16** | | | | 38 | | 12 | | 9 |
| **1534** | | **17** | | | | 39 | | 13 | | 10 |
| **1533** | | **18** | | | | 40 | | 14 | | **11** |
| **1532** | | **19** | | | Moses (Apopi) | **1** | | 15 | Kamose | 1 |
| **1531** | | **20** | | | *(Sinai Exodus)* | 2 | | 16 | | 2 |
| **1530** | | **21** | | | | 3 | Šamšî-Adad III | 1 | | 3 |
| **1529** | Samsuditana | 1 | | | | 4 | | 2 | Ahmose | 1 |
| **1528** | | 2 | | | | 5 | | 3 | | 2 |
| | | | | | | | | 4 | | |
| **1527** | | 3 | | | | 6 | | 5 | | 3 |
| **1526** | | 4 | | | | 7 | | 6 | | 4 |
| **1525** | | 5 | | | | 8 | | 7 | | 5 |
| **1524** | | 6 | | | | 9 | | 8 | | 6 |
| **1523** | | 7 | | | | 10 | | 9 | | 7 |
| **1522** | | 8 | | | | 11 | | 10 | | 8 |
| **1521** | | 9 | | | | 12 | | 11 | | 9 |
| **1520** | | 10 | | | | 13 | | 12 | | 10 |
| **1519** | | 11 | | | | 14 | | 13 | | 11 |
| **1518** | | 12 | | | | 15 | | 14 | | 12 |
| **1517** | | 13 | | | | 16 | | 15 | | 13 |
| **1516** | | 14 | | | | 17 | | 16 | | 14 |
| **1515** | | 15 | | | | 18 | Aššur-nêrârî I | 1 | | 15 |
| **1514** | | 16 | | | | 19 | | 2 | | 16 |
| **1513** | | 17 | | | | 20 | | 3 | | 17 |
| **1512** | | 18 | | | | 21 | | 4 | | 18 |
| **1511** | | 19 | | | | 22 | | 5 | | 19 |
| **1510** | | 20 | | | | 23 | | 6 | | 20 |
| **1509** | | 21 | | | | 24 | | 7 | | 21 |
| **1508** | | 22 | | | | 25 | | 8 | | 22 |
| **1507** | | 23 | | | | 26 | | 9 | | 23 |
| **1506** | | 24 | | | | 27 | | 10 | | 24 |
| **1505** | | 25 | | | | 28 | | 11 | | 25 |
| **1504** | | 26 | | | | 29 | | 12 | Amenhotep I | 1 |
| **1503** | | 27 | | | | 30 | | 13 | | 2 |
| **1502** | | 28 | | | | 31 | | 14 | | 3 |
| **1501** | | 29 | | | | 32 | | 15 | | 4 |
| **1500** | *Fall of Alep* | 30 | | | | 33 | | 16 | | 5 |
| **1499** | *Fall of Babylon* | 31 | | | | 34 | | 17 | | 6 |
| **1498** | | 0 | | | | 35 | | 18 | | 7 |
| **1497** | *recovery of* | 1 | | | | 36 | | 19 | | 8 |
| **1496** | *Babylon* | 2 | | | | 37 | | 20 | | 9 |
| **1495** | | 3 | | | | 38 | | 21 | | 10 |
| **1494** | | 4 | | | | 39 | | 22 | | 11 |
| | | | | | | | | 23 | | |
| **1493** | | 5 | | | | 40 | | 24 | | 12 |
| **1492** | | 6 | | | Joshua | 1 | | 25 | | 13 |
| **1491** | | 7 | | | (*in Canaan*) | 2 | | 26 | | 14 |
| **1490** | | 8 | | | | 3 | Puzur-Aššur III | 1 | | 15 |
| **1489** | | 9 | | | | 4 | | 2 | | 16 |
| **1488** | | 10 | | | | 5 | | 3 | | 17 |
| **1487** | | 11 | | | | 6 | | 4 | | 18 |
| **1486** | | 12 | | | | 7 | | 5 | | 19 |
| **1485** | | 13 | | | | 8 | | 6 | | 20 |
| **1484** | | 14 | | | | 9 | | 7 | | 21 |
| **1483** | | 15 | | | | 10 | | 8 | Thutmose I | 1 |
| **1482** | | 16 | | | | 11 | | 9 | | 2 |
| **1481** | | 17 | | | | 12 | | 10 | | 3 |
| **1480** | | 18 | | | | 13 | | 11 | | 4 |
| **1479** | | 19 | | | | 14 | | 12 | | 5 |
| **1478** | | 20 | | | | 15 | | 13 | | 6 |
| **1477** | | 21 | | | | 16 | | 14 | | 7 |
| **1476** | | 22 | | | | 17 | | 15 | | 8 |
| **1475** | | 23 | | | | 18 | | 16 | | 9 |
| **1474** | | 24 | | | | 19 | | 17 | | 10 |
| **1473** | | 25 | | | | 20 | | 18 | | 11 |
| **1472** | | 26 | | | | 21 | | 19 | | 12 |
| **1471** | | 27 | | | | 22 | | 20 | Thutmose II | **1** |
| **1470** | | 28 | | | | 23 | | 21 | | 2 |
| **1469** | | 29 | | | | 24 | | 22 | Hatshepsut | 3 |
| **1468** | | 30 | | | | 25 | | 23 | *[Thoutmosis III]* | [4] |
| **1467** | | 31 | | | | 26 | | 24 | | [5] |
| **1466** | | 32 | | | | 27 | Enlil-nâṣir I | 1 | | [6] |
| **1465** | | 33 | | | | 28 | | 2 | | [7] |
| **1464** | | 34 | | | | 29 | | 3 | | [8] |
| **1463** | | 35 | | | | 30 | | 4 | | [9] |
| **1462** | | 36 | | | (*without judge*) | 1 | | 5 | | [10] |
| **1461** | | 37 | | | | 2 | | 6 | | [11] |
| **1460** | | **38** | | | | 3 | | 7 | | [12] |
| **1459** | | 39 | | | | 4 | | 8 | | [13] |
| **1458** | | 40 | | | | 5 | | 9 | | [14] |
| **1457** | | 41 | | | | 6 | | 10 | | [15] |
| | | | | | | | | 11 | | |
| **1456** | Ulam-Buriaš | *1* | | | | 7 | | 12 | | [16] |
| **1455** | | *2* | | | | 8 | | 13 | | [17] |
| **1454** | | *3* | | | | 9 | Nûr-ili | 1 | | [18] |
| **1453** | | *4* | | | | 10 | | 2 | | [19] |
| **1452** | | *5* | | | | 11 | | 3 | | [20] |
| **1451** | | *6* | | | Kušan-rišataïm | 1 | | 4 | | [21] |
| **1450** | | *7* | | | (*Mitanni*) | 2 | | 5 | | [22] |
| **1449** | | *8* | | | | 3 | | 6 | Thutmose III | 23 |
| **1448** | | *9* | | | | 4 | | 7 | | 24 |
| **1447** | | *10* | | | | 5 | | 8 | | 25 |
| **1446** | | *11* | | | | 6 | | 9 | | 26 |
| **1445** | | *12* | | | | 7 | | 10 | | 27 |
| **1444** | | *13* | | | | 8 | | 11 | | 28 |
| **1443** | | *14* | | | Othniel | 1 | Aššur-šadûni | 12 | | 29 |

The death of Elamite king Chedorlaomer in year 36 of his reign, at the end of the period called Awan, coinciding with the death of Babylonian king Shulgi in the year 48 of his reign, can be dated with some accuracy. The biblical chronology according to the Masoretic text is as follows (King Hoshea died at the fall of Samaria in 720 BCE, King Josias died at the battle of Haran in 609 BCE):

| Event | Period | # | | | | Reference |
|---|---|---|---|---|---|---|
| **Abraham** in Ur | 2038-**1963** | 75 | From birth to departure into Canaan | | | Genesis 12:4-5 |
| Israelites as foreigners | 1963-1533 | 430 | From Canaan stay to Egypt deliverance | | | Exodus 12:40-41 |
| Exodus in Sinai | 1533-1493 | 40 | From Egypt deliverance to entering Canaan | | | Exodus 16:35 |
| Israelites in Canaan | 1493-**1013** | 480 | From entering Canaan to year 4 of Solomon | | | 1Kings 6:1 |
| King of Judah | Reign | | King of Israel | Reign | | |
| Solomon | **1017** - 977 | 40 | | | | 1Kings 11:42 |
| Rehoboam | 977-960 | 17 | Jeroboam I | 10/977 - -05/955 | 22 | 1Kings 14:20-21 |
| Abiyam | 960-957 | 3 | | | | |
| Asa | 957 - | 41 | Nadab | 06/955-05/954 | 2 | 1Kings 15:10,25 |
| | | | Baasha | 06/954-04/931 | 24 | 1Kings 15:28,33 |
| | | | Elah | 05/931-04/930 | 2 | 1Kings 16:8 |
| | | | Zimri | 05/930 | *7 d.* | 1Kings 16:10-16 |
| | | | Omri/ | 06/930-05/919/ | 12 | 1Kings 16:21-23 |
| | -916 | | [Tibni] | [06/930-01/925] | 6 | |
| Jehoshaphat | 916 - | 25 | Ahab | 06/919-01/898 | 22 | 1Kings 16:29 |
| | -891 | | Ahaziah | 02/898-01/897 | 2 | 1Kings 22:51 |
| Jehosaphat/Jehoram | [893-891] | [2] | Jehoram Ahab's son | 02/897-09/886 | 12 | 2Kings 3:1 |
| Jehoram | 893 - | 8 | [Ahaziah]/ Joram | [07/887-09/886] | 1 | 2Kings 9:29 |
| | -885 | | Ahaziah | 10/886-09/885 | 1 | 2Kings 9:24,27 |
| [Athaliah] Jehoyada | 885-879 | 6 | Jehu | 10/885-03/856 | 28 | 2Kings 10:36 |
| Joash | 879 - | 40 | Jehoahaz | 04/856-09/839 | 17 | 2Kings 10:35; 13:1 |
| | -839 | | Jehoahaz/ Jehoash | [01/841-09/839] | 2 | 2Kings 13:10 |
| Amasiah | 839 - | 29 | Jehoash | 09/839-01/823 | 16 | 2Kings 13:10 |
| | -810 | | Jeroboam II | 01/823-05/782 | 41 | 2Kings 14:23 |
| Uzziah | 810 - | 52 | [Zechariah] | 06/782-02/771 | [11] | 2Kings 14:29 |
| [Azariah] | [796 - | | Zechariah | 03/771-08/771 | *6 m.* | 2Kings 15:8 |
| | | | Shallum | 09/771 | *1 m.* | 2Kings 15:13 |
| | | | Menahem | 10/771-03/760 | 10 | 2Kings 15:17 |
| | -758 | | Peqayah | 04/760-03/758 | 2 | 2Kings 15:23 |
| Jotham | 758-742 | 16 | Peqah | 04/758-05/738 | 20 | 2Kings 15:27 |
| Ahaz | 742-726 | 16 | [Hoshea] | 06/738-01/729 | 9 | 2Kings 15:27-30 |
| Hezekiah | 726-697 | 29 | **Hoshea** | 02/729-09/**720** | 9 | 2Kings 17:1,3 |
| Manasseh | 697-642 | 55 | | | | 2Kings 21:1 |
| Amon | 642-640 | 2 | | | | 2Kings 21:19 |
| **Josias** | 640-**609** | 31 | | | | 2Kings 22:1 |
| Jehoachaz | -609 | *3 m.* | | | | 2Chronicles 36:2 |
| Jehoiaqim | 609-598 | 11 | | | | 2Chronicles 36:5 |
| Jehoiachin | -598 | *3 m.* | | | | 2Chronicles 36:9 |
| Zedekiah | 598-587 | 11 | | | | 2Chronicles 36:11 |
| Jehoiachin (exile) | 587-561 | 26 | | | | 2Kings 25:27-28 |

According to the text of Genesis 14:1-17, a coalition of Mesopotamian kings would raided the Negev: *[For] 12 years they [the kings of Sodom and Gomorrah] served Chedorlaomer, and in the 13th year they rebelled. In the 14th year came Chedorlaomer and the kings that were with him (...) when Abram heard that his brother was taken captive, he armed his trained servants, born in his own house, 318, and pursued them unto Dan. And he divided himself against them, he and his servants, by night, and smote them, and pursued them unto Hobah, which is on the left hand of Damascus. And he brought back all the goods, and also brought again his brother Lot, and his goods, and the women also, and the people. And the king of Sodom went out to meet him [Abraham] after his return from the slaughter of Chedorlaomer, and of the kings that were with him, at the valley of Shaveh, which is the king's dale.* Although the kings of the biblical account have not been discovered, onomastics confirms the existence of such a name in the early second millennium BCE[105]. Kedor-Lagomer (the Hebrew 'ayn' coming from a former 'g') corresponds to the Akkadian Kudur-Lagarma which comes from the Elamite Kutir-Lagamar "bearer (servant) of Lagamar[106]" (Lagamar being an Elamite deity[107]). According to the biblical text, a powerful king of Elam would be associated with three other kings, two of Sumer and one of Gutium, to make reprisals against several wealthy cities of Transjordan, the latter ones having refused to pay their tribute[108]. The operation is described as a simple police operation, not as a war of conquest. Likewise, retaliation against Abraham's Mesopotamian kings is not presented as a war but as a night ambush to shoot down enemies (Genesis 14:15-17). The text speaks of "slaughter[109]" and not of conquest, which involves the death of all these kings.

| | [A] | [B] | [C] | [D] | |
|---|---|---|---|---|---|
| **1967** | 1 | 35 | 71 | | [A] **Kudur-Lagamar** king of Elam |
| **1966** | 2 | 36 | 72 | | [B] **Šulgi** king of Ur |
| **1965** | 3 | 37 | 73 | | |
| **1964** | 4 | 38 | 74 | | |
| **1963** | 5 | 39 | 75 | | [C] **Abraham** *arrived in Canaan from Ur at age 75 (Genesis 12:4-5)* |
| **1962** | 6 | 40 | 76 | 1 | |
| **1961** | 7 | 41 | 77 | 2 | |
| **1960** | 8 | 42 | 78 | 3 | |
| **1959** | 9 | 43 | 79 | 4 | |
| **1958** | 10 | 44 | 80 | 5 | |
| **1957** | 11 | 45 | 81 | 6 | |
| **1956** | 12 | 46 | 82 | 7 | |
| **1955** | 13 | 47 | 83 | 8 | [A] *Revolt of Transjordan kings against* **Kudur-Lagamar** *(Genesis 14:4)* |
| **1954** | 14 | 48 | 84 | 9 | [A] **Kudur-Lagamar** *killed by* **Abraham** *(Genesis 14:5-17)* |
| **1953** | | 1 | 85 | 10 | [D] *10 years later (Genesis 16:4)* |
| **1952** | | 2 | 86 | | [B] **Amar-Sîn** king of Ur |

According to this reconstruction, Kudur-Lagamar would have intervened in Transjordan in the last years of the reign of Šulgi. In addition, Abraham left his hometown of Ur (in -1963) 51 years prior this City State collapsed (in -1912) due to Amorite and Elamite invasions[110]. According to the Middle Chronology, he would have left this city 41 years after its destruction (in 2004 BCE)!

105 R. DE VAUX - Histoire ancienne d'Israël des origines à l'installation en Canaan
Paris 1986 Éd. Gabalda pp. 208-212.
106 R. ZADOK - The Elamite Onomasticon
in: *Annali Supplementi* 44:1 (1984) Ed. Istituto orientale di Napoli pp. 24-26.
107 M.J. STEVE - Mémoires de la délégation archéologique en Iran, tome III
Nice 1987 Éd. Serre p. 30.
108 L. PIROT, A. CLAMER – La Sainte Bible Tome 1, 1ère partie
Paris 1953 Éd. Letouzey et Ané pp. 247-256.
109 The Hebrew verbal form *hakot* means "to strike" (1Samuel 17:57, 18:6, 2Chronicles 25:14) or "kill" (Genesis 4:15, 8:21).
110 P. GARELLI, J.M. DURAND, H. GONNET, C. BRENIQUET - Le Proche-Orient Asiatique
Paris 1997 Éd. P.U.F. pp. 73-95.
G. ROUX - La Mésopotamie
Paris 1995 Éd. Seuil pp. 207-220.

Elamite chronology[111] of this period can be fixed through numerous synchronisms with Mesopotamian chronology. Some anomalies arise: 1) there is a gap of at least 60 years between Puzur-Inšušinak the king of Awan and Girname the king of Simaški, it seems to miss at least three kings at the end of the Awan dynasty (where should have appeared Kutir-Lagamar); 2) the first kings of Simaški also seem to miss the call because Puzur-Inšušinak mentions at least one in his inscriptions, again, he declares himself governor of Susa and viceroy of the country of Elam or sometimes king of Awan[112]; 3) the first three kings of the dynasty of Simaški were contemporaries, they probably ruled over the different capitals of Elam (Susa, Anšan, Simaški); 4) information on Elam, almost exclusively from the kings of Sumer, is incomplete, biased and difficult to interpret because the reversals of alliances between coalitions of kings, even in very remote areas were apparently common during the period 2000-1700[113]. Mesopotamian kings had indeed used to intervene west until the Mediterranean. Sargon of Akkad and Naram-Sîn had marched to Taurus Mountains, Elam sent its armies into Syria (to Qatna), Mari went regularly to at Hazor. In fact, King Gilgamesh (2450-2400) had already reached the Mount Hermon (*Saria*) in Lebanon from his city of Uruk, flying in just 3 days (sic), a journey of 1620 kilometers usually traveled in a month and a half (the Epic of Gilgamesh IV:12).

About twenty Economic tablets, dated year 47 of Šulgi (in -1955), mention an incident and loot taken by the Elamites[114], indirectly confirming a raid or a tribute, but the region of origin is not mentioned. The date of death of Šulgi is known precisely, on 1/XI/48 (around February -1954) it happened in a dramatic atmosphere, because there would have been violent death of a part of the royal family[115], but we do not know why.

Thus, the death of Elamite king Chedorlaomer at the end of the Awan period, coinciding with the death of Babylonian king Shulgi in the year 48 of his reign agrees with the fall of Ur dated 1912 BCE. In addition, the fall of Babylon dated 1499 BCE agrees also better with the Elamite chronology, because as F. Vallat noticed: *different texts show that five generations of the same family have occupied the space between the reign of Kutir-Nahhunte to that of Kuk-Našur III, the last Sukkalmah. By assigning thirty years to each generation, the interval between the two kings is about one century and a half. As two sovereigns rules as sukkalmahat between Kutir-Nahhunte and Kuk-Nasur II (Temti-Agun and Kutir-Silhaha), we can estimate that one century [not two] separates the reign of Kuk-Našur II from that of Kuk-Našur III.*

To conclude, the set of Assyrian reigns (which are without intercalation before Aššur-Dan I), combined with the construction length between temples, enable us to date precisely the death of Šamšî-Adad I in 1680 BCE, which fixes the reign of Hammurabi (1697-1654) and therefore those of Ibbi-Sîn (1936-1912) and Ammisaduqa (1551-1530). The lunar eclipse at the end of Ibbi-Sîn's reign and at the end of Shulgi's reign, the risings and settings of Venus dated according to a lunar calendar during Ammisaduqa's reign, allow to obtain absolute astronomical dating that anchoring Mesopotamian chronology (synchronisms are highlighted, astronomical dating are highlightedin blue sky, underlined dates are adjusted from one year to take account of the absence of intercalation, framed dates represent exact values deduced from synchronisms and dates in italic represent the average values deduced from synchronisms):

---

111 F. VALLAT, H.GASCHE - Suse
in: *Supplément au dictionnaire de la Bible.* Fascicule 73 (2002) pp. 374-391.
112 E. SOLLBERGER, J.-R. KUPPER – Inscriptions royales sumériennes et akkadiennes
in: *Littératures Anciennes du Proche-Orient* n°3, Cerf, 1971, pp. 124-127.
113 K.A. KITCHEN - On the Reliability of the Old Testament
Cambridge 2003 Ed. W.B. Eerdmans pp. 319-324.
114 F. VALLAT, H.GASCHE - Suse
in: *Supplément au dictionnaire de la Bible.* Fascicule 73 (2002) pp. 374-391, 433-434.
115 F. JOANNES - Dictionnaire de la civilisation mésopotamienne
Paris 2001 Éd. Robert Laffont pp. 68,69,822-824.

| ASSYRIA | Reign | | BABYLONIA | Reign | | ELAM (AWAN) | Reign | |
|---|---|---|---|---|---|---|---|---|
| | | | Ur-Nammu | 2020-2002 | 18 | [-]-lu | *2015-1990* | 25 |
| | | | Šulgi | 2002-**1954** | 48 | Chedorlaomer | 1990-**1954** | 36 |
| | | | Amar-Sîn | 1954-1945 | 9 | | | |
| | | | Šu-Sîn | 1945-1936 | 9 | ISIN | Reign | |
| | | | Ibbi-Sîn | 1936-**1912** | 24 | Išbi-Erra | 1923 - | 33 |
| Puzur-Aššur I | **1913**-1900 | *14* | *Collapse of Ur* | | | | | |
| Šalim-ahum | 1900-1886 | *14* | | | | | -1890 | |
| Ilu-šumma | 1886-1873 | *14* | | | | Šû-ilišu | 1890-1880 | 10 |
| Êrišu I | 1873 - | 40 | | | | Iddin-Dagân | 1880-1859 | 21 |
| | -1834 | | | | | Išme-Dagân | 1859-1839 | 20 |
| Ikunum | 1834 - | 14 | | | | Lipit-Eštar | 1839-1828 | 11 |
| | -1821 | | | | | Ur-Ninurta | 1828-1800 | 28 |
| Sargon I | 1821-1782 | 40 | Sûmû-abum | 1799-1785 | 14 | Bûr-Sîn | 1800-1779 | 21 |
| Puzur-Aššur II | 1782-1774 | 8 | Sûmû-la-Il | 1785 - | 36 | Lipit-Enlil | 1779-1774 | 5 |
| Naram-Sîn | 1774 - | 54 | | | | Erra-imittî | 1774-1767 | 7 |
| | | | | -1749 | | Enlil-Bâni | 1767-1743 | 24 |
| | | | Sâbium | 1749 - | 14 | Zambîya | 1743-1740 | [3] |
| | | | | | | Iter-piša | 1740-1736 | [4] |
| | -1722 | | | -1735 | | Ur-dukuga | 1736-1732 | [4] |
| Êrišu II | 1722-1712 | 10 | Apil-Sîn | 1735-1717 | 18 | Sîn-mâgir | 1732-1721 | 11 |
| Šamšî-Adad I | 1712 - | 33 | Sîn-muballiṭ | 1717-1697 | 20 | Damiq-ilîšu | 1721-**1698** | 23 |
| | **-1680** | | Hammurabi | **1697-1680** | 17 | *Isin annexed* | | |
| Išme-Dagan I | 1680-1670 | 11 | | 1680 - | 26 | | | |
| Aššur-dugul | 1670-1664 | 6 | | | | | | |
| Aššur-apla-idi | 1664 | 0 | | | | | | |
| Nâṣir-Sîn | 1664 | 0 | | | | | | |
| Sîn-namir | 1664 | 0 | | | | | | |
| Ipqi-Ištar | 1664 | 0 | | | | | | |
| Adad-ṣalûlu | 1664 | 0 | | | | | | |
| Adasi | 1664 | 0 | | | | | | |
| Bêlu-bâni | 1664-1654 | 10 | | -1654 | | | | |
| Libbaya | 1654 - | 17 | Samsu-iluna | 1654 - | 38 | ELAM | Reign | |
| | -1638 | | | | | Kutir-Nahhunte I | *1645 -* | *25* |
| Šarma-Adad I | 1638-1626 | 12 | | | | | | |
| Puzur-Sîn | 1626-1615 | 12 | | -1616 | | | *-1620* | |
| Bazaya | 1615-1588 | 28 | Abi-ešuḫ | 1616-1588 | 28 | Temti-Agun II | *1620-1595* | *25* |
| Lullaya | 1588-1582 | 6 | Ammiditana | 1588 - | 37 | Kutir-Silhaha | *1595-1570* | *25* |
| Šû-Ninûa | 1582-1568 | 14 | | | | | | |
| Šarma-Adad II | 1568-1565 | 3 | | | | | | |
| Êrišu III | 1565-1553 | 13 | | -1551 | | Kuk-Našur II | *1570 -* | *25* |
| Šamšî-Adad II | 1553-1547 | 6 | Ammiṣaduqa | **1551 -** | 21 | | *-1545* | |
| Išme-Dagan II | 1547-1531 | 16 | | **-1530** | | Kudu-zuluš II | *1545-1525* | *20* |
| Šamšî-Adad III | 1531-1516 | 16 | Samsuditana | 1530 - | 31 | Tan-Uli | *1525-1505* | *20* |
| Aššur-nêrârî I | 1516-1491 | 26 | *Fall of Babylon* | **-1499** | | Temti-halki | *1505 -* | *20* |
| Puzur-Aššur III | 1491 - | 24 | Agum II | *1503-1487* | *16* | | *-1485* | |
| | -1467 | | Burna-Buriaš I | *1487-1471* | *16* | Kuk-Našur III | *1485-1465* | *20* |
| Enlil-nâṣir I | 1467-1455 | 13 | Kaštiliaš III | *1471-1455* | *16* | Kidinu | *1465-1450* | *15* |
| Nûr-ili | 1455-1443 | 12 | Ulam-Buriaš | *1455 -* | *16* | Inšušinak-sunkir- | *1450 -* | *10* |
| Aššur-šadûni | 1443-1443 | 0 | | *-1439* | | nappipir | *-1440* | |
| Aššur-rabi I | 1443-1433 | [10] | Agum III | *1439 -* | *16* | Tan-Ruhuratir II | *1440-1435* | *5* |
| Aššur-nâdin-aḫḫe I | 1433-1424 | [10] | | *-1423* | | Šalla | *1435-1425* | *10* |
| Enlil-naṣir II | 1424-1418 | 6 | Kadašman-Harbe I | *1423 -* | *16* | Tepti-ahar | 1425 - | *20* |
| Aššur-nêrârî II | 1418-1411 | 7 | | *-1407* | | | | |
| Aššur-bêl-nišešu | 1411-1403 | 9 | Kara-indaš | *1407 -* | *16* | | -1405 | |
| Aššur-rê'im-nišešu | 1403-1395 | 8 | | *-1391* | | Igi-halki | 1405 - | *20* |
| Aššur-nâdin-aḫḫe II | 1395-1385 | 10 | Kurigalzu I | *1391 -* | *16* | | -1385 | |
| Erîba-Adad I | 1385 - | 27 | | *-1375* | | Pahir-iššan | 1385-1375 | *10* |
| | -1358 | | Kadašman-Enlil I | 1375-1360 | 15 | Attar-Kittah | 1375-1365 | *10* |

Most experts of Hittite history[116] reject this chronology, because they considered it too short of a century against theirs (which does not based on any synchronism dated by astronomy in the period prior -1350!). Synchronisms according to Freu[117] are highlighted and corrected reigns are reconstituted from an average (#) smaller for Hittite reigns:

| n° | HITTITE | (Freu) | # | Reign | # | MITANNIAN | Reign | BABYLONIAN | Reign |
|---|---|---|---|---|---|---|---|---|---|
| 1 | [Ḫuzziya I ?] | *1670-1650* | *20* | *1605-1585* | *15* | | | | |
| | [Tudḫaliya ?] | | | *1585-1565* | *20* | | | | |
| | [PU-Šarruma ?] | | | *1565-1550* | *15* | | | | |
| 2 | Labarna | *1650-1625* | *25* | *1550-1530* | *20* | | | | |
| 3 | Ḫattušili I | *1625-1600* | *25* | *1530-1510* | *20* | | | | |
| | average period 1 | | **23** | | **18** | | | | |
| 4 | Muršili I | *1600-1585* | *15* | *1510-1500* | *10* | | | Samsuditana | 1530-**1499** |
| 5 | Ḫantili I | *1585-1570* | *15* | *1500-1495* | *5* | Kirta | *1500 -* | | |
| 6 | Zidanta I | *1570* | *0* | *1495* | *0* | | | KASSITE | Reign |
| 7 | Ammuna | *1570-1550* | *20* | *1495-1485* | *10* | | | Agum II | *1503-1487* |
| 8 | Ḫuzziya II | *1550* | *0* | *1485* | *0* | | *-1485* | | |
| 9 | Telipinu | *1550-1530* | *20* | *1485-1480* | *5* | Šutarna I | *1485-1480* | Burna-Buriaš I | *1487 -* |
| 10 | Alluwamna | *1530-1515* | *10* | *1480-1475* | *5* | Barattarna I | *1480 -* | | |
| 11 | Ḫantili II | *1515-1505* | *10* | *1475-1470* | *5* | | | | *-1471* |
| 12 | Taḫurwaili I | *1505-1500* | *5* | *1470* | *0* | | | Kaštiliaš III | *1471 -* |
| 13 | Zidanza (II) | *1500-1485* | *15* | *1470-1465* | *5* | | | | |
| 14 | Ḫuzziya III | *1485-1470* | *15* | *1465-1460* | *5* | | | | |
| | average period 2 | | **11** | | **5** | | | | |
| 15 | Muwatalli I | *1470-1465* | *5* | *1460-1455* | *5* | | *-1455* | | *-1455* |
| 16 | Tutḫaliya I | *1465-1440* | *25* | *1455-1435* | *20* | Šauštatar I | *1455-1435* | | |
| 17 | Ḫattušili II | *1440-1425* | *15* | *1435-1425* | *10* | Paršatatar | *1435-1425* | | |
| 18 | Tutḫaliya II | *1425-1390* | *35* | *1425-1395* | *30* | Šauštatar II | *1425-1395* | | |
| 19 | Arnuwanda I | *1400 -* | *30* | *1395 -* | *25* | Barattarna II | *1395-1390* | | |
| | | *-1370* | | *-1370* | | Artatama I | *1390-1373* | | |
| 20 | Tutḫaliya III | *1370-1350* | *20* | *1370-1355* | *15* | Šutarna II | *1373-1355* | | |
| | average period 3 | | **23** | | **18** | | (**17**) | | |
| 21 | Šuppiluliuma I | 1353-1322 | 31 | 1353-1322 | 31 | | | | |
| 22 | Arnuwanda II | 1322 | <1 | 1322 | <1 | | | | |
| 23 | Muršili II | 1322-1295 | 27 | 1322-1295 | 27 | | | | |
| 24 | Muwatalli II | 1295-1275 | 20 | 1295-1275 | 20 | | | | |
| 25 | Urhi-Teshub | 1275-1268 | 7 | 1275-1268 | 7 | | | | |
| 26 | Ḫattušili III | 1268-1241 | 27 | 1268-1241 | 27 | | | | |
| 27 | Tutḫaliya IV | 1241-1209 | 32 | 1241-1209 | 32 | | | | |
| 28 | Arnuwanda III | 1209-1207 | 2 | 1209-1207 | 2 | | | | |
| 29 | Šuppiluliyama II | 1207-1185 | 22 | 1207-1185 | 22 | | | | |
| | average period 4 | | **18** | | **18** | | | | |

Early Empire (n° 1 to 9); Middle Empire (n° 10 to 20); Late Empire (n° 21 to 29)

This Hittite chronology contains four periods. The first and oldest period consists of three kings, whose reign would average 23 years, this period does not include any synchronism. The second, which begins with the fall of Babylon, consists of 11 kings whose average reign would be 11 years. The third comprises six kings whose average reign would be 23 years and the fourth includes nine kings whose average reign would be 18 years. However, the average length of these reigns is arbitrary, except for the last period which can be dated precisely thanks to synchronisms from El Amarna letters (anchored by the total solar eclipse dated year 10 of Muršili II). Instead of the average duration of 23 years for the third period, it is more logical to maintain the 18 years since the third period precedes and looks like the fourth. It is therefore necessary lowering 5 years (= 23 - 18) the reigns given by Freu for that period. The second period was very disturbed by numerous assassinations of kings, which has arguably led Freu to choose a lower average (11 years instead of 23 for the next period). It is therefore necessary lowering 12 (= 23 - 11) years the reigns given by Freu for the third period. Finally, Freu has once again selected a 23-year average for the first period. It seems more logical to maintain the average duration of 18 years from the fourth period which was relatively stable. The reconstruction based on the corrected duration of Hittite reigns provides a chronology that agrees with the fall of Babylon in 1499 BCE. There is no precise synchronism with Egyptian chronology that can anchor Kassite and Mitannian chronologies. The kingdom of Mitanni appears for the first time on the Theban stele of the astronomer Amenemhat, which mentions the name as "land of Meten" (*ḫ3st Mṯn*), indicating that it was an enemy against which the pharaoh Thoumosis I had launched an expedition in the year 4 of his reign (in -1481). Now the king of Mitanni during this period is Šutarna I (1485-1480). It is possible that the disappearance of the Babylonian kingdom have favored the rise of the Mitanni[118] and the Pharaoh had wanted to stop a possible westward extension of this new kingdom founded by Kirta.

[116] http://www.hittites.info/history.aspx?text=history%2fEarly+Empire.htm
[117] J. FREU – Note sur les sceaux des rois de Mitanni/Mittani
in: *NABU* (mars 2008) pp. 5-8.

The triple synchronism between kings Agum II (Kassite), Kirta (Mitannian) and Ammuna (Hittite) requires setting reign of those kings over a period covering the reign of Agum II (1503-1487). The corrected duration of the reign of Muršili I (1510-1500 instead of 1600-1585), the Hittite king who overthrew the city of Babylon, is consistent with the date of 1499 BCE. The $^{14}C$ dating of strata corresponding to the period of the Old Hittite Empire (1565-1510) gives well 1600-1500[119] instead of 1670-1530 proposed by Freu, who also refuses dating the fall of Babylon in 1499 BCE because there would have been "too many" kings of Hana[120] during the period 1600-1500 BCE called «dark ages».

Hana, which means "Bedouins[121]", was a confederation of Syrian cities (at least 6) located in the region north of Mari, however many political events in this area remain sketchy[122]. Moreover its kings did not belong to a classic dynasty but it was generally used as a honorific title, as Yahdun-Lim (1716-1700) who was "King of Mari, Tutul and the land of Hana". The chronology of these kings[123] is difficult to obtain because there are few synchronisms, in addition, the sequence of these kings is uncertain:

- Zimri-Lim (1680-1667) was "King of Mari and the land of Hana".
- Yadiḫ-Abu I, overseer of Hana (*ugula Ḫana*), had fought Samsuiluna, a Babylonian king, in the latter's 27th year (1627 BCE). Afterwards the kingdom of Hana is likely under the influence of Kassites since the following king of Terqa was Kaštiliaš I (1613-1591).
- From Ammiditana (1588-1551) the kingdom of Hana was in fact headed by Babylonian kings up to Samsuditana (1530-1499). Kings of Terqa were probably vassals of Babylon.
- The Hurrians were enemies of the Hittite kings Ḫattušili I (1530-1510) and Muršili I (1510-1500), and their strengh is shown by records of their conquest of much of the Hittite kingdom in the time of Ḫattušili I who seems to have retaliated late in his career, attacking Aleppo (Halab). However, Kuwari, a king of Hana, managed to defeat an attack led by the warriors of Hatti (Ḫatte). Conceivably[124], the Hittite expedition of

118 According to the Israelite chronology (Judges 3:8-15), there was also a Mitannian domination over Syria by a king called Kushan-rishataïm (1452-1444), who was likely Šauštatar I (1455-1435).
119 R.L. GORNY – Çadir Höyük
in: 2006-2007 Annual Report, The Oriental Institute pp. 18-33.
120 J. FREU -Des origines à la fin de l'Ancien royaume hittite
Paris 2007, Éd. L'Harmattan pp. 111-117.
121 D. CHARPIN – Le «pays de Mari et des Bédouins» à l'époque de Samsu-iluna de Babylone
in: *Revue d'Assyriologie et d'archéologie orientale* volume CV (2011) pp. 41-59.
122 Some eras are well understood, but many others remain almost unknown.
123 A.H. PODANY –The Land of Hana: Kings, Chronology, Scribal Tradition
2002 Ed. CDL Press pp. 1-74.
124 T. BRYCE – The Kingdom of the Hittites
Oxford 2005 Ed. Oxford University Press pp. 99-100.

Muršili I in arose from an alliance between the Hittites and the Kassites, the incentive for the Hittites being the rich spoils of Babylon, and for the Kassites the prospect of creating a new ruling dynasty in Babylonia.

- Qiš-Addu, a king of Bidah?, was a vassal of both Barattarna I and Šauštatar I. The kingdom of Hana (Terqa) became independent afterwards. Although several Hanean reigns are controversial its chronology is as follows[125] (synchronisms highlighted):

| HANEAN | Reign | # | KASSITE | Reign | # | BABYLONIAN | Reign | # |
|---|---|---|---|---|---|---|---|---|
| Yahdun-Lim | 1716-1700 | 16 | | | | Sîn-muballiṭ | 1717-1697 | 20 |
| (kings of Mari) | 1700-1680 | 20 | | | | Hammurabi | **1697** - | 43 |
| Zimri-Lim | 1680-1667 | 13 | | | | | | |
| Yâpaḫ-Šumu-Abu | *1667-1654* | *[13]* | | | | | -1654 | |
| Iṣi-Šumu-Abu | *1654-1641* | *[13]* | Gandaš | *1661* - | [2]6 | Samsu-iluna | 1654-1645 | 38 |
| Yadiḫ-Abu I | *1641-1627* | *[14]* | | *-1635* | | | 1945-1927 | |
| [Muti-Huršana ?] | *1627-1613* | *[14]* | Agum I | *1635-1613* | 22 | | 1627-1616 | |
| Kaštiliaš | *1613-1591* | 22 | Kaštiliaš I | *1613-1591* | 22 | Abi-ešuḫ | 1616-1588 | 28 |
| Šunuḫru-Ammu | *1591-1575* | *[16]* | Ušši | *1591-1583* | 8 | Ammiditana | 1588 - | 37 |
| Ammi-madar | *1575* - | *[16]* | Abirattaš | *1583-1567* | *[16]* | | | |
| | *-1559* | | Kaštiliaš II | *1567-1551* | *[16]* | | -1551 | |
| Yadiḫ-Abu II | *1559-1543* | *[16]* | Urzigurumaš | *1551-1535* | *[16]* | Ammiṣaduqa | 1551 - | 21 |
| Zimri-Lim II | *1543-1527* | *[16]* | Harbašihu | *1535* - | *[16]* | | -1530 | |
| Kasap-ilî | *1527-1511* | *[16]* | | *-1519* | | Samsuditana | 1530 - | 31 |
| Kuwari | *1511-1495* | *[16]* | Tiptakzi | *1519-1503* | *[16]* | | **-1499** | |
| Ya'usa / Hanaya | *1495* - | *[15]* | Agum II | *1503* - | *[16]* | MITANNIAN | Reign | # |
| | *-1480* | | | *-1487* | | Kirta | *1500-1485* | *15* |
| Qiš-Addu | *1480* - | *[25]* | Burna-Buriaš I | *1487* - | *[16]* | Šutarna I | *1485-1475* | *10* |
| | | | | *-1471* | | Barattarna I | *1475* - | *20* |
| | *-1455* | | Kaštiliaš III | *1471-1455* | *[16]* | | *-1455* | |
| Iddin-Kakka | *1455-1435* | *[20]* | Ulam-Buriaš | *1455-1439* | *[16]* | Šauštatar I | *1455-1435* | *20* |
| Išar-Lim | *1435-1415* | *[20]* | Agum III | *1439-1423* | *[16]* | Paršatatar | *1435-1425* | *10* |
| Iggid-Lim | *1415-1395* | *[20]* | Kadašman-Harbe I | *1423-1407* | *[16]* | Šauštatar II | *1425-1395* | *30* |
| Išiḫ-Dagan | *1395* - | *[20]* | Kara-indaš | *1407-1391* | *[16]* | Barattarna II | *1395-1390* | *5* |
| | *-1375* | | Kurigalzu I | *1391-1375* | *[16]* | Artatama I | *1390-1373* | *17* |
| Ahuni | *1375-1355* | *[20]* | Kadašman-Enlil I | **1375**-1360 | 15 | Šutarna II | *1373-1355* | *18* |
| Hammurapi | *1355-1335* | *[20]* | Burna-Buriaš II | 1360-1333 | 27 | Tušratta | *1353-1339* | *14* |
| Pagiru | *1335-1315* | *[20]* | Kurigalzu II | 1333-1308 | 25 | Artatama II | *1339-1325* | *14* |

The numerous synchronisms during the Late Empire confirm the chronology of Assyrian reigns without intercalation through Egyptian chronology:

- The death of Hattušili III is dated in year 42 of Ramses II[126].
- The Nihriya battle involved the Hittite king Tudhaliya IV, Hattušili III successor, and the Assyrian king Tukulti-Ninurta I in the first two years of his reign[127].
- The reign of Hattušili III[128] is located within Shalmaneser I's reign. Hattušili III died shortly before Shalmaneser I, and the successor of Hattušili, Tuthaliya IV, has been at war with the successor of Shalmaneser I, Tukulti-Ninurta I, in the first two years of the latter, which gives: Year 42 of Ramses II = death of Hattušili III = death of Shalmaneser I +/- 1 year. Thus the accession of Tukulti-Ninurta I (year 0) matches the year 42 of Ramses II. Tuthaliya IV began to rule from this year, but it is possible that his father (Hattušili III), feeling old and sick, associated him to kingship as crown prince

[125] S. YAMADA – An Adoption Contract from Tell Taban, the Kings of the Land of Hana, and the Hana-style Scribal Tradition in: *Revue d'Assyriologie et d'archéologie orientale* volume CV (2011) pp. 61-84.
[126] C. DESROCHES NOBLECOURT – Ramsès II la véritable histoire Paris 1996 Ed. Pygmalion p. 365.
[127] T. BRYCE – The Kingdom of the Hittites (tablette KBo IV 14). Oxford 2005 Ed. Oxford University Press p. XV, 375-382.
[128] G. BECKMAN – Hittite Chronology in: *Akkadica* 119-120 (2000) p. 24.

(appearing under the name of Hišmi-šarruma). Two synchronisms are therefore particularly important to examine in order to verify the chronology:

1) accession (year 0) of Kadašman-Enlil II (in -1264) = year 19 of Ramses II.
2) accession (year 0) of Tukulti-Ninurta I$^{er}$ (in -1242) = year 42 of Ramses II.

The first synchronism "year 0 of Kadašman Enlil II = year 19 of Ramses II" derives from the sequence of the following events[129]:

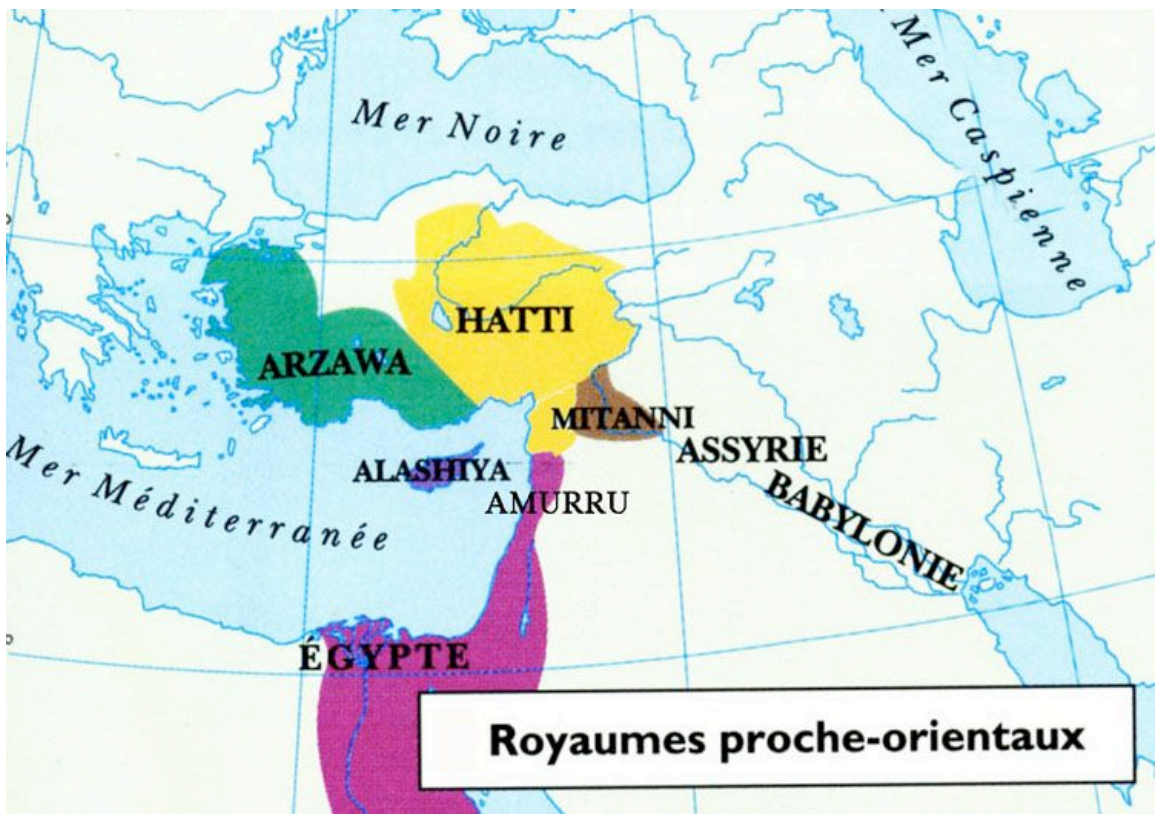

1. To expand his empire, the Hittite king Šuppiluliuma I engaged [in -1352] a process of conquest, which came at the expense of Mitanni and Amurru, a vassal kingdom of Egypt
2. To reconquer Amurru, Ramses II attacked the Hittite king Muwatalli II. The Battle of Kadesh is presented as a victory by Ramses, although he actually faced disaster because of over-optimism. This battle is dated III Shemu 9 Year 5 (ie at the extreme end of year 5).
3. Ramses II 'took advantage' of Muwatalli II's death and of accession of young king Urhi-Teshub [Muršili III] to launch a new conquest of Amurru. He temporarily conquered 18 cities (including Dapur and Tunip). This war is dated towards the end of year 8 (April/May) that implies to place the death of the Hittite king in the 1$^{st}$ half of year 8.
4. After 7 years of reign[130], Hattušili III expelled his nephew Urhi-Tešub who took refuge in Egypt. Hattušili III demanded his extradition to Ramses II, who refused it.
5. Fearing a possible coup fomented by Egypt hosting Urhi-Tešub, his rival, Hattušili III combines with Babylonian king Kadašman-Turgu to face Egypt. The epithets of Ramses on a stele at Beth-Shean, dated IV Peret 1 of his year 18 (2$^{nd}$ half of the year), have a strong military flavour and attest to the frenetic activity that prevailed in this region[131].
6. Having learned the collapse of the kingdom of Mitanni, annexed by Assyrian king Shalmaneser I, Ramses II preferred to stabilize the volatile situation with Hittite king Hattušili III by a peace treaty, dated I Peret 21, year 21 of Ramses II (1$^{st}$ half of year 21).
7. A letter of Hattušili III[132] sent to Babylonian king Kadašman-Enlil II to justify his shifting alliances, tells us that the latter's father, Kadašman-Turgu, had died shortly before the peace treaty[133] of the year 21.

The synchronism "year 0 of Tukulti-Ninurta I [in -1242] = year 42 of Ramses II" results of the sequence of following events:

1. The Treaty of the year 21 led to an era of stability, which pushed Ramses II to boost his ties by suggesting Hattušili III to marry one of his daughters. The Hittite king accepted and proposed to send his daughter [Maathorneferure] for his 2$^{nd}$ jubilee (year 33). Negotiations for the marriage began IV Akhet of year 33 but were without result III until III Peret of year 34 because of the reluctance of Puduhepa, Hattušili III's wife.

---

[129] K.A. KITCHEN – RAMSES II le pharaon triomphant
Monaco 1985 2003 Éd. du Rocher pp. 82-134.
C. DESROCHES NOBLECOURT – Ramsès II la véritable histoire
Paris 1996 Éd. Pygmalion pp. 257-294, 329-365.
[130] T. BRYCE – The Kingdom of the Hittites
Oxford 2005 Ed. Oxford University Press p. 261.
[131] C. R. HIGGINBOTHAM – Egyptianization and Elite Emulation in Ramesside
Leuven 2000 Ed. Brill pp. 31-34.
[132] G. BECKMAN – Hittite Diplomatic Texts
Atlanta 1999 Ed. Society of Biblical Literature pp. 138-143.
[133] T. Bryce notes that after the year 21 Hattušili could have appealed of extradition clauses that would contain this treaty.

2. Relations between the two kings became so cordial that Ramses II, after his 4th jubilee year 39, proposed to the Hittite king to meet him in person. Hattušili III appears to have accepted and proposed, as a pledge, another of his daughters to Ramses II to seal this agreement at the top. Nothing is known of the name and fate of the girl who followed her sister into the harem of Ramses II. There is also no more information on later relationships between the two courts. The wedding date is not specified, but presumably it intervened in the year 42, because it is from this time that Ramses II assumed his new title of "Sovereign God of Heliopolis" (found in cuneiform texts).

The general study of special epithets shows that they were adopted at a given time and in under certain circumstances to define and consecrate forever a theological aspect of the royal person[134]. The title "Ruler of Heliopolis" appears for the first time on the ostracon Louvre 2262, dated IV Peret, year 42 of Ramses II. This title might appear shortly before, but this is unlikely because no special circumstances mentioned in connection with Ramses except he married the daughter of the Hittite king. Hattušili III had offered her first daughter, plus a rich dowry, for the 2nd jubilee of Ramses II year 33, one can assume that he proceeded the same way for his second daughter at the 5th jubilee year 42. The fact that relationships are interrupted just after the marriage can be explained only by the disappearance of Hattušili III. This death has probably pushed the new Assyrian king Tukulti-Ninurta I to attack Tuthaliya IV the young successor of Hattušili III, who lost his Tarhuntassa region. This defeat pushed the Hittite king to bind to the Babylonian king (not named) by a wedding with one of his daughters. Ramses II wrote to Tuthaliya IV to discourage such a connection, but in vain[135] (Ramses celebrated his 14th jubilee year 66[136]).

The year 42 of Ramses was chosen by Wennufer, high priest of Osiris at Abydos, to praise Ramses II and to thank him for having appointed several members of his family to high office. It is likely that this special year was one where it was decided to build the temple of Wadi es-Seboua dedicating the new function of Ramses II as Ruler of Heliopolis. This temple was completed after the year 44 (stele of the officer Ramose).

The reign of Ramses II is fixed by two astronomical phenomena: 1) a helical rising of Sirius during the 11-year reign of Sety I, dated I Akhet 1, year 4[137], which fixes[138] his accession around -1294 +/- 4. It is indeed a Sothic rising because the astronomical ceiling of Sety I actually started by a Sothic rising and according to his Cenotaph: *All these stars begin on 1st Akhet when Sirius appears*[139]; 2) the 1st day of the egyptian lunar calendar (called *psḏntyw* "shining ones") dated II Peret 27 in the year 52 of Ramses II[140] (December 20, 1232 BCE) actually coincides with a full moon[141] (such coincidence occurs only every 25 years).

Chronological reconstruction[142] of all the Egyptian, Hittite, Babylonian and Assyrian reigns over the period 1295-1215 is as follows (synchronisms are highlighted):

---

134 J. YOYOTTE – Le nom de Ramsès "Souverain d'Héliopolis"
in: Mit Rahineh 1956 Philadelphia Ed. The University Museum pp. 66-70.

135 T. BRYCE – The Kingdom of the Hittites
Oxford 2005 Ed. Oxford University Press pp. 297-298.

136 C. DESROCHES NOBLECOURT – Ramsès II la véritable histoire
Paris 1996 Éd. Pygmalion pp. 361-376.

137 K. SETHE - Sethos I und die Erneuerung der Hundssternperiode
in: *Zeitschrift für Ägyptische Sprache* 66 (1931) pp. 1-7.

138 At Thebes (Longitude 32°39' Latitude 25°42') with an *arcus visionis* of 8.7 the Sothiac rising is dated 12 July on the period 1370-600 (see http://www.imcce.fr/fr/grandpublic/phenomenes/sothis/index.php ) and I Akhet 1 = 12 July only for 4 years 1293-1290 (see http://www.chronosynchro.net/wordpress/convertisseur )

139 O. NEUGEBAUER, R.A. PARKER – Egyptian Astronomical Texts I
London 1960 Ed. Brown University Press pp. 44, 54 (Text $T_2$ plate 47).
K. SETHE - Sethos I und die Erneuerung der Hundssternperiode
in: *Zeitschrift für Ägyptische Sprache* 66 (1931) pp. 1-7.

140 J.J. JANSSEN – Two Ancient Egyptian Ship's Logs
Leiden 1961 Ed. E.J. Brill p. 12.

141 http://www.imcce.fr/fr/grandpublic/phenomenes/phases_lune/index.php

142 Years of Ramses II go from June to May accession dated III Shemu 27) and years of Babylonian reigns run from April to March.

| EGYPT | HATTI | BABYLONIA | | ASSYRIA |
|---|---|---|---|---|
| **Ramses I** | [-]/**Muwatalli II** | 13 | **1295** | 9 |
| 2/**Sethy I** | [1] | 14 | **1294** | 10 |
| 1/2 | [2] | 15 | **1293** | 11 |
| 2/3 | [3] | 16 | **1292** | 12 |
| 3/**4** | [4] | 17 | **1291** | 13 |
| 4/5 | [5] | 18 | **1290** | 14 |
| 5/6 | [6] | 19 | **1289** | 15 |
| 6/7 | [7] | 20 | **1288** | 16 |
| 7/8 | [8] | 21 | **1287** | 17 |
| 8/9 | [9] | 22 | **1286** | 16 |
| 9/10 | [10] | 23 | **1285** | 17 |
| 10/11 | [11] | 24 | **1284** | 18 |
| 11/**Ramses II** | [12] | 25 | **1283** | 19 |
| 1/2 | [13] | 26/**Kadašman-Turgu** | **1282** | 20 |
| 2/3 | [14] | 1 | **1281** | 21 |
| 3/4 | [15] | 2 | **1280** | 22 |
| 4/5 | [16] | 3 | **1279** | 23 |
| **5**/6 | [*Battle of Kadesh*] | 4 | **1278** | 24 |
| 6/7 | [18] | 5 | **1277** | 25 |
| 7/8 | [19] | 6 | **1276** | 26 |
| **8**/9 | **Urhi-Tešub [Muršili III]** | 7 | **1275** | 27 |
| 9/10 | 1 | 8 | **1274** | 28 |
| 10/11 | 2 | 9 | **1273** | 29 |
| 11/12 | 3 | 10 | **1272** | 30 |
| 12/13 | 4 | 11 | **1271** | 31/**Shalmaneser I** |
| 13/14 | 5 | 12 | **1270** | 1 |
| 14/15 | 6 | 13 | **1269** | 2 |
| **15**/16 | 7/**Ḫattušili III** | 14 | **1268** | 3 |
| 16/17 | [1] | 15 | **1267** | 4 |
| 17/**18** | [*alliance*] | 16 | **1266** | 5 |
| **18**/19 | [3] | 17 | **1265** | 6 |
| **19**/20 | [4] | 18/**Kadašman-Enlil II** | **1264** | 7 [*Collapse of Mitanni*] |
| 20/**21** | [*Peace treaty*] | 1 | **1263** | 8 |
| 21/22 | [6] | 2 | **1262** | 9 |
| 22/23 | [7] | 3 | **1261** | 10 |
| 23/24 | [8] | 4 | **1260** | 11 |
| 24/25 | [9] | 5 | **1259** | 12 |
| 25/26 | [10] | 6 | **1258** | 13 |
| 26/27 | [11] | 7 | **1257** | 14 |
| 27/28 | [12] | 8 | **1256** | 15 |
| 28/29 | [13] | 9/**Kudur-Enlil** | **1255** | 16 |
| 29/**30** | [1st *Jubilee*] [14] | 1 | **1254** | 17 |
| 30/31 | [15] | 2 | **1253** | 18 |
| 31/32 | [16] | 3 | **1252** | 19 |
| 32/**33** | [2nd *Jubilee, 1st wedding*] | 4 | **1251** | 20 |
| 33/34 | [18] | 5 | **1250** | 21 |
| 34/35 | [19] | 6 | **1249** | 22 |
| 35/**36** | [3rd *Jubilee*] [20] | 7 | **1248** | 23 |
| 36/37 | [21] | 8 | **1247** | 24 |
| 37/38 | [22] | 9/**Šagarakti-Šuriaš** | **1246** | 25 |
| 38/**39** | [4th *Jubilee*] [23] | 1 | **1245** | 26 |
| 39/40 | [24] | 2 | **1244** | 27 |
| 40/41 | [25] | 3 | **1243** | 28 |
| 41/**42** | [5th *Jubilee, 2nd wedding*] | 4 | **1242** | 29/**Tukulti-Ninurta I** |
| 42/43 | [27]/**Tutḫaliya IV** | 5 | **1241** | 1 |
| 43/44 | [1] | 6 | **1240** | 2 |
| 44/**45** | [6th *Jubilee*] [2] | 7 | **1239** | 3 |
| 45/46 | [3] | 8 | **1238** | 4 |
| 46/47 | [4] | 9 | **1237** | 5 |
| 47/**48** | *[7$^{th}$ Jubilee]* [5] | 10 | **1236** | 6 |
| 48/49 | [6] | 11 | **1235** | 7 |
| 49/50 | [7] | 12 | **1234** | 8 |
| 50/**51** | *[8$^{th}$ Jubilee]* [8] | 13/**Kaštiliašu IV** | **1233** | 9 |
| 51/**52** | [9] | 1 | **1232** | 10 |
| 52/53 | [10] | 2 | **1231** | 11 |
| 53/**54** | *[9$^{th}$ Jubilee]* [11] | 3 | **1230** | 12 |
| 54/55 | [12] | 4 | **1229** | 13 |
| 55/56 | **[Kurunta]** | 9 | **1228** | 14 |
| 56/**57** | *[10$^{th}$ Jubilee]* [14] | 10 | **1227** | 15 |
| 57/58 | [15] | 11 | **1226** | 16 |
| 58/59 | [16] | 12/**Enlil-nâdin-šumi** | **1225** | 17 |
| 59/**60** | *[11$^{th}$ Jubilee]* [17] | 1/**Kadašman-Harbe II** | **1224** | 18 |
| 60/**61** | *[12$^{th}$ Jubilee]* [18] | **Adad-šuma-iddina** | **1223** | 19 |
| 61/62 | [19] | 1 | **1222** | 20 |
| 62/**63** | *[13$^{th}$ Jubilee]* [20] | 2 | **1221** | 21 |
| 63/64 | [21] | 3 | **1220** | 22 |
| 64/65 | [22] | 4 | **1219** | 23 |
| 65/**66** | *[14$^{th}$ Jubilee]* [23] | 5 | **1218** | 24 |
| 66/67 | [24] | 6/**Adad-šuma-uṣur** | **1217** | 22 |
| **67/Merenptah** | [25] | 1 | **1216** | 23 |
| 1/2 | [26] | 2 | **1215** | 24 |

The agreement with all these dates is excellent. However this Egyptian chronology is generally not accepted by Egyptologists who prefer to set the accession of Ramses II in 1279, based on the lunar cycle proposed by Parker[143] (in 1950). Egyptian chronology of this period (1300-1200) must be reviewed precisely.

If the dates obtained from $^{14}C$ method, calibrated by dendrochronology, still remain imprecise, however, they set out values (in 2010) with a precision of +/- 13 years over the period 1300-1000[144]. Durations of reigns according to synchronisms are calculated taking into account accession and highest dates in the reign (see next page):

| | | Reign | Length according to: | | |
|---|---|---|---|---|---|
| No. | XIX$^{th}$ dynasty | ($^{14}C$) | $^{14}C$ | synchronisms | discrepancy |
| 1 | Ramses I | 1302-1302 | 0 year | 1 year 4 months | -1 |
| 2 | Sethy I | 1302-1285 | 17 years | 11 years | +6 |
| 3 | Ramses II | 1285-1219 | 66 years | 67 years 2 months | -1 |
| 4 | Merenptah | 1219-1206 | 13 years | 9 years 3 months | +4 |
| 5 | Sethy II | 1206-1200 | 6 years | 5 years | +1 |
| 6 | [Amenmes] | -1209 | - | [4 years] | |
| 7 | Siptah | 1200-1194 | 6 years | 6 years | 0 |
| | -*Tausert* | 1194-1192 | 2 years | 1 year 6 months | 0 |
| | XX$^{th}$ dynasty | | | | |
| 1 | Sethnakht | 1192-1189 | 3 years | 3 years 5 months | 0 |
| 2 | Ramses III | 1189-1158 | 31 years | 31 years 1 month | 0 |
| 3 | Ramses IV | 1158-1152 | 6 years | 6 years 8 months | -1 |
| 4 | Ramses V | 1152-1148 | 4 years | 3 years 2 months | +1 |
| 5 | Ramses VI | 1148-1140 | 8 years | 7 years | +1 |
| 6 | Ramses VII | 1140-1133 | 7 years | 7 years 1 month | 0 |
| 7 | Ramses VIII | 1133-1130 | 3 years | 3 months ? | +3 |
| 8 | Ramses IX | 1130-1112 | 18 years | 18 years 4 months | 0 |
| 9 | Ramses X | 1112-1103 | 9 years | 2 years 5 months | +7 |
| 10 | Ramses XI | 1103-1073 | 30 years | 26 years 1 month ? | +4 |

143 L.W. CASPERSON – The Lunar Date of Ramesses II
in: *Journal of Near Eastern Studies* 47 (1988) pp. 181-184.
144 C.B. RAMSEY, M.W. DEE, J.M. ROWLAND, T.F. G. HIGHAM, S.A. HARRIS, F. BROCK, A. QUILES, E.M. WILD, E.S. MARCUS, A.J. SHORTLAND - Radiocarbon - Based Chronology for Dynastic Egypt in: *Science* Vol 328 (10 june 2010) pp. 1554-1557.
http://www.sciencemag.org/cgi/data/328/5985/1554/DC1/1

## DETERMINING EGYPTIAN CHRONOLOGY ON 1295-1155 PERIOD

As the lunar day *psḏntyw*, II Peret 27, year 52 of Ramses II is astronomically dated December 20, 1232 BCE (full moon) the accession of Ramses II (III Shemu 27) has to be dated June 1, 1283 BCE. This date is confirmed by the accession of Sety I in 1294 BCE, determined by the Sothic rising of I Akhet 1 year 4. In fact, the reign of Sety I lasted 11 years (actually 11 years and a few days) as shown in the autobiography of the priest Bakenkhons[145] (the 11 years of Sety I are all represented, except 10, which confirms the 11 years reign)[146], his accession must be dated in 1294 (= 1283 + 11). Furthermore, the accession of Kadašman-Enlil II (1264-1255) is dated in year 19 of Ramses II[147], implying again dating the accession of Ramses II in 1283 (= 1264 + 19). Chronology of dynasties based on years of reign and accession dates[148]:

| | Length of reign | accession date | highest date | Reign |
|---|---|---|---|---|
| Ramses I | 1 year 4 months | III Peret ? | **2** II Peret 20 | 01/1295-05/1294 |
| Sethy I | 11 years | III Shemu 24 ? | **11** IV Shemu 13 | 06/**1294**-06/1283 |
| Ramses II | 67 years 2 months | III Shemu 27 | **67** I Akhet 18 | 06/**1283**-07/1216 |
| Merenptah | 9 years 3 months | II Akhet 5-13? | **10** IV Akhet 7 | 08/1216-10/1207 |
| Sethy II | 5 years | I Peret ? | **6** I Peret 19 | 11/1207-10/1202 |
| [Amenmes] | [4 years] | [II Shemu ?] | [ **4** III Shemu 29 ?] | [04/1206-03/1202] |
| Siptah | 6 years | I Peret 2? | **7** IV Akhet 22 | 11/**1202**-10/1196 |
| (Siptah)-Tausert | 1 year 6 months | " | **8** II Shemu 29 | 11/1196-04/1194 |
| Sethnakht | 3 years 5 months | " | **4** | 11/1196-03/1192 |
| Ramses III | 31 years 1 month | I Shemu 26 | **32** III Shemu 14 | 04/**1192**-04/1161 |
| Ramses IV | 6 years 8 months | III Shemu 15 | **7** III Akhet 29? | 05/**1161**-12/1155 |

The reign durations fit quite well with those of Manetho (via Flavius Josephus). However, because of the uncertainty on some accession dates three of these reigns may have an additional year if we place it at the end of the last year of reign instead of the beginning. Thus, Sety II may have reigned 6 years instead of 5 (the most likely)[149] and Ramses II may have reigned 67 years and 2 months instead of 66 years and 2 months. In his stele dated beginning of year 4, Ramses IV compares his 3 years of reign with the 67 years (not 66) of Ramses II, which involves a death of Ramses II at the beginning of his year 68 in accordance with the number of his jubilees (*sed* festivals). In fact, 14 jubilees are attested, the first one being celebrated in year 30 and the others every 3 years: the 11th in year 60 (=2x30), the 12th in year 61 and the 14th in year 66. The case most delicate being the 4 years reign of pharaoh Amenmes, that some place between Merenptah and Sety II, and others in parallel with Sety II (and delay it of approximately 5 months). Several synchronisms and lunar dates, dated by astronomy, can resolve these uncertainties.

The violent crisis that hit the eastern Mediterranean caused the ruin of the great empires of the Bronze Age, which the Trojan War is the most famous episode, is exactly dated year 8 of Ramesses III. Thebes, Lefkandi, Tiryns, Mycenae and Pylos in mainland

---

145 *Bakenkhonsu states that he spent 4 years as an excellent youngster, 11 years as a youth, as a trainee stable-master for king Men[maat]re (Sety I), wab priest of Amun for 4 years, god's father of Amun for 12 years, third pries of Amun for 15 years, second priest of Amun for 12* (E. FLOOD – Biographical Texts from Ramessid Egypt Atlanta 2007 Ed. Society of Biblical Literature p. 41).

146 E. HORNUNG – The New Kingdom
in: Ancient Egyptian Chronology (Leiden 2006) Ed. Brill pp. 210-211.

147 W.A. WARD - The Present Status of Egyptian Chronology
in: *Bulletin of the American Schools of Oriental Research* 288 (1991) pp. 55,56.

148 E. HORNUNG – The New Kingdom
in: Ancient Egyptian Chronology. Leiden 2006 Ed. Brill pp. 208-211.
C. VANDERSLEYEN - L'Egypte et la vallée du Nil Tome 2
Paris 1995 Éd. Presses Universitaires de France pp. 467-512.
J. VON BECKERATH – Chronologie des pharaonischen ägypten
1997 Ed. Verlag Philipp von Zabern pp. 201-202.

149 H. ALTENMÜLLER – Bemerkunden zu den königsgräbern des neuen reiches
in: *Studien zur Altägyptischen Kultur* 10 (1983) pp. 43-61.

Greece and Chania in Crete, were ransacked and sometimes completely destroyed. Most of these cities and their palaces were burned. In Anatolia, among the most important sites, archaeological levels similarly destroyed are found and which dates from the same period. Hattusha, the Hittite capital, was sacked and burned just like the major cities of Cyprus. On the north coast of Syria, the flourishing city of Ugarit was destroyed and never inhabited thereafter. Mesopotamia was preserved as the wave of devastation did not extend to the east[150], and it was the Egyptians who alone could stop it. The temple of Ramses III at Medinet Habu contains an account of this victory over the Sea Peoples. The identification of these peoples as their reasons for migration are poorly understood, however, these events are precisely dated. The great Alexandrian scholar Eratosthenes (276-193), for example, dated the famous Trojan War in -1184. Manetho[151], while confirming the 7-year reign of Queen [Siptah]/Tausert (1202-1194) states: *Thouôris, (...) at the time when Troy was taken, reigned 7 years* (Tausert actually reigned, from 1195 to 1194, at the beginning of the war, 10 years before the destruction of Troy)[152]. This destruction coincides with the fall of the Hittite Empire dated indirectly in year 8 of Ramses III and in year 2 of Meli-Shipak (the last texts from Emar are dated [-]/$VI_2$/**2** and 6/VII/[**2**] of Meli-Shipak)[153], in October 1185 BCE. This war led by the Sea Peoples had to be spread over less than one year because, according to the inscription of Ramses III, all countries (Hatti, the coast of Cilicia, Carchemish, Cyprus, etc.) were "destroyed all at once" and, according to the text of Homer (Odyssey XIV:240-280), the sacking of the city of Priam [Troy], after 10 years of fighting, was followed "in less than 1 month" by the cruise of Achaeans to Egypt and the sacking of its wonderful fields. As year 2 of Meli-Shipak is dated in 1185 BCE, Ramses III's accession has to be dated in 1192 (= 1185 +8 – 2 +1)[154]. This date is consistent with the accession of Ramses II in 1283 (= 1192 + 3+5 m. +6 + 5 + 9+3 m. + 67+2 m.).

The reign of Tausert is well known[155]. Wife of Sety II, she exercised after his death a strong influence on his son Siptah (Regency?) then, at the latter's death, she continued his reign instead of inaugurating a new one (Sethnakht also began his reign from Siptah's death)[156]. Egyptian women, as wife or daughter of Pharaoh, could access the deity, which authorized them to embody and so prolong the reign of a dead pharaoh without successor, but not to begin a new reign. This case occurred three times over the period 1500-1200: 1) Tausert, wife of Sety II, continuing the reign of his son Siptah, 2) Ankhkheperure continuing the reign of Semenkhkare her husband and 3) Hatshepsut continuing the reign of her husband Thutmose II (which was in turn extended by Thutmose III at Hatshepsut's death). These extended reigns were interpreted by some as co-regencies, that distorts the chronology. Another source of error comes from the change of name by some pharaohs, interpreted as the reign of new sovereign. In fact it is not the case, since for no apparent reason Ramses-Siptah (Sekhâenre-Meryamon) was then called Merenptah Siptah (Akhenrê-Setepenre) from the year 3 of his reign. It is possible to anchor Tausert's reign, and consequently the one of Ramses III, thanks to a graffito scribe named Thotemhab left at the Theban temple of Deir el-Bahari, in memory of his participation in the Festival of the Valley. During this annual celebration, the processional statue of Amon passed two nights at the funerary temple of the reigning monarch. The graffito of Thotemhab tells us that in the II Shemu 28 Year 7 of Tausert, the statue of Amon was transported to the mortuary temple. The Beautiful Festival of the Valley was celebrated the day after the 1st lunar day, which implies a to date that day 1 (*psḏntyw*) to II Shemu 27 Year 7 of Tausert[157].

---

150 R. MORKOT – Atlas de la Grèce antique
Paris 1996 Éd. Autrement pp. 33-34.

151 W.G. WADDELL – Manetho
Massachusetts 1956 Ed. Harvard University Press pp. 101-119.

152 According to Thucydides, the Trojan War was the result of an expedition of disparate tribes of pirates (see Odyssey III:71-74), living on islands around Achaia, who were united by King Agamemnon of Mycenae. This expedition against the Trojans was the culmination of 10 years of battle (The Peloponnesian War I:8-12). For example, a battle in Egypt is mentioned in the year 5 of Rameses III.

153 Y. COHEN, I. SINGER – A Late Synchronism between Ugarit and Emar
in: Essays on Ancient Israel in Its Near Eastern Context (Eisenbrauns 2006) Indiana p. 134.

154 Year 2 of Meli-Shipak beginning on Nisan 1, or on April 4, 1185 BCE, and year 8 of Ramesses III starts at I Shemu 26 or so in April at that time. The accession is counted as year 0 by the Babylonians and as a year 1 by the Egyptians.

155 V.G. CALLENDER – Queen Tausret and the End of Dynasty 19
in: *Studien zur Altägyptischen Kultur* 32 (2004) pp. 81-104.

156 C. VANDERSLEYEN - L'Egypte et la vallée du Nil Tome 2
Paris 1995 Éd. Presses Universitaires de France pp. 591-593.

The reign of Pharaoh Amenmes[158] can not be placed between that of Merenptah and Sety II, but only in parallel with the one of Sety II, as can be deduced from the lunar dates (see table hereafter for dates), because the insertion of 4-year reign of Amenmes would push the lunar date, either in II Peret 21 in -1236 if the reign of Sety II is 5 years long, either in II Peret 2 in -1237 if this reign is 6 years long, yet the only possibility is that of II Peret 27 in -1232.

The reign of Ramses III began at I Shemu 26 year 1, or March 9, -1192. This reconstruction also confirms the 2-year reign of the pharaoh Sethnakht because the duration of 3 years[159] would imply a lunar date II Shemu 7 (April -1196), incompatible with that of II Shemu 27 from the graffito. This date 27 Shemu II Year 7 corresponds to April 10 in -1195 and actually coincides with a full moon[160]. A good indication of the rivalry between the two kings, Setnakht and Amenmes (later considered as usurper), comes from their cartouche, each having made erase the name of the other. Year 4 of Sethnakht (Al-Ahram Weekly 11-17 January 2007 No. 827) involves at least 3 years of reign, but as this reign began with the death of Siptah, Tausert's reign (1 year 6 months) must be subtracted.

We also note that the two lunar dates (*psḏntyw*) of Ramses III (I Shemu 11 and IV Peret 1)[161] fall at the beginning and end of year 5. Moreover, the *beautiful feast of the valley*[162] [probably at the end of year 5], celebrated just after the lunar day 1(*psḏntyw*), is dated II Shemu 1 and 2, which implies to date this lunar day I Shemu at 30 or March 12 in -1187 (full moon). The lunar day *psḏntyw* has always played an important role in Egyptian cult. On the stele from Abydos dated Year 4 of Ramses IV, Pharaoh says: *My heart has not forgotten the day of my psḏntyw feast*[163] and this stele is dated 10 Akhet III, which implies a connection with this lunar day. The year 4 of Ramses IV begins at III Shemu 15 (the day of his accession)[164] and in 1158 according to the previous scheme, one can also verify that the 4 year of Ramses IV begins with a lunar day 1 dated III Shemu 16, which explains the choice of the year 4 for this inscription. The III Shemu 15 corresponds to April 19 in -1158, full moon day, as the III Akhet 10 which corresponds to August 16, -1158.

The complete reconstruction of all Egyptian reigns on the period 1295-1155, based on the lunar cycle of 25 years, allows to check the coincidences of dates which occur only every 25 years, if there is no error, or every 11/14 years if there is an error of 1 day.

---

157 R. KRAUSS – Moïse le pharaon
Monaco 2005 Éd. Rocher pp. 125-127.
158 T. SCHNEIDER – Conjectures about Amenmesse
in: Ramesside Studies in Honour of K.A. itchen (Rutherford Press, 2011) pp. 445-451.
159 If the Elephantine Stele (KRI V:671-672) states that all the enemies of Egypt were eliminated on II Shemu 1 in year 2 of Sethnakht, there is no explicit link with a accession date, but it could correspond to the time of the disappearance of Tausert (whose highest date is the II Shemu 29 year 8 of Siptah).
160 http://www.imcce.fr/fr/grandpublic/phenomenes/phases_lune/index.php
161 A. SPALINGER – Egyptian Festival Dating and the Moon
in: Under One Sky (Münster 2002) Ed. Ugarit-Verlag pp. 385-389.
162 S. EL-SABBAN – Temple Festival Calendars of Ancient Egypt
Liverpol 2000 Ed. Liverpool University Press pp. 67,68.
163 A.J. PEDEN – The Reign of Ramesses IV
Warminster 1994 Ed. Aris & Phillips Ltd pp. 91-94.
164 C. VANDERSLEYEN - L'Egypte et la vallée du Nil Tome 2
Paris 1995 Éd. Presses Universitaires de France p. 616.

| | |
|---|---|
| | Legend of colours: |
| | Year 1 of Ramses I from IV Peret, June in -1294, to III Peret, May in -1293 (-1293 = 1293 BCE) |
| | Synchronism with the Sothic rising dated I Akhet 1 in year 4 of Sety I (July 12, -1291). |
| | Synchronism with Babylonian chronology:<br>years 19 and 42 of Ramses II (in -1264 and -1241); year 8 of Ramesses III (in -1185). |
| | Lunar dates: year 52 de Ramses II (in -1232); year 7 de Siptah (in -1195); year 4 de Ramses IV (in -1158). |

| SEASON | | | | AKHET | | | | PERET | | | | SHEMU | | | | |
|---|---|---|---|---|---|---|---|---|---|---|---|---|---|---|---|---|
| | | | | I | II | III | IV | I | II | III | IV | I | II | III | IV | 5 |
| month | | | | Jul. | Aug. | Sept. | Oct. | Nov. | Dec. | Jan. | Feb. | Mar. | Apr. | May | Jun. | |
| **Ramses I** | 1295 | **1** | 1 | 1 | 30 | 29 | 29 | 28 | 28 | 27 | 27 | 26 | 26 | 25 | 25 | |
| **Sety I** | 1294 | **2** | 2 | 19 | 19 | 18 | 18 | 18 | 17 | 17 | 16 | 16 | 15 | 15 | 14 | |
| | 1293 | **1** | 3 | 9 | 8 | 8 | 7 | 7 | 6 | 6 | 6 | 5 | 5 | 4 | 4 | 3 |
| | 1292 | **2** | 4 | 28 | 27 | 27 | 26 | 26 | 25 | 25 | 24 | 24 | 23 | 23 | 23 | |
| | 1291 | **3** | 5 | 17 | 17 | 16 | 16 | 15 | 15 | 14 | 14 | 13 | 13 | 12 | 12 | |
| | 1290 | **4** | 6 | 6 | 6 | 6 | 5 | 5 | 4 | 4 | 3 | 3 | 2 | 2 | 1 | 1 |
| | 1289 | **5** | 7 | 25 | 25 | 24 | 24 | 23 | 23 | 23 | 22 | 22 | 21 | 21 | 20 | |
| | 1288 | **6** | 8 | 15 | 14 | 14 | 13 | 13 | 12 | 12 | 11 | 11 | 10 | 10 | 10 | |
| | 1287 | **7** | 9 | 4 | 4 | 3 | 3 | 2 | 2 | 1 | 1 | 30 | 29 | 29 | 28 | |
| | 1286 | **8** | 10 | 23 | 23 | 22 | 22 | 21 | 21 | 20 | 20 | 19 | 19 | 18 | 18 | |
| | 1285 | **9** | 11 | 12 | 12 | 11 | 11 | 10 | 10 | 10 | 9 | 9 | 8 | 8 | 7 | |
| | 1284 | **10** | 12 | 2 | 1 | 1 | 30 | 30 | 29 | 28 | 28 | 27 | 27 | 27 | 26 | |
| **Ramses II** | 1283 | **11** | 13 | 21 | 20 | 20 | 19 | 19 | 18 | 18 | 17 | 17 | 16 | 16 | 15 | |
| | 1282 | **1** | 14 | 10 | 9 | 9 | 9 | 8 | 8 | 7 | 7 | 6 | 6 | 5 | 5 | 4 |
| | 1281 | **2** | 15 | 29 | 28 | 28 | 27 | 27 | 27 | 26 | 26 | 25 | 25 | 24 | 24 | |
| | 1280 | **3** | 16 | 18 | 18 | 17 | 17 | 16 | 16 | 15 | 15 | 14 | 14 | 14 | 13 | |
| | 1279 | **4** | 17 | 8 | 7 | 7 | 6 | 6 | 5 | 5 | 4 | 4 | 3 | 3 | 2 | 2 |
| | 1278 | **5** | 18 | 26 | 26 | 26 | 25 | 25 | 24 | 24 | 23 | 23 | 22 | 22 | 21 | |
| | 1277 | **6** | 19 | 16 | 15 | 15 | 14 | 14 | 14 | 13 | 13 | 12 | 12 | 11 | 11 | |
| | 1276 | **7** | 20 | 5 | 5 | 4 | 4 | 3 | 3 | 2 | 2 | 1 | 1 | 1 | 30 | |
| | 1275 | **8** | 21 | 24 | 24 | 23 | 23 | 22 | 22 | 21 | 21 | 20 | 20 | 19 | 19 | |
| | 1274 | **9** | 22 | 13 | 13 | 13 | 12 | 12 | 11 | 11 | 10 | 10 | 9 | 9 | 8 | |
| | 1273 | **10** | 23 | 3 | 2 | 2 | 1 | 1 | 30 | 30 | 29 | 29 | 28 | 28 | 27 | |
| | 1272 | **11** | 24 | 22 | 21 | 21 | 20 | 20 | 19 | 19 | 18 | 18 | 18 | 17 | 17 | |
| | 1271 | **12** | 25 | 11 | 11 | 10 | 10 | 9 | 9 | 8 | 8 | 7 | 7 | 6 | 6 | |
| | 1270 | **13** | 1 | 1 | 30 | 29 | 29 | 28 | 28 | 27 | 27 | 26 | 26 | 25 | 25 | |
| | 1269 | **14** | 2 | 19 | 19 | 18 | 18 | 18 | 17 | 17 | 16 | 16 | 15 | 15 | 14 | |
| | 1268 | **15** | 3 | 9 | 8 | 8 | 7 | 7 | 6 | 6 | 6 | 5 | 5 | 4 | 4 | 3 |
| | 1267 | **16** | 4 | 28 | 27 | 27 | 26 | 26 | 25 | 25 | 24 | 24 | 23 | 23 | 23 | |
| | 1266 | **17** | 5 | 17 | 17 | 16 | 16 | 15 | 15 | 14 | 14 | 13 | 13 | 12 | 12 | |
| | 1265 | **18** | 6 | 6 | 6 | 6 | 5 | 5 | 4 | 4 | 3 | 3 | 2 | 2 | 1 | 1 |
| | 1264 | **19** | 7 | 25 | 25 | 24 | 24 | 23 | 23 | 23 | 22 | 22 | 21 | 21 | 20 | |
| | 1263 | **20** | 8 | 15 | 14 | 14 | 13 | 13 | 12 | 12 | 11 | 11 | 10 | 10 | 10 | |
| | 1262 | **21** | 9 | 4 | 4 | 3 | 3 | 2 | 2 | 1 | 1 | 30 | 29 | 29 | 28 | |
| | 1261 | **22** | 10 | 23 | 23 | 22 | 22 | 21 | 21 | 20 | 20 | 19 | 19 | 18 | 18 | |
| | 1260 | **23** | 11 | 12 | 12 | 11 | 11 | 10 | 10 | 10 | 9 | 9 | 8 | 8 | 7 | |
| | 1259 | **24** | 12 | 2 | 1 | 1 | 30 | 30 | 29 | 28 | 28 | 27 | 27 | 27 | 26 | |
| | 1258 | **25** | 13 | 21 | 20 | 20 | 19 | 19 | 18 | 18 | 17 | 17 | 16 | 16 | 15 | |
| | 1257 | **26** | 14 | 10 | 9 | 9 | 9 | 8 | 8 | 7 | 7 | 6 | 6 | 5 | 5 | 4 |
| | 1256 | **27** | 15 | 29 | 28 | 28 | 27 | 27 | 27 | 26 | 26 | 25 | 25 | 24 | 24 | |
| | 1255 | **28** | 16 | 18 | 18 | 17 | 17 | 16 | 16 | 15 | 15 | 14 | 14 | 14 | 13 | |
| | 1254 | **29** | 17 | 8 | 7 | 7 | 6 | 6 | 5 | 5 | 4 | 4 | 3 | 3 | 2 | 2 |
| | 1253 | **30** | 18 | 26 | 26 | 26 | 25 | 25 | 24 | 24 | 23 | 23 | 22 | 22 | 21 | |
| | 1252 | **31** | 19 | 16 | 15 | 15 | 14 | 14 | 14 | 13 | 13 | 12 | 12 | 11 | 11 | |
| | 1251 | **32** | 20 | 5 | 5 | 4 | 4 | 3 | 3 | 2 | 2 | 1 | 1 | 1 | 30 | |
| | 1250 | **33** | 21 | 24 | 24 | 23 | 23 | 22 | 22 | 21 | 21 | 20 | 20 | 19 | 19 | |
| | 1249 | **34** | 22 | 13 | 13 | 13 | 12 | 12 | 11 | 11 | 10 | 10 | 9 | 9 | 8 | |
| | 1248 | **35** | 23 | 3 | 2 | 2 | 1 | 1 | 30 | 30 | 29 | 29 | 28 | 28 | 27 | |
| | 1247 | **36** | 24 | 22 | 21 | 21 | 20 | 20 | 19 | 19 | 18 | 18 | 18 | 17 | 17 | |
| | 1246 | **37** | 25 | 11 | 11 | 10 | 10 | 9 | 9 | 8 | 8 | 7 | 7 | 6 | 6 | |
| | 1245 | **38** | 1 | 1 | 30 | 29 | 29 | 28 | 28 | 27 | 27 | 26 | 26 | 25 | 25 | |
| | 1244 | **39** | 2 | 19 | 19 | 18 | 18 | 18 | 17 | 17 | 16 | 16 | 15 | 15 | 14 | |
| | 1243 | **40** | 3 | 9 | 8 | 8 | 7 | 7 | 6 | 6 | 6 | 5 | 5 | 4 | 4 | 3 |
| | 1242 | **41** | 4 | 28 | 27 | 27 | 26 | 26 | 25 | 25 | 24 | 24 | 23 | 23 | 23 | |
| | 1241 | **42** | 5 | 17 | 17 | 16 | 16 | 15 | 15 | 14 | 14 | 13 | 13 | 12 | 12 | |
| | 1240 | **43** | 6 | 6 | 6 | 6 | 5 | 5 | 4 | 4 | 3 | 3 | 2 | 2 | 1 | 1 |
| | 1239 | **44** | 7 | 25 | 25 | 24 | 24 | 23 | 23 | 23 | 22 | 22 | 21 | 21 | 20 | |
| | 1238 | **45** | 8 | 15 | 14 | 14 | 13 | 13 | 12 | 12 | 11 | 11 | 10 | 10 | 10 | |
| | 1237 | **46** | 9 | 4 | 4 | 3 | 3 | 2 | 2 | 1 | 1 | 30 | 29 | 29 | 28 | |
| | 1236 | **47** | 10 | 23 | 23 | 22 | 22 | 21 | 21 | 20 | 20 | 19 | 19 | 18 | 18 | |
| | 1235 | **48** | 11 | 12 | 12 | 11 | 11 | 10 | 10 | 10 | 9 | 9 | 8 | 8 | 7 | |
| | 1234 | **49** | 12 | 2 | 1 | 1 | 30 | 30 | 29 | 28 | 28 | 27 | 27 | 27 | 26 | |
| | 1233 | **50** | 13 | 21 | 20 | 20 | 19 | 19 | 18 | 18 | 17 | 17 | 16 | 16 | 15 | |
| | 1232 | **51** | 14 | 10 | 9 | 9 | 9 | 8 | 8 | 7 | 7 | 6 | 6 | 5 | 5 | 4 |
| | 1231 | **52** | 15 | 29 | 28 | 28 | 27 | 27 | 27 | 26 | 26 | 25 | 25 | 24 | 24 | |
| | 1230 | **53** | 16 | 18 | 18 | 17 | 17 | 16 | 16 | 15 | 15 | 14 | 14 | 14 | 13 | |
| | 1229 | **54** | 17 | 8 | 7 | 7 | 6 | 6 | 5 | 5 | 4 | 4 | 3 | 3 | 2 | 2 |
| | 1228 | **55** | 18 | 26 | 26 | 26 | 25 | 25 | 24 | 24 | 23 | 23 | 22 | 22 | 21 | |
| | 1227 | **56** | 19 | 16 | 15 | 15 | 14 | 14 | 14 | 13 | 13 | 12 | 12 | 11 | 11 | |
| | 1226 | **57** | 20 | 5 | 5 | 4 | 4 | 3 | 3 | 2 | 2 | 1 | 1 | 1 | 30 | |
| | 1225 | **58** | 21 | 24 | 24 | 23 | 23 | 22 | 22 | 21 | 21 | 20 | 20 | 19 | 19 | |
| | 1224 | **59** | 22 | 13 | 13 | 13 | 12 | 12 | 11 | 11 | 10 | 10 | 9 | 9 | 8 | |
| | 1223 | **60** | 23 | 3 | 2 | 2 | 1 | 1 | 30 | 30 | 29 | 29 | 28 | 28 | 27 | |
| | 1222 | **61** | 24 | 22 | 21 | 21 | 20 | 20 | 19 | 19 | 18 | 18 | 18 | 17 | 17 | |
| | 1221 | **62** | 25 | 11 | 11 | 10 | 10 | 9 | 9 | 8 | 8 | 7 | 7 | 6 | 6 | |
| | 1220 | **63** | 1 | 1 | 30 | 29 | 29 | 28 | 28 | 27 | 27 | 26 | 26 | 25 | 25 | |
| | 1219 | **64** | 2 | 19 | 19 | 18 | 18 | 18 | 17 | 17 | 16 | 16 | 15 | 15 | 14 | |
| | 1218 | **65** | 3 | 9 | 8 | 8 | 7 | 7 | 6 | 6 | 6 | 5 | 5 | 4 | 4 | 3 |
| | 1217 | **66** | 4 | 28 | 27 | 27 | 26 | 26 | 25 | 25 | 24 | 24 | 23 | 23 | 23 | |
| | 1216 | **67** | 5 | 17 | 17 | 16 | 16 | 15 | 15 | 14 | 14 | 13 | 13 | 12 | 12 | |
| **Merenptah** | 1215 | **1** | 6 | 6 | 6 | 6 | 5 | 5 | 4 | 4 | 3 | 3 | 2 | 2 | 1 | 1 |
| | 1214 | **2** | 7 | 25 | 25 | 24 | 24 | 23 | 23 | 23 | 22 | 22 | 21 | 21 | 20 | |
| | 1213 | **3** | 8 | 15 | 14 | 14 | 13 | 13 | 12 | 12 | 11 | 11 | 10 | 10 | 10 | |
| | 1212 | **4** | 9 | 4 | 4 | 3 | 3 | 2 | 2 | 1 | 1 | 30 | 29 | 29 | 28 | |
| | 1211 | **5** | 10 | 23 | 23 | 22 | 22 | 21 | 21 | 20 | 20 | 19 | 19 | 18 | 18 | |
| | 1210 | **6** | 11 | 12 | 12 | 11 | 11 | 10 | 10 | 10 | 9 | 9 | 8 | 8 | 7 | |
| | 1209 | **7** | 12 | 2 | 1 | 1 | 30 | 30 | 29 | 28 | 28 | 27 | 27 | 27 | 26 | |
| | 1208 | **8** | 13 | 21 | 20 | 20 | 19 | 19 | 18 | 18 | 17 | 17 | 16 | 16 | 15 | |
| | 1207 | **9** | 14 | 10 | 9 | 9 | 9 | 8 | 8 | 7 | 7 | 6 | 6 | 5 | 5 | 4 |
| **Sety II** | 1206 | **1** | 15 | 29 | 28 | 28 | 27 | 27 | 27 | 26 | 26 | 25 | 25 | 24 | 24 | |
| | 1205 | **2** | 16 | 18 | 18 | 17 | 17 | 16 | 16 | 15 | 15 | 14 | 14 | 14 | 13 | |
| | 1204 | **3** | 17 | 8 | 7 | 7 | 6 | 6 | 5 | 5 | 4 | 4 | 3 | 3 | 2 | 2 |
| | 1203 | **4** | 18 | 26 | 26 | 26 | 25 | 25 | 24 | 24 | 23 | 23 | 22 | 22 | 21 | |
| | 1202 | **5** | 19 | 16 | 15 | 15 | 14 | 14 | 14 | 13 | 13 | 12 | 12 | 11 | 11 | |
| **Siptah** | 1201 | **1** | 20 | 5 | 5 | 4 | 4 | 3 | 3 | 2 | 2 | 1 | 1 | 1 | 30 | |
| | 1200 | **2** | 21 | 24 | 24 | 23 | 23 | 22 | 22 | 21 | 21 | 20 | 20 | 19 | 19 | |
| | 1199 | **3** | 22 | 13 | 13 | 13 | 12 | 12 | 11 | 11 | 10 | 10 | 9 | 9 | 8 | |
| | 1198 | **4** | 23 | 3 | 2 | 2 | 1 | 1 | 30 | 30 | 29 | 29 | 28 | 28 | 27 | |
| | 1197 | **5** | 24 | 22 | 21 | 21 | 20 | 20 | 19 | 19 | 18 | 18 | 18 | 17 | 17 | |
| | 1196 | **6** | 25 | 11 | 11 | 10 | 10 | 9 | 9 | 8 | 8 | 7 | 7 | 6 | 6 | |
| **Tausert** | 1195 | **7** | 1 | 1 | 30 | 29 | 29 | 28 | 28 | 27 | 27 | 26 | 26 | 25 | 25 | |
| **Sethnakht** | 1194 | **2** | 2 | 19 | 19 | 18 | 18 | 18 | 17 | 17 | 16 | 16 | 15 | 15 | 14 | |
| | 1193 | **3** | 3 | 9 | 8 | 8 | 7 | 7 | 6 | 6 | 6 | 5 | 5 | 4 | 4 | 3 |
| **Ramses III** | 1192 | **1** | 4 | 28 | 27 | 27 | 26 | 26 | 25 | 25 | 24 | 24 | 23 | 23 | 23 | |
| | 1191 | **2** | 5 | 17 | 17 | 16 | 16 | 15 | 15 | 14 | 14 | 13 | 13 | 12 | 12 | |
| | 1190 | **3** | 6 | 6 | 6 | 6 | 5 | 5 | 4 | 4 | 3 | 3 | 2 | 2 | 1 | 1 |
| | 1189 | **4** | 7 | 25 | 25 | 24 | 24 | 23 | 23 | 23 | 22 | 22 | 21 | 21 | 20 | |
| | 1188 | **5** | 8 | 15 | 14 | 14 | 13 | 13 | 12 | 12 | 11 | 11 | 10 | 10 | 10 | |
| | 1187 | **6** | 9 | 4 | 4 | 3 | 3 | 2 | 2 | 1 | 1 | 30 | 29 | 29 | 28 | |
| | 1186 | **7** | 10 | 23 | 23 | 22 | 22 | 21 | 21 | 20 | 20 | 19 | 19 | 18 | 18 | |
| | 1185 | **8** | 11 | 12 | 12 | 11 | 11 | 10 | 10 | 10 | 9 | 9 | 8 | 8 | 7 | |
| | 1184 | **9** | 12 | 2 | 1 | 1 | 30 | 30 | 29 | 28 | 28 | 27 | 27 | 27 | 26 | |
| | 1183 | **10** | 13 | 21 | 20 | 20 | 19 | 19 | 18 | 18 | 17 | 17 | 16 | 16 | 15 | |
| | 1182 | **11** | 14 | 10 | 9 | 9 | 9 | 8 | 8 | 7 | 7 | 6 | 6 | 5 | 5 | 4 |
| | 1181 | **12** | 15 | 29 | 28 | 28 | 27 | 27 | 27 | 26 | 26 | 25 | 25 | 24 | 24 | |
| | 1180 | **13** | 16 | 18 | 18 | 17 | 17 | 16 | 16 | 15 | 15 | 14 | 14 | 14 | 13 | |
| | 1179 | **14** | 17 | 8 | 7 | 7 | 6 | 6 | 5 | 5 | 4 | 4 | 3 | 3 | 2 | 2 |
| | 1178 | **15** | 18 | 26 | 26 | 26 | 25 | 25 | 24 | 24 | 23 | 23 | 22 | 22 | 21 | |
| | 1177 | **16** | 19 | 16 | 15 | 15 | 14 | 14 | 14 | 13 | 13 | 12 | 12 | 11 | 11 | |
| | 1176 | **17** | 20 | 5 | 5 | 4 | 4 | 3 | 3 | 2 | 2 | 1 | 1 | 1 | 30 | |
| | 1175 | **18** | 21 | 24 | 24 | 23 | 23 | 22 | 22 | 21 | 21 | 20 | 20 | 19 | 19 | |
| | 1174 | **19** | 22 | 13 | 13 | 13 | 12 | 12 | 11 | 11 | 10 | 10 | 9 | 9 | 8 | |
| | 1173 | **20** | 23 | 3 | 2 | 2 | 1 | 1 | 30 | 30 | 29 | 29 | 28 | 28 | 27 | |
| | 1172 | **21** | 24 | 22 | 21 | 21 | 20 | 20 | 19 | 19 | 18 | 18 | 18 | 17 | 17 | |
| | 1171 | **22** | 25 | 11 | 11 | 10 | 10 | 9 | 9 | 8 | 8 | 7 | 7 | 6 | 6 | |
| | 1170 | **23** | 1 | 1 | 30 | 29 | 29 | 28 | 28 | 27 | 27 | 26 | 26 | 25 | 25 | |
| | 1169 | **24** | 2 | 19 | 19 | 18 | 18 | 18 | 17 | 17 | 16 | 16 | 15 | 15 | 14 | |
| | 1168 | **25** | 3 | 9 | 8 | 8 | 7 | 7 | 6 | 6 | 6 | 5 | 5 | 4 | 4 | 3 |
| | 1167 | **26** | 4 | 28 | 27 | 27 | 26 | 26 | 25 | 25 | 24 | 24 | 23 | 23 | 23 | |
| | 1166 | **27** | 5 | 17 | 17 | 16 | 16 | 15 | 15 | 14 | 14 | 13 | 13 | 12 | 12 | |
| | 1165 | **28** | 6 | 6 | 6 | 6 | 5 | 5 | 4 | 4 | 3 | 3 | 2 | 2 | 1 | 1 |
| | 1164 | **29** | 7 | 25 | 25 | 24 | 24 | 23 | 23 | 23 | 22 | 22 | 21 | 21 | 20 | |
| | 1163 | **30** | 8 | 15 | 14 | 14 | 13 | 13 | 12 | 12 | 11 | 11 | 10 | 10 | 10 | |
| | 1162 | **31** | 9 | 4 | 4 | 3 | 3 | 2 | 2 | 1 | 1 | 30 | 29 | 29 | 28 | |
| | 1161 | **32** | 10 | 23 | 23 | 22 | 22 | 21 | 21 | 20 | 20 | 19 | 19 | 18 | 18 | |
| **Ramses IV** | 1160 | **1** | 11 | 12 | 12 | 11 | 11 | 10 | 10 | 10 | 9 | 9 | 8 | 8 | 7 | |
| | 1159 | **2** | 12 | 2 | 1 | 1 | 30 | 30 | 29 | 28 | 28 | 27 | 27 | 27 | 26 | |
| | 1158 | **3** | 13 | 21 | 20 | 20 | 19 | 19 | 18 | 18 | 17 | 17 | 16 | 16 | 15 | |
| | 1157 | **4** | 14 | 10 | 9 | 9 | 9 | 8 | 8 | 7 | 7 | 6 | 6 | 5 | 5 | 4 |
| | 1156 | **5** | 15 | 29 | 28 | 28 | 27 | 27 | 27 | 26 | 26 | 25 | 25 | 24 | 24 | |
| | 1155 | **6** | 14 | 18 | 18 | 17 | 17 | 16 | 16 | 15 | 15 | 14 | 14 | 14 | 13 | |
| | 1154 | | 15 | 8 | 7 | 7 | 6 | 6 | 5 | 5 | 4 | 4 | 3 | 3 | 2 | 2 |

The table above may be used to check possible coincidences of dates. For example, the helical rising of Sirius is dated July 12 at the time of Sety I (around -1300)[165], which corresponds to the I Akhet 3 in 1284 BCE[166], year 10 of Sety I. The lunar day *psḏntyw* for Egyptian month called Akhet has to be dated I Akhet 2 in year 10 of Sety I, which corresponds to July 11, 1284 BCE (full moon)[167].

| | | | | Full moon | | Sothic rising | |
|---|---|---|---|---|---|---|---|
| **Sety I** | 1294 | | 2 | | | | |
| | 1293 | **1** | 3 | I Akhet 9 | July 20 | I Akhet 1 | July 12 |
| | 1292 | **2** | 4 | I Akhet 28 | August 8 | I Akhet 1 | July 12 |
| | 1291 | **3** | 5 | I Akhet 17 | July 28 | I Akhet 1 | July 12 |
| | 1290 | **4** | 6 | I Akhet 6 | July 17 | I Akhet 1 | July 12 |
| | 1289 | **5** | 7 | I Akhet 25 | August 4 | I Akhet 2 | July 12 |
| | 1288 | **6** | 8 | I Akhet 15 | July 25 | I Akhet 2 | July 12 |
| | 1287 | **7** | 9 | I Akhet 4 | July 14 | I Akhet 2 | July 12 |
| | 1286 | **8** | 10 | I Akhet 23 | August 2 | I Akhet 2 | July 12 |
| | 1285 | **9** | 11 | I Akhet 12 | July 21 | I Akhet 3 | July 12 |
| | 1284 | **10** | 12 | I Akhet 2 | July 11 | I Akhet 3 | July 12 |
| | 1283 | **11** | 13 | I Akhet 21 | July 30 | I Akhet 3 | July 12 |

Synchronisms (below) between Egyptian, Babylonian and Israelite chronologies are in perfect agreement (highlighted). Astronomical dates have been highlighted in blue sky

165 At Thebes (Longitude 32°39' Latitude 25°42') with an *arcus visionis* of 8.7 the Sothic rising is dated 12 July on the period 1370-600 http://www.imcce.fr/fr/grandpublic/phenomenes/sothis/index.php
166 http://www.chronosynchro.net/wordpress/convertisseur
167 http://www.imcce.fr/fr/grandpublic/phenomenes/phases_lune/index.php

| Assyria | Reign | Babylon | Reign | Egypt | Reign |
|---|---|---|---|---|---|
| Erîba-Adad I | 1385-1358 | Kadašman-Enlil I | 1375-1360 | Amenhotep III | 1383-1345 |
| Aššur-uballiṭ I | 1358 - | Burna-Buriaš II | 1360 - | Akhenaton | 1356-1340 |
| | | | | Semenkhkare | 1340-1338 |
| | | | -1333 | -*Ankhkheperure* | 1338-1336 |
| | | Kara-ḫardaš | 1333 | Tutankhamon | 1336 - |
| | | Nazi-Bugaš | 1333 | | -1327 |
| | -1323 | Kurigalzu II | 1333 - | Aÿ | 1327-1323 |
| Enlil-nêrârî | 1323-1313 | | -1308 | Horemheb | 1323-1309 |
| Arik-dên-ili | 1313-1302 | Nazi-Maruttaš | 1308 - | | 1309-1295 |
| Adad-nêrârî I | 1302 - | | | Ramses I | 1295-1294 |
| | | | -1282 | Sethy I | **1294**-1283 |
| | -1271 | Kadašman-Turgu | 1282 - | Ramses II | **1283** - |
| Shalmaneser I | 1271 - | | -1264 | | |
| | (-1264) | Kadašman-Enlil II | **1264**-1255 | | (-1264) |
| | -1242 | Kudur-Enlil | 1255-1246 | | |
| Tukultî-Ninurta I | **1242** - | Šagarakti-šuriaš | 1246-1233 | | (-1242) |
| | | Kaštiliašu IV | 1233-**1225** | | |
| | | Enlil-nâdin-šumi | 1225-1224 | | |
| | | Kadašman-Harbe II | 1224-1223 | | |
| | | Adad-šuma-iddina | 1223-1217 | | -1216 |
| | -1206 | Adad-šuma-uṣur | 1217 - | Merenptah | 1216-1207 |
| Aššur-nâdin-apli | 1206 -1202 | | | Sethy II | 1207-1202 |
| Aššur-nêrârî III | 1202-1196 | | | Siptah | **1202**-1196 |
| Enlil-kudurri-uṣur | 1196 - | | | -*Tausert* | 1196-1194 |
| | -1191 | | -1187 | Sethnakht | 1196-1192 |
| Ninurta-apil-Ekur | 1191-1179 | Meli-Šipak | **1187**-1172 | Ramses III | **1192** - |
| Aššur-dân I | 1179 - | Marduk-apla-iddina | 1172-1159 | | -1161 |
| | | Zababa-šuma-iddina | 1159-1158 | Ramses IV | **1161** - |
| | | Enlil-nâdin-ahi | 1158-1155 | | -1155 |
| | | Marduk-kabit-aḫḫešu | 1155 - | Ramses V | 1154-1151 |
| | | | -1141 | Ramses VI | 1151-1144 |
| | | Itti-Marduk-balaṭu | 1141 - | Ramses VII | 1144-1137 |
| | | | | Ramses VIII | 1137 |
| | **-1133** | | **-1133** | Ramses IX | 1137-1119 |
| | | | | Ramses X | 1119-1116 |
| | | | | Ramses XI | 1116-1090 |
| | | | | Smendes | 1090-1064 |
| | | Israel | Reign | [Amenemnesut] | [1064-1060] |
| | | David | 1057-1017 | Psusennes I | 1064-1018 |
| | | Solomon | 1017 - | Amenemope | 1018-1009 |
| | | | | Osorkon the Elder | 1009-1003 |
| | | | (-993) | Siamun | 1003 - 984 |
| | | | -977 | Psusennes II/III | 994-980 |
| | | Rehoboam | 977-960 | Shoshenq I | 980-959 |
| | | Asa | 957 - | Osorkon I | 959-924 |
| | | | | Shoshenq II | 924-922 |
| | | | | Shoshenq IIb | -922 |
| | | | -916 | Takelot I | 922-909 |
| | | Jehoshaphat | 916-891 | Osorkon II | 909 - |
| | | Jehoram | 893-885 | | |
| | | [Athaliah] Jehoyada | 885-879 | | -865 |
| | | Joash | 879-839 | Takelot II | **865**-840 |
| | | Amasiah | 839-810 | Shoshenq III | 840-800 |
| | | Uzziah [Azariah] | 810 - | Shoshenq IV | 800-788 |
| Aššur-dân III | **773**-755 | | -758 | Pamiu | 788-782 |
| Aššur-nêrârî V | 755-745 | Jotham | 758-742 | Shoshenq V | 782-745 |
| Tiglath-pileser III | 745-727 | Ahaz | 742-726 | Osorkon IV | 745 - |
| Shalmaneser V | 727-722 | Hezekiah | 726 - | | -712 |
| Sargon II | 722-705 | | -697 | Chabataka/Taharqa | 712-690 |

### LUNAR DAY 1 (*PSḎNTYW*) MATCHES FULL MOON

As lunar day 1 (*psḏntyw*) has played a major role in Egyptian religious celebrations, it is regularly quoted in ancient documents, which sometimes also date it in the civil calendar. This double-dating then allows an absolute dating, on condition that provided proper identification of the moon phase for that particular day. Present specialists rely on the work of Parker (in 1950) who defined this day as a first invisibility, that is to say the day (invisible!) just before the first lunar crescent. In fact, the findings of Parker are only based on a single document: the papyrus Louvre 7848 containing a double date (lunar and civil) in the year 44 of Amasis. Parker's work was validated later thanks to the Elephantine papyri (5th century BCE) containing several double dates in agreement with the Babylonian chronology (from king lists) for the reigns of Darius, Xerxes and Artaxerxes. However, this official Babylonian chronology is incorrect since it ignores 10 years of co-regency of Xerxes (496-475) with Darius (522-486) and 8 years of Darius B (434-426), the analysis of the Louvre Papyrus 7848 has to be redone.

The year 44 of Amasis, the last of his reign should be dated 526 BCE, and therefore the year 12 to be dated 528 BCE. Double-dated documents are rare, they are all the more valuable since they allow absolute dating, which is the case of the following papyrus (pap. Louvre 7848)[168] both dated II Shemu 13 / I Shemu 15, Year 12 of Amasis (line 5):

| | | |
|---|---|---|
| 1 | *Ḥ3.t-sp 12.t ibd-1 šmw (sw) 21*<br>*⸢n⸣ Pr-ꜥ3 Iꜥḥ-ms ꜥ.w.s.*<br>*Ḏd w3ḥ-mw P3-dj-Wsir s3*<br>*Ir.t.w-rṯ (a) s3 Ir.t-Ḥr-r=w*<br>*[mw.t.ṯ]⸢=f⸣ Ir.t.w-r⸢=w⸣* | *Year 12, 1st month of Shemu, (day) 21*<br>*under Pharaoh Amasis life-prosperity-health (...)*<br>*Has said the choachyte Petosiris son of*<br>*Itourodj son of Inarou, his*<br>*[mother] being Itourou,*<br>(choachyte = mummies guardian) |
| 4 | *T3-Ḫrw (d) ḥnꜥ w3ḥ-mw Ḏd-ḥj*<br>*s3 Dj-s-Mnṯ dmḏ s (c) 3:*<br>*inn dj ⸢ꜥrḳ⸣ n=n w3ḥ-mw (e)*<br>*P3-dj-⸢Wsir⸣ (f) s3 Ir.t.w-rṯ* | *Tacherou and the choachyte Djechy*<br>*son of Tesmont, total 3 men:*<br>*"It is we who have caused the choachyte*<br>*Petosiris son of Itourodj to swear for us* |
| 5 | *m-b3ḥ Ḫnsw-m-W3s.t-Nfr-ḥtp n*<br>*ḥ3.t-sp 12.t ibd-2 šmw (sw) 13*<br>*n 15.t (g) ibd-1 šmw ḏd:*<br><br>*t3 s.t p3 ḏw r.ḏd(=j) šp(=j)* | *in the presence of Chonsemwasneferhotep, in*<br>*year 12, 2nd month of Shemu, (day) 13,*<br>*on the 15th day of the 1st month of Shemu, saying:*<br><br>*'The place of the mountain, of which I said: «I have received* |

Parker assumed that the first date was from the civil calendar and the second from the lunar calendar, but it is illogical for the following reasons:

- Egyptian lunar dates being exceptional they should be specified in the civil calendar and not the opposite. Among the twenty papyrus from Elephantine in southern Egypt, which contain double dates, all begin with the date of the lunar calendar followed by that of the Egyptian civil calendar, but never the reverse.
- "*It is we who have caused the choachyte to swear for us*" refers to the past not to the future ("*It is we who will cause the choachyte to swear for us*"). If this vow was recorded and dated, it is logical to assume that it was written relatively soon after having been delivered, otherwise one would admit the existence of a "prophetic vow", but the document being dated I Shemu 21 in the civil calendar, the vow had to be made on I Shemu 15, actually 6 days before.

[168] K. DONKER VAN HEEL – Abnormal Hieratic and Early Demotic Texts collected by the Theban Choachytes in the reign of Amasis: Papyrus from the Louvre Eisenlohr Lot (Thesis). Leiden 1996 Ed. Rijksuniversiteit pp. 93-99.

- As the lunar year is shorter than the solar year (the lunar month being 29 or 30 days when the Egyptian civil month is always 30 days), dating in a lunar calendar goes faster than in the civil calendar, thus the lunar dates are more advanced (II Shemu) than those of the civil calendar (I Shemu).

According to these logical arguments, the first date (II Shemu 13) is lunar and the second (I Shemu 15) is civil. As the civil date *I Shemu 15* fell in -558 on September 21, the lunar date *II Shemu 1* fell on September 9 (= 21 – 12), which was a full moon day according to astronomy[169]. However, there are two difficulties in reckoning the days:

- The Babylonians counted the new day after sunset (around 18 pm) while the Egyptians counted it after the disappearance of the stars (around 5 am). If a scribe wrote on 17 Thoth around 16 pm, for example, he dated his document on 18 Kislev, but if he wrote about 20 pm he dated it on 19 Kislev.

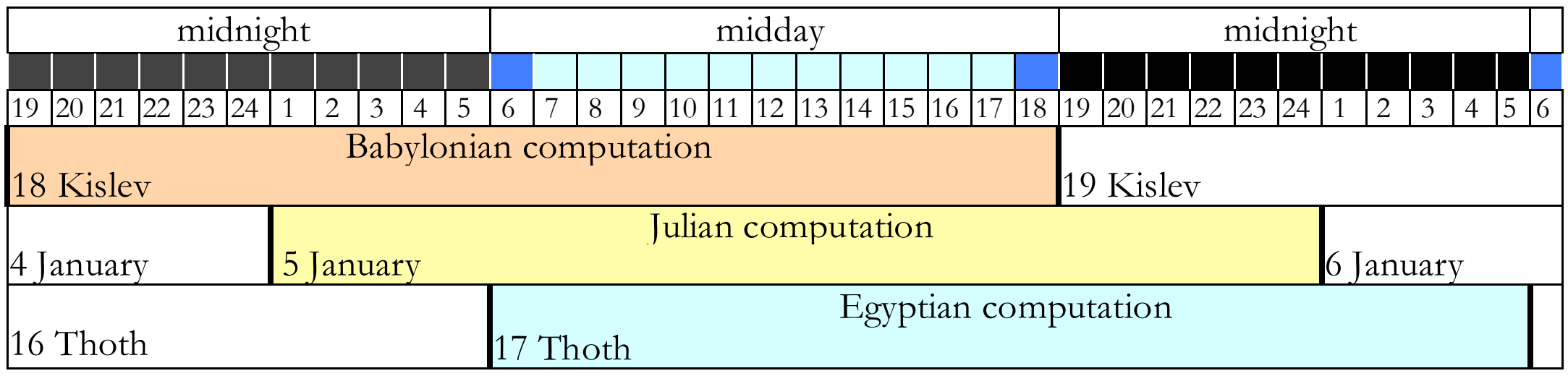

- Astronomical observations being made by night, at the beginning of the day for the Babylonians, but at the end for the Egyptians. At last, the observation of the first crescent can be delayed by one day (due to bad weather, for example) while watching the full moon can be shifted more or less one day.

According to this lunar calendar, the two papyrus double dated years 15 and 21 of Xerxes[170] involve an accession in 496 BCE (the full moon of 1st Elul fell on August 29 in 481 BCE at Elephantine and the full moon of 1st Kislev fell on December 20 in 475 BCE):

| **Year** | Xerxes I | Civil Egyptian | Julian | Lunar Egyptian | Julian | Gap |
|---|---|---|---|---|---|---|
| | | | | 1st Elul | 29 August | (full moon) |
| **481** | **15** | 28 Pakhons | 14 September | 18 Elul | 15 September | 1 |
| | | | | 1st Kislev | 20 December | (full moon) |
| **474** | **21** | 17 Thoth | 5 January | 18 Kislev | 5 January | 0 |

When Porten published the Elephantine papyri he wrote: *The language, religion, and names of the Jews differed from their Egyptian neighbours, but their legal procedures and formulary bear striking similarity. Though we cannot explain the phenomenon of "Who gave to whom" we must conclude that in matters legal the Jews and Arameans fit into their Egyptian environment rather snugly. Whereas the demotic contracts constitute a little over 20% of the thirty-seven demotic texts here published, the Aramaic contracts constitute almost 60% of the total Aramaic selection of fifty-two documents. If thirty documents are ample material to ascertain schemata and verify formulae, eight may not be, particularly if they are of different types. Comparison, nonetheless, shows how much the demotic and Aramaic conveyances had in common. Both followed an identical schema (...) Variations were slight. As indigenous documents,*

[169] http://www.imcce.fr/fr/grandpublic/phenomenes/phases_lune/index.php
[170] B. PORTEN - The Elephantine Papyri in English
Leiden 1996 Ed E.J. Brill pp. 18, 153-161.

*the demotic contracts noted only the Egyptian calendar, whereas the Jewish/Aramean scribes, writing in the* lingua franca *of the Persian Empire, added for most of the fifth century a synchronous Babylonian date.* This last remark contradicts what was said at the beginning because the Egyptians never used a Babylonian calendar in Egypt. In addition, Porten fails to mention that several Babylonian dates have a gap of 2 days (which is difficult to explain by errors of scribes), or even a month apart (B32 and B42 for example), and that lunar calendar was closer to the Jewish or Aramaic calendar than its Babylonian counterpart[171]. Stern[172] noted: *This explanation has been fully endorsed by Porten, but it is problematic in more than one respect. In the ancient world, where artificial lighting was often expensive and/or inadequate, scribes would have been reluctant to write legal documents at night: legal documents, indeed, had to be written with precision and care. Although such a practice was possible — as Porter points out, the Mishna refers to legal documents written at night (M. Gittin 2:2), and further evidence could conceivably be found — it seems unlikely that the majority of contracts at Elephantine would have been written at night (...) In order to account for this high incidence of discrepancies, it seems more plausible to argue that the Babylonian calendar at Elephantine was reckoned differently from the standard Babylonian calendar. How it was reckoned, however, remains somewhat unclear. The inconsistent relationship between document dates and visibility of the new moon (nil, 1 day, or 2 days) suggest perhaps that at Elephantine, visibility of the new moon was not used as a criterion to determine when the new month began.* The solution was at hand, but Stern did not know that the problem stemmed from the wrong interpretation of Parker. This is particularly more regrettable that Parker had given all the elements to find it.

Parker refused to consider a lunar reckoning starting at full moon, as proposed by Macnaughton[173], for three reasons:

- He felt that Macnaughton was an eccentric[174] (no comment!).
- This type of calendar was not well known during his time. Parker was unaware that the Hindu lunar calendar, for example, is equally divided between *amanta* versions (8 states in southern India) which start on new moon and *purnimanta* versions (10 states in the northern India) starting on full moon. In addition, it is likely that some ancient lunar calendars began on the full moon, like the Old Persian calendar whose 30[th] day is called *jiyamna* "decreasing", that would be inexplicable if the lunar cycle began on 1[st] crescent.
- Lunar phases being symbolized at Dendera (around -50) by 14 deities climbing stairs toachieve the filling of the eye Wedjat[175] (safe eye) the 15[th] day at the full moon, the lunar day 1 (*psḏntyw*) must match the 1[st] invisibility. But this cycle of 15 days is only a ½ month, the next full month had to begin at the end of this cycle, that is at the full moon.

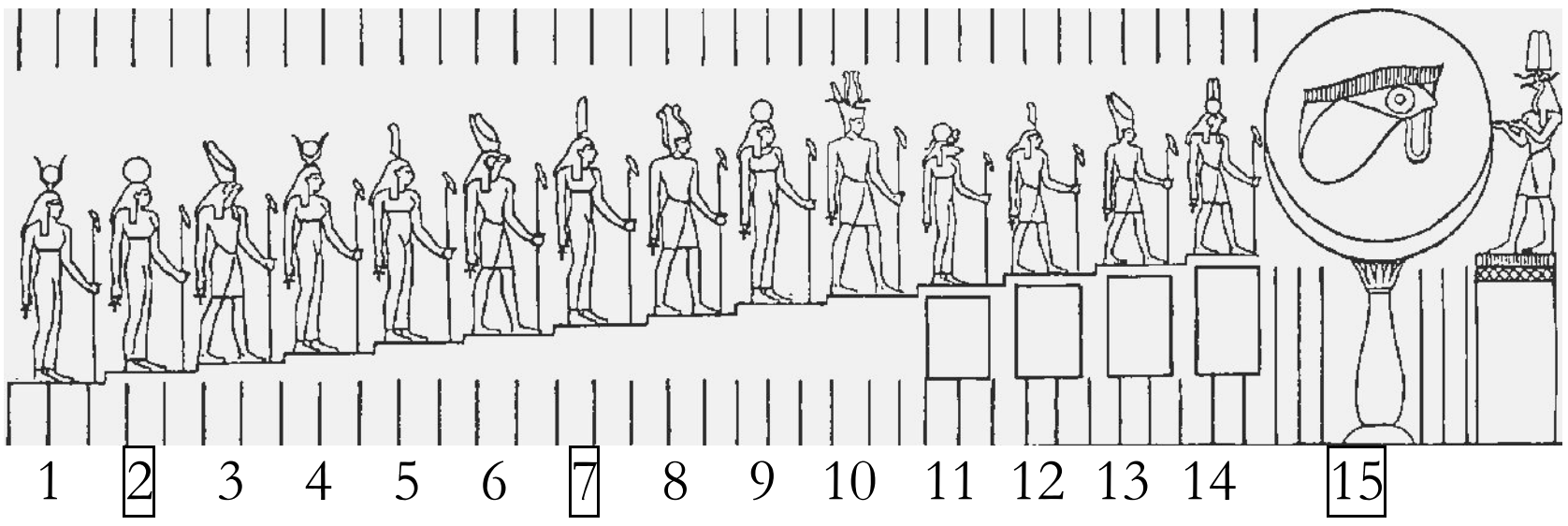

[171] S. STERN - The Babylonian Calendar at Elephantine
in: *Zeitschrift für Papyrologie und Epigraphik* 130 (2000) pp. 159-171.
[172] S.H. HORN, L.H. WOOD - The Fifth-Century Jewish Calendar at Elephantine
in: *Journal of Near Eastern Studies* XIII/1 (1954) pp. 1-20.
[173] D. MACNAUGHTON - A Scheme of Egyptian Chronology
London 1932 Ed. Luzac and co. pp. 145-151.
[174] R.A. PARKER - The Calendars of Ancient Egypt
in: *Studies in Ancient Oriental Civilization* 26 (1950) Ed. University of Chicago p. 9.
[175] E.A.W. BUDGE - Gods of the Egyptian Vol II
1969 Ed. Dover Publications p. 321.

| Babylonian lunar cycle | | Egyptian lunar cycle according to: | | | |
|---|---|---|---|---|---|
| astro | | ½ month | **Parker** | **Macnaughton** | astro |
| ○ | 14 full moon | | | **1** shining ones [day] | ○ |
| ○ | 15 | | | 2 month [day] | ○ |
| | 16 | | | 3 | |
| | 17 | | | 4 | |
| | 18 | | | 5 | |
| | 19 | | | 6 | |
| | 20 | | | 7 quarter [day] | |
| ◑ | 21 last quarter | | | 8 | ◑ |
| | 22 | | | 9 | |
| | 23 | | | 10 | |
| | 24 | | | 11 | |
| | 25 | | | 12 | |
| | 26 | | | 13 | |
| ☾ | 27 last crescent | | | 14 perceptions [day] | ☾ |
| ● | 28 | | | 15 subordinate [day] | ● |
| ● | 29 new moon | | | 16 | ● |
| ● | 30 1st invisibility | 1 *tp 3bd* | 1 shining ones [day] ϟ | 17 perceptions [day] | ● |
| ☽ | 1 1st crescent | **2** *3bd* | **2** month [day] | **18** Moon [day] | ☽ |
| | 2 | 3 | 3 | 19 | |
| | 3 | 4 | 4 | 20 | |
| | 4 | 5 | 5 | 21 | |
| | 5 | 6 *snt* | 6 | 22 | |
| ◐ | 6 first quarter | **7** *dnit* | **7** quarter [day] | 23 quarter [day] | ◐ |
| | 7 | 8 | 8 | 24 | |
| | 8 | 9 | 9 | 25 | |
| | 9 | 10 | 10 | 26 | |
| | 10 | 11 | 11 | 27 | |
| | 11 | 12 | 12 | 28 | |
| | 12 | 13 | 13 | 29 | |
| ○ | 13 | 14 | 14 perceptions [day] | 30 Min rise [day] | ○ |
| ○ | 14 full moon | **15** *smdt* | **15** subordinate [day] | | ○ |
| ○ | 15 *šapattu* | | 16 | | ○ |
| | 16 | | 17 perceptions [day] ϟ | | |
| | 17 | | **18** Moon [day] ϟ | | |
| | 18 | | 19 | | |
| | 19 | | 20 | | |
| | 20 | | 21 | | |
| ◑ | 21 last quarter | | 22 | | ◑ |
| | 22 | | **23** quarter [day] | | |
| | 23 | | 24 | | |
| | 24 | | 25 | | |
| | 25 | | 26 | | |
| | 26 | | 27 | | |
| ☾ | 27 last crescent | | 28 | | ☾ |
| ● | 28 | | 29 | | ● |
| ● | 29 new moon | | 30 Min rise [day] ϟ | | ● |
| ● | 30 | | **1** | | ● |

Parker has compiled and explained the 30 days of the Egyptian lunar month, which shows that several days do not fit at all with their Moon phases.

| ½ month | n° | Day of the month | | Moon phase according to: | |
|---|---|---|---|---|---|
| | | **Name** | **Meaning** | **Macnaughton** | **Parker** |
| *(15)* | 1 | *psḏntyw* | Shining ones | Full moon | First invisibility |
| | 2 | *3bd* | Month | After full moon | First crescent |
| | 7 | *dnit* | Quarter | Last quarter | First quarter |
| | 14 | *si3w* | Perceptions | Last crescent | Before full moon |
| | 15 | *smdt* | Subordinate | Before new moon | Full moon |
| *1* | 17 | *si3w* | Perceptions | Before first crescent | - |
| *2* | 18 | *i'ḥ* | Moon | First crescent | - |
| *7* | 23 | *dnit* | Quarter | First quarter | Last quarter |
| *14* | 30 | *prt Mn* | Min going-forth | Before full moon | New moon |

In Parker's lunar cycle it is obvious that the meaning of days 1 (*psḏntyw*) and 18 (*i'ḥ*) has nothing to do and even opposed to the lunar phase that corresponds to them. The Egyptian word *psḏntyw* literally means "shining ones" which is opposed to its moon phase (after the new moon) called "first invisibility". In addition the day 18 which literally means "moon" would have no link with the lunar cycle, which would be the last straw. According to Depuydt[176]: *There is little doubt as to what ancient Egyptians saw of the moon on the day they called* psḏntyw *the first of the lunar month (...) Parker has done the most to consolidate the theory of* psḏntyw *outlined above. Yet the view that Egyptian lunar months began with the observation of nothing has met with resistance. Černy and Posener believed that the passage from Theban Tomb 57 quoted above "shows that it was possible to depict* psḏntyw *... For the Egyptians,* psḏntyw *was therefore something visible ... Indeed, it would be difficult to understand how the Egyptians could have conceived of 'moon on* psḏntyw*' ... if* psḏntyw *was an invisible celestial phenomenon." This remark disregards the fact, however, that "moon on* psḏntyw*" is modified by "whose brightness has illuminated the netherworld" (...) "you set like Re on the day of* psḏntyw*".* To summarize his arguments, the Egyptian day 1 (*psḏntyw*) would represent both the invisibility of the moon for the living ones and the sun illuminating the netherworld, but this explanation is more theological than scientific.

Year 10 of Amasis (in -560) that began on I Akhet 1 (January 10) coincided with a full moon, which involved the starting equivalence I Akhet 1 (lunar) = I Akhet 1 (civil). It is noteworthy that the observation of the full moon is more difficult than the 1st lunar crescent, because depending on the time of day or night the 1st astronomical crescent may be seen with a day late (but never in advance) so that the full astronomical moon can be seen frequently with one day difference (delay or advance).

| **Amasis year** | | Lunar calendar (day 1) | Civil calendar | Julian day | Full moon (astronomy) |
|---|---|---|---|---|---|
| **10** | **-560** | **I Akhet 1** | **I Akhet 1** | 10 January | 9 January |
| | | II Akhet **1** | **I Akhet 30** | 8 February | 7 February |
| | | III Akhet **1** | II Akhet **30** | 10 March | 9 March |
| | | IV Akhet **1** | III Akhet **29** | 8 April | 8 April |
| | | I Peret **1** | IV Akhet **29** | 8 May | 7 May |
| | | II Peret **1** | I Peret **28** | 6 June | 6 June |
| | | III Peret **1** | II Peret **28** | 6 July | 6 July |
| | | IV Peret **1** | III Peret **27** | 4 August | 4 August |
| | | I Shemu **1** | IV Peret **27** | 3 September | 2 September |
| | | II Shemu **1** | I Shemu **26** | 2 October | 2 October |
| | | III Shemu **1** | II Shemu **25** | 1 November | 1 November |
| | | IV Shemu **1** | III Shemu **25** | 30 November | 30 November |
| | | I Akhet **1** | IV Shemu **25** | 30 December | 29 December |

176 L. DEPUYDT - The Hieroglyphic Representation of the Moon's Absence (*Psḏntyw*) in: Ancient Egyptian and Mediterranean Studies (1998) Ed. L.H. Lesko pp. 71-89.

| | | | | | |
|---|---|---|---|---|---|
| **11** | **-559** | II Akhet **1** | I Akhet **19** | 28 January | 28 January |
| | | III Akhet **1** | II Akhet **19** | 27 February | 26 February |
| | | IV Akhet **1** | III Akhet **18** | 28 March | 28 March |
| | | I Peret **1** | IV Akhet **18** | 27 April | 27 April |
| | | II Peret **1** | I Peret **17** | 26 May | 26 May |
| | | III Peret **1** | II Peret **17** | 25 June | 25 June |
| | | IV Peret **1** | III Peret **16** | 24 July | 24 July |
| | | I Shemu **1** | IV Peret **16** | 23 August | 23 August |
| | | II Shemu **1** | I Shemu **15** | 21 September | 21 September |
| | | III Shemu **1** | II Shemu **15** | 21 October | 21 October |
| | | IV Shemu **1** | III Shemu **14** | 19 November | 19 November |
| | | I Akhet **1** | IV Shemu **14** | 19 December | 19 December |
| **12** | **-558** | II Akhet **1** | I Akhet **8** | 17 January | 17 January |
| | | III Akhet **1** | II Akhet **8** | 16 February | 16 February |
| | | IV Akhet **1** | III Akhet **7** | 17 March | 17 March |
| | | I Peret **1** | IV Akhet **7** | 16 April | 16 April |
| | | II Peret **1** | I Peret **6** | 15 May | 15 May |
| | | III Peret **1** | II Peret **6** | 14 June | 14 June |
| | | IV Peret **1** | III Peret **5** | 13 July | 13 July |
| | | I Shemu **1** | IV Peret **5** | 12 August | 12 August |
| | | II Shemu **1** | I Shemu **4** | 10 September | 10 September |
| | | II Shemu 13 | I Shemu 16 | 22 September | |
| | | III Shemu **1** | II Shemu **4** | 10 October | 10 October |
| | | IV Shemu **1** | III Shemu **3** | 8 November | 9 November |
| | | I Akhet **1** | IV Shemu **3** | 8 December | 8 December |

According to this table: II Shemu 13 (Egyptian lunar calendar) = I Shemu 16 (Egyptian civil calendar) = 22 September (Julian calendar). If the full moon was seen on September 9, instead of 10, we have: II Shemu 13 (Egyptian lunar calendar) = I Shemu 15 (Egyptian civil calendar) = 21 September (Julian calendar).

### ELEPHANTINE CALENDARS

The calendar at Elephantine with its system of double dates (Egyptian and Babylonian) was used by Persians officials and Jewish scribes only during a short period from -500 to -400. For example, a Persian official erected a votive stele stating: *This temple, (W)id(arnaga) head of the garrison at Syene was done in the month of Siwan, that is to say Mehir, year 7 of King Artaxerxes, (to) Osirnaḥty, the god. Peace*[177].

After the conquest of Egypt by Cambyses it became a Persian satrapy but most of the scribes were Egyptians or Jews. According to Herodotus (The Histories II:152-154), Psammetichus I, dynasts of Sais, called on foreign mercenaries, including Ionians and Carians, to consolidate his power in Egypt. The pharaoh then installed these mercenary garrisons in Daphne west of Delta, and Elephantine, on the border in the south (The Histories II:30-31). The Letter of Aristeas to Philocrates III:13 states that among these mercenaries there were Jews. According to the biblical text, the massive emigration of Jews into Egypt began shortly after the pharaoh Necho II established King Jehoiakim (in -609) on the throne in Jerusalem (2 Kings 23:34, Jeremiah 26:21-23, 42:14). After the murder of Gedaliah, many of these Jews emigrated to Egypt (Jeremiah 43:7, 44:1) especially in the country of Patros (meaning "the Land of the South" in Egyptian) the southern province in which Elephantine was the main town.

[177] A. LEMAIRE – Recherches d'épigraphie araméenne en Asie mineure et en Égypte in: *Achaemenid History* V (1991) Ed. Nederlands Instituut Leiden pp.199-201.